\documentclass[fleqn,usenatbib]{mnras}
\usepackage{newtxtext,newtxmath}
\usepackage[T1]{fontenc}

\usepackage{multirow}
\usepackage{makecell} 
\usepackage{subcaption}
\usepackage{ulem}
\usepackage{soul}
\usepackage{graphicx}	%
\usepackage{amsmath}	
\usepackage{subcaption} 

\newcommand{\msun}{\,M$_{\odot}$}

\title[X-ray cavities in TNG-Cluster: a direct comparison to observations]{X-ray cavities in TNG-Cluster: a direct comparison to observations}

\author[M. Prunier et al.]{
Marine Prunier,$^{1,2,3}$\thanks{E-mail: marine.claude.anne.prunier@umontreal.ca}
Annalisa Pillepich,$^{3}$, Julie Hlavacek-Larrondo,$^{1,2}$ and Dylan Nelson$^{4}$
\\
$^{1}$Département de Physique, Université de Montréal, Succ. Centre-Ville, Montréal, Québec, H3C 3J7, Canada\\
$^{2}$Centre de recherche en astrophysique du Quebec (CRAQ)\\
$^{3}$Max-Planck-Institut f{\"u}r Astronomie, K{\"o}nigstuhl 17, D-69117 Heidelberg, Germany\\
$^{4}$Universit\"{a}t Heidelberg, Zentrum f\"{u}r Astronomie, ITA, Albert-Ueberle-Str. 2, 69120 Heidelberg, Germany} 

\date{}
\pubyear{\the\year{}}

\begin{document}
\label{firstpage}
\maketitle

\begin{abstract} 
The TNG-Cluster magnetohydrodynamic cosmological simulations, produce a diverse population of X-ray cavities in the intracluster medium (ICM) of simulated galaxy clusters. These arise from episodic, high velocity, kinetic energy injections from the central active supermassive black hole (AGN, SMBH). Here, we present the first comprehensive comparative analysis of X-ray cavities in TNG-Cluster with observational data. First, we select a volume-limited sample of 35 real clusters ($z \leq 0.071$, M$_\text{500c}$ = 10$^{14-14.8}$ M$_\odot$) observed with the Chandra X-ray Observatory, identify 3 analogs for each in TNG-Cluster (total of 105) and generate mock Chandra images using same exposure times as their observed counterparts. We identify X-ray cavities and measure their properties in both datasets using identical techniques, ensuring a direct, apples-to-apples comparison. Our analysis reveals that both samples have a similar fraction of X-ray cavities (35-43 per cent). They exhibit comparable sizes and morphologies, although the sizes of simulated X-ray cavities still attached to the SMBH are somewhat larger in TNG-Cluster -- a scarcity at $< 10$ kpc. The area of TNG X-ray cavities increases as they rise in the ICM, consistent with the trend of the observational sample. The cavity powers, estimated using observational techniques, show good agreement between the two samples (10$^{42-45}$ erg~s$^{-1}$), suggesting that X-ray cavities in the simulation are an important heating mechanism in cluster cores. Overall, the rather simple AGN feedback model of TNG, with no model choices made to reproduce X-ray morphological features, and without cosmic rays, creates a quantitatively realistic population of X-ray cavities at cluster scales. \end{abstract}


\section{Introduction} \label{sec:intro}

Galaxy clusters -- hereafter clusters -- consist of hundreds to thousands of galaxies. Complex physical processes occur within the hot plasma that fills the space between them, known as the intracluster medium (ICM). This dense and extreme environment leaves a lasting imprint on galaxies, driving morphological transformations, environmental quenching, and distinctive patterns in the galaxy distribution \citep[e.g.,][]{2021CorteseReview}. In fact, multi-wavelength observations have shown that the thermodynamic properties of clusters are influenced not only by gravitational dynamics but also by feedback from powerful supermassive black holes (SMBHs), known as active galactic nuclei (AGN). Located in the brightest cluster galaxy (BCG), the active SMBH impacts the cluster-wide gaseous halo through relativistic jets, winds, and radiation, and is often associated with prominent X-ray depleted regions in the ICM known as X-ray cavities \citep[also called bubbles; see][for a recent review]{hlavacek-larrondo_agn_2022}.

Numerous observational studies have investigated the properties of X-ray cavities seen in massive elliptical galaxies and at the centers of galaxy groups and clusters. The Chandra X-ray Observatory -- hereafter Chandra -- with its high-resolution imaging, has been instrumental in revealing their characteristics. Observationally, X-ray cavities appear as prominent, X-ray surface brightness depressions projected along the line of sight, with sizes ranging from a few kiloparsecs \citep[kpc; e.g., A4059,][]{2008Reynolds} to hundreds of kpc \citep[e.g., MS0735+7421,][]{2005McNamara}. They are often observed in symmetric pairs around the central SMBH, and some systems, such as Perseus \citep{2000Fabian_radio_erseus}, host multiple generations of X-ray cavities on different spatial scales. Clusters with both radio-filled and ghost X-ray cavities -- i.e. no longer emitting in the high-frequency radio -- suggest that the central galaxy has undergone several feedback events, with the ghost X-ray cavities resulting from buoyantly rising lobes plausibly inflated during past outburst episodes. Some X-ray cavities are observed with bright rims at their edges, where the gas is thought to be compressed \citep{2008Salome}. Strong shocks at their boundaries are rarely detected \citep[only two known systems: Centaurus A with Mach number $\sim$8 and NGC 3801 with Mach number $\sim$4,][]{2003Kraft,2007Croston}, but some systems, such as Hydra A, exhibit weak shock fronts with Mach number $< 2$ \citep[][]{Nulsen2005}. 

Population studies \citep[such as][]{2008diehl,2010Dong,2016Shin} suggest that X-ray cavities rise and expand in the ICM adiabatically, as inferred from the observed correlation between X-ray cavity sizes and their distance from the central SMBH. X-ray cavities are also believed to play a critical role in regulating star formation. Observations show that star formation rates in cluster cores are much lower than expected from ICM cooling, referred to as the ``cooling catastrophe'' \citep[e.g.,][]{peterson_x-ray_2006, bohringer_x-ray_2010,fabian_observational_2012}. AGN feedback, particularly through X-ray cavities, is thought to be the primary mechanism balancing the cooling of the intracluster gas (and effectively of the gas within the BCG), with the power of these cavities strongly correlating with the rate of radiative cooling \citep[e.g.,][]{2004Birzan,2006Rafferty,hlavacek-larrondo_agn_2022}. However, recent work challenges this view \citep[e.g.][]{Fabian_hiddend1,Fabian_hiddend2,Fabian_hiddend3,Fabian_hiddend4}, where ``hidden'' cooling flows in ellipticals, groups and clusters imply mass cooling rates $20-50$ per cent higher than previously inferred from X-ray studies. These findings offer a fresh perspective on the cooling-flow problem, indicating that AGN feedback may not require the precise fine-tuning previously assumed.

In fact, how the energy inferred from X-ray cavities couples with the ICM remains uncertain and may involve the dissipation of sound waves or weak shocks generated during their formation \citep[e.g.,][]{Fabian2008_perseusWS,2018Tang,2019Bambic}. Additionally, the volume-filling gas, if hot and composed of relativistic particles, might heat via thermal mixing and/or cosmic ray flux escaping and exciting plasma waves transfer energy to the gas \citep{2018Ehlert}. However, X-ray cavities are not the sole heating source within clusters: other processes, such as sloshing \citep{2010ZuHone_slosh_heat}, shocks from the AGN outbursts \citep[e.g.,][]{2023Ubertosi_rbs}, turbulence \citep{2014Zhuravleva}, and thermal conduction from hot outer layers \citep{2001Narayan,2004Dolag} can all contribute. Understanding how efficiently these different processes couple to the ICM, and the precise role of X-ray cavities therein, remains a challenge.

Theoretical and numerical models have been used over the last decades to understand the complex physics of galaxy clusters, SMBH feedback, and ICM. They provide an interpretation of observable manifestations. 
On the one hand, idealized high-resolution simulations of isolated clusters have been instrumental in shedding light on how AGN mechanical feedback can carve out X-ray cavities, heat the ICM, and prevent gas cooling \citep[e.g.,][]{2010Pope,2017Weinberger_jet_ICM,2018Prasad,2019Chen_Heinz,2022Beckman}. 
However, because of their idealized setup, these simulations lack a full cosmological context, excluding interactions such as mergers between clusters and between clusters and smaller systems, gas accretion from the inter-galactic medium or from the farthest regions of the ICM, and typically also lack galaxy-formation processes, other types of feedback, and magnetic fields. 
On the other hand, cosmological simulations of galaxy formation \citep[e.g.,][]{Eagle,2018Pillepich,2019Dave_SIMBA, Astrid, FLAMINGO, MilleniumTNG}, where clusters evolve over billions of years along with their galaxies, offer a more realistic cluster environment and assembly history to study AGN-induced X-ray cavities.
The combination of a large volume, numerical resolution, and a comprehensive galaxy formation model are all essential to quantitatively model and hence study X-ray cavities. 

Recently, in \cite{Prunier2024}, we demonstrated that the IllustrisTNG cosmological magnetohydrodynamic model of galaxy formation \citep[TNG hereafter,][]{2017Rainer,2018Pillepich} naturally produces a diverse range of X-ray cavities through its low-accretion AGN feedback mechanism, which consists of wind-like directional momentum injections. In particular, there we showcased the X-ray cavity population within TNG-Cluster \citep{2024Nelson}, a recent extension project of the TNG suite\footnote{\url{www.tng-project.org/cluster}}, with 352 zoom-in magnetohydrodynamic cosmological simulations of massive galaxy clusters (M$_\text{500c} = 10^{14.0}-10^{15.3}$ \msun) \footnote{M$_\text{500c}$ denotes the mass enclosed in a sphere whose mean density is $500$ times the critical density of the Universe, at the time the cluster is considered.}.

Similarly, recent work by \cite{2024Jennings} on the Hyenas suite of cosmological zoom simulations of galaxy groups searched and characterized X-ray cavities using mock Chandra observations, employing a methodology similar to that presented in this paper for the simulated sample. The Hyenas simulations feature better mass resolution than TNG-Cluster ($m_\text{gas} = 1.5 \times 10^{6}$ \msun$/h$ vs. $ 1.8 \times 10^{7}$ \msun$/h$) and employ the SIMBA model, where kinetic AGN feedback is implemented in a bipolar fashion, with jet axes aligned with the angular momentum of the inner gas disk. Their study demonstrated that the SIMBA models too produces a diverse population of X-ray cavities whose properties are qualitatively consistent with several observed scaling relations. Moreover, \cite{2024Jennings} compared the cavity enthalpies inferred from X-ray observations with the energy injection in the simulations and found that the cavities tend to be overpowered relative to the energy injected by recent AGN feedback. By comparing the relation between X-ray cavity area and distance to the central SMBH in the Hyenas simulations \citep{2024Jennings} with those from TNG-Cluster \citep[see Fig. 3 and section 5.1 in][]{Prunier2024}, we find that, although Hyenas focuses on the galaxy group mass range, both cavity populations populate the similar loci of the scaling relation. However, they differ in their slopes, and the Hyenas groups exhibit a larger proportion of cavities with areas greater than $10^3$ kpc$^2$.

Finally, X-ray cavities are also visible in the RomulusC low-mass cluster, with yet another implementation of AGN feedback: those result from feedback based on delayed cooling thermal energy injection \citep{2019Tremmel}. 

But how realistic are simulated X-ray cavities, whether from idealized or cosmological frameworks? Are they truly comparable to those observed in the real Universe? In fact, possible (dis)agreements might impact our ability to use current models to understand the physics of AGN feedback. Conversely, the characteristics of simulated X-ray cavities may serve as a crucial tool for evaluating the realism of current models, while also refining them and discriminating among their diverse physical and numerical implementations. For example, the aforementioned results based on cosmological galaxy simulations suggest that different AGN feedback models, such as those in TNG, SIMBA, and RomulusC, can all produce X-ray cavities that qualitatively resemble observed ones. Similar conclusions may be derived from the literature of different idealized simulations. All this seems to imply that these structures are a common outcome, rather than a definitive test of any single model. However, it remains uncertain whether the X-ray cavities simulated so far are, not only qualitatively, but also quantitatively realistic when compared to actual observations. 

Comparing simulation outcomes to astronomical observations is particularly challenging, especially when the comparison can and should involve both visual features in images and summary statistics that reflect the underlying physical properties. Stringent comparisons between simulations and observations of X-ray cavities have arguably never been conducted. Idealized simulations of individual clusters often lack the statistical diversity needed to assess how well they capture the observed variety of X-ray cavities, as a single model can in principle produce a wide range of manifestations. Moreover, observational biases, including instrumental and selection effects, are critical for meaningful and unbiased comparisons between simulations and reality. A consistent procedure to produce mock observations must account for instrumental limitations, observational strategies, and sample selection biases, as observational X-ray studies necessarily return only a partial view of the underlying observed systems and often probe only a subset of clusters with specific properties.

In the realm of cosmological X-ray cavities, the case of RomulusC, with only a single albeit notable cluster, lacks statistical significance. The Hyenas X-ray cavities are plentiful and their study \citep{2024Jennings} has been conducted via mocking Chandra images, getting closer to the observational testbeds: however, no observational selection criteria have been applied and the proposed comparisons throughout the group and cluster mass scales offer only qualitative insights.
A more thorough evaluation of the properties of X-ray cavities over a large sample is necessary, as discrepancies in their sizes, energetics, and scaling relations can all constrain simulation models. 

In this second paper of our series on X-ray cavities using TNG-Cluster, we follow up on \cite{Prunier2024} and propose a quantitative and rigorous comparison between populations of observed X-ray cavities and those of our simulation to explore a key question: Are TNG X-ray cavities comparable to those observed in the Universe? To address this, we perform a careful comparison between X-ray cavities that we identify in a volume-limited sample of observed clusters, which we build and analyze based on Chandra archival data, and a corresponding subset in TNG-Cluster, closely replicating the observational methods. 

This paper is organized as follows. In Section~\ref{sec:meth}, we present the sample selection and data reduction of our novel sample of observed clusters. We then detail the methodology for selecting matched TNG-Cluster systems and creating their mock Chandra observations. Finally, we describe the approach used to identify and characterize X-ray cavities in these two unique samples. Section~\ref{sec:results} presents our main findings, including a comparative analysis of the statistics, demographics, morphologies, and energetics of the two X-ray cavity populations. In Section~\ref{sec:discussion}, we interpret our results and their implications before concluding in Section~\ref{sec:conclusion}.

\section{Method}\label{sec:meth}

\subsection{Comparing simulated and observed X-ray cavities}\label{subsec:meth_howto}

We aim to evaluate whether X-ray cavities produced by the TNG SMBH feedback in TNG-Cluster exhibit quantitatively similar properties to those observed in real clusters, in particular with Chandra. 

To ensure a fair, ``apples-to-apples'' comparison, we conduct the analysis from scratch. Our methodology minimizes biases related to cluster sample selection, which could skew results if only specific types of clusters are chosen, as well as instrumental factors affecting observational data, such as the resolution and limitations of Chandra, and human decisions when detecting and measuring X-ray cavity properties. Our approach is three-pronged, as outlined below.

\begin{enumerate} 
\item \textbf{Selection of cluster samples:}  

\begin{itemize}
\item \textbf{Observational sample (Prunier+2025, Section~\ref{subsec:meth_selection_obs})}: First and foremost, we select a subsample of clusters observed with Chandra and available in the Chandra Data Archive. These clusters are drawn from a parent, volume-limited sample of \( z \leq 0.071 \) clusters in the Northern ROSAT All-Sky Survey \citep[NORAS,][]{REFLEX} and the ROSAT-ESO Flux Limited X-ray Galaxy Cluster Survey \citep[REFLEX,][]{REFLEX}, and include X-ray luminous clusters down to a flux limit of \( 3 \times 10^{-12} \, \mathrm{erg \, s^{-1} \, cm^{-2}} \) in the \( 0.1{-}2.4 \, \mathrm{keV} \) energy band. Our subsample, totaling 35 clusters, is a priori not strongly biased toward bright or strong cool-core clusters, even though implicit biases within the Chandra Data Archive may be lingering. We refer to these selected observed clusters as the \textit{observational Prunier+2025 sample} (or \textit{Chandra} (Prunier+2025) in the Figures).\\

\item \textbf{Simulated sample (TNG-Cluster matched, Section~\ref{subsec:meth_selection_tng}):} We then identify analog clusters in the TNG-Cluster simulation (3 simulated clusters per observed one, i.e. 105 TNG-Cluster systems in total), ensuring they possess similar X-ray luminosity, mass, and central cooling time as the observed ones. We detail this matching procedure below (Section~\ref{subsec:meth_selection_tng}) and we refer to this subsample of analog TNG-Cluster systems as the \textit{TNG-Cluster matched sample}.

\end{itemize}

\item \textbf{X-ray images:} For the observational Prunier+2025 sample, Chandra event files are available from the Chandra Data Archive and processed as described in Section~\ref{subsec:meth_selection_obs}. For the TNG-Cluster matched sample, we create mock Chandra X-ray images. In this way, we ensure consistency in instrumental characteristics (point spread function and instrumental response), maintaining the same field of view and exposure time as their observed Prunier+2025 counterparts. More details are in Section~\ref{subsec:meth_mocks}.\\

\item \textbf{X-ray cavity identification and characterization (Sections~\ref{subsec:_meth_det} and \ref{subsec:met_power}):} We apply the same techniques to both sets of images -- mock and real Chandra observations -- following established observational methodologies to identify and measure the properties of X-ray cavities (see Figure~\ref{fig:chandra_ex}). Consequently, for the observational Prunier+2025 sample, this study is new and offers a new identification and evaluation of X-ray cavities compared to previous similar works. Additionally, it provides a means to assess the reliability of our method by comparing it with values from the literature.

\end{enumerate}

\subsection{Observational Prunier+2025 sample} \label{subsec:meth_selection_obs}

Several previous studies of X-ray cavities have primarily focused on the brightest or most massive sources in the sky \citep[e.g.,][]{2006Dunn,2006Rafferty,2012Hlavacek}. However, such an approach can introduce a strong selection bias, affecting the detection rate and the overall properties of the identified X-ray cavity population. To mitigate this, we adopt the volume-limited sample of clusters of \cite{2014Panagoulia1,2014Panagoulia_vol} where clusters and groups were systematically selected within a distance of 300 Mpc ($z\leq 0.071$) from the Northern ROSAT All-Sky Survey (NORAS) and the ROSAT-ESO Flux Limited X-ray galaxy cluster survey (REFLEX) with cuts in X-ray luminosity: this is meant to achieve a statistically complete sample \citep[see][for details]{2014Panagoulia1}. 

From the \cite{2014Panagoulia1,2014Panagoulia_vol} list of 65 objects, we further refine the selection by including only the clusters with masses M$_\text{500c} > 10^{13.75}$ \msun \hspace{0.1cm} that have been observed with Chandra. Our final selection comprises 35 clusters, which are listed in Table~\ref{tab:obs_sample} along with their location, redshift, X-ray luminosity as provided by the parent surveys, cooling time calculated by \cite{2014Panagoulia_vol}, and corresponding ObsIDs and total exposure times. 

\begin{figure}
\centering
    \includegraphics[width=0.5\textwidth]{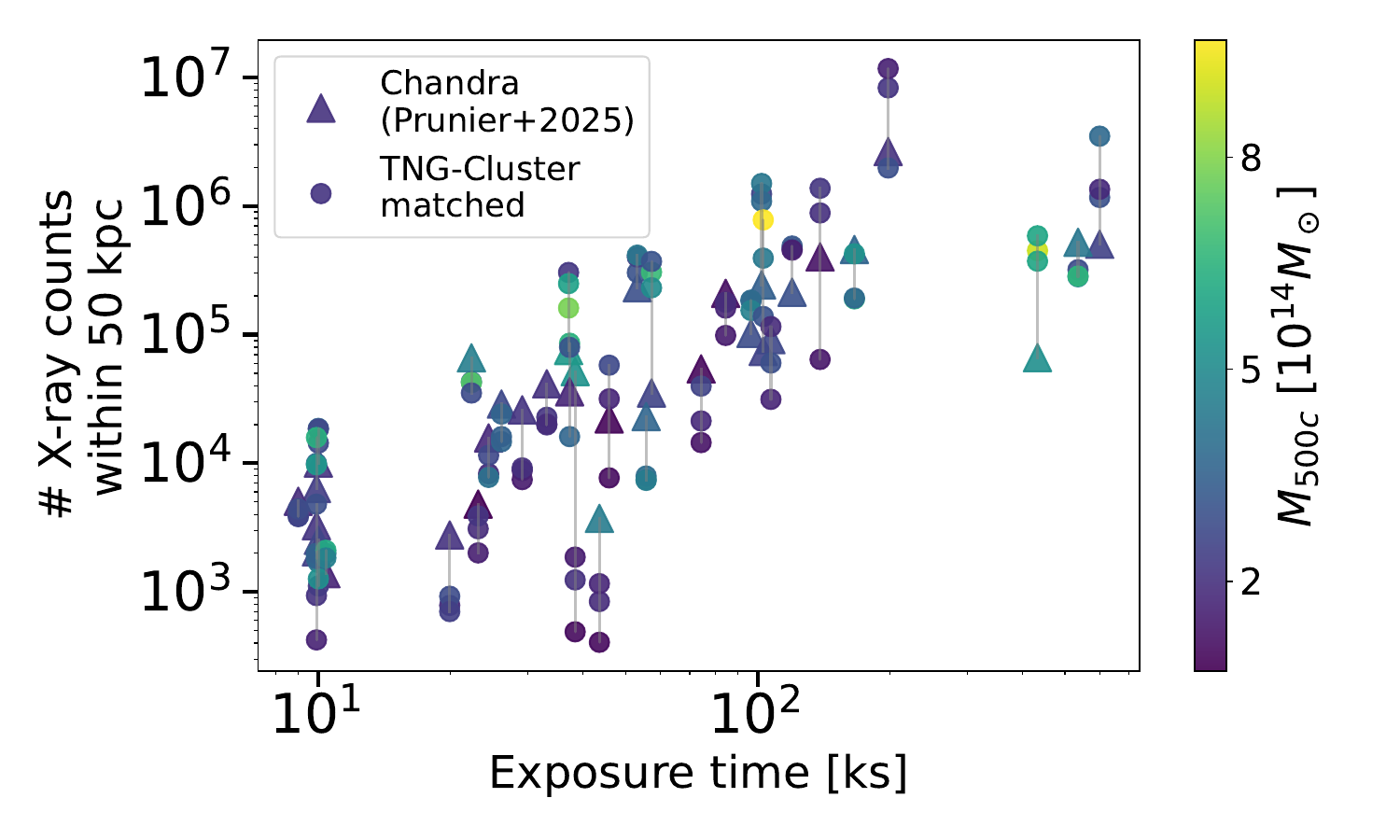}
    \caption{X-ray photon count (within a 50 kpc aperture) vs. exposure time of all observed and simulated clusters studied in this work, from the observational Prunier+2025 sample based on Chandra data (triangles) and the corresponding TNG-Cluster matched sample (circles). Each point is color-coded according to the cluster's mass M$_\text{500c}$. Grey lines connect matching clusters between the samples, which have the same exposure time.}
    \label{fig:exp_count_mass}
\end{figure}

\begin{figure*}
  \begin{subfigure}[b]{\textwidth}
    \centering
    \vspace{-0.4cm}
    \includegraphics[width=0.78\textwidth]{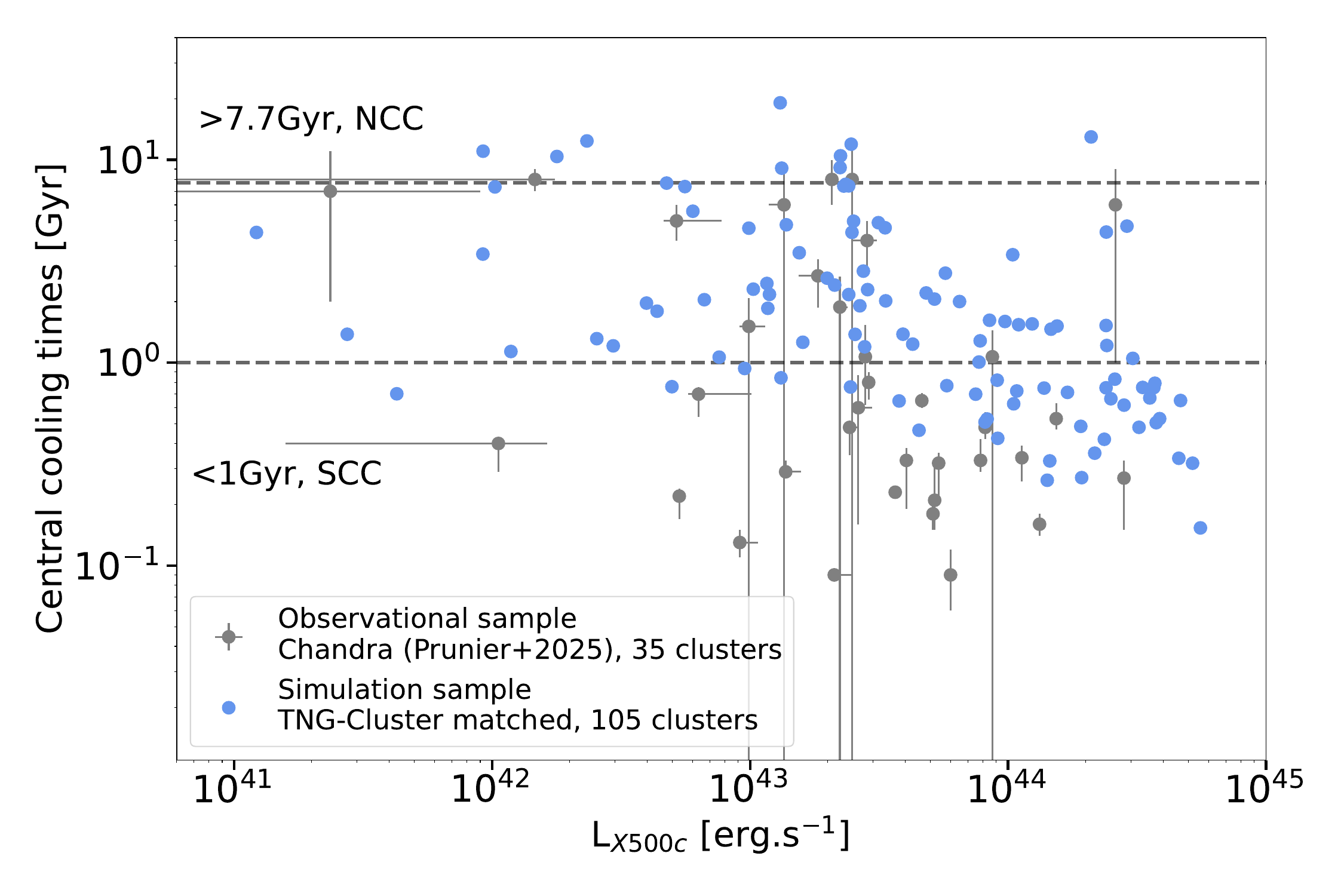}
  \end{subfigure}
  \begin{subfigure}[b]{0.32\textwidth}
    \includegraphics[width=\textwidth]{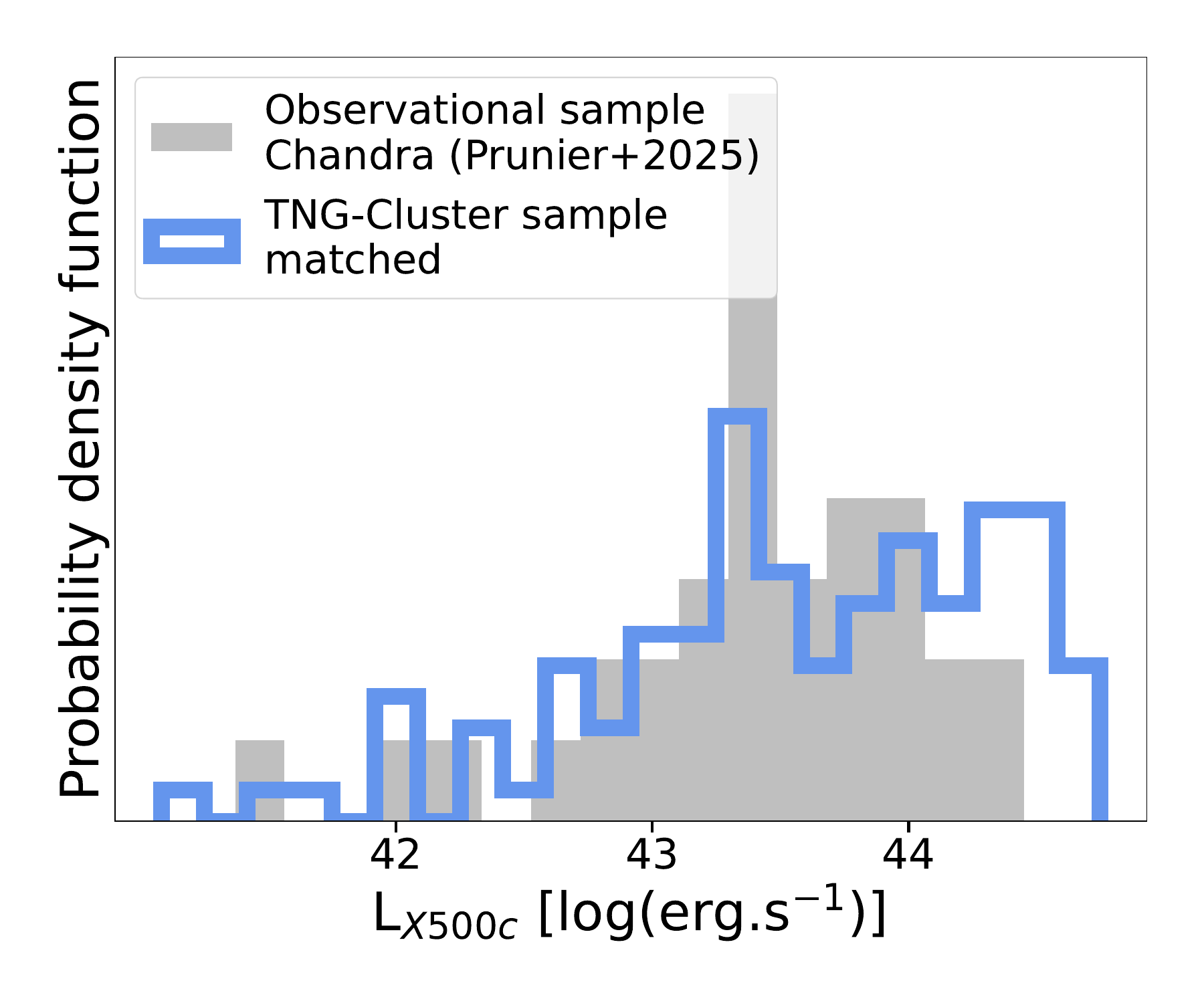}
  \end{subfigure}
    \hfill
  \begin{subfigure}[b]{0.32\textwidth}
    \includegraphics[width=\textwidth]{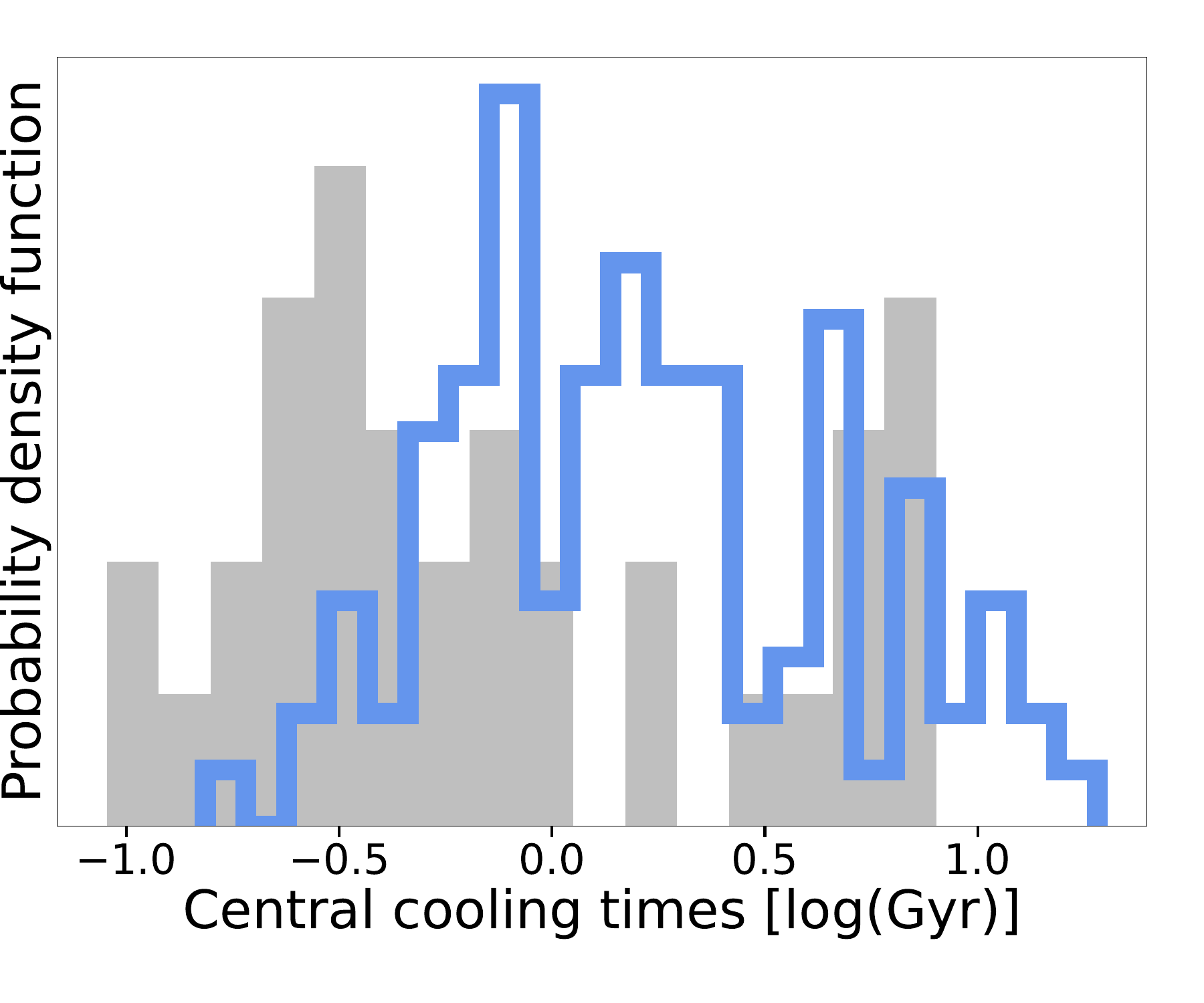}
  \end{subfigure}
    \hfill
      \begin{subfigure}[b]{0.32\textwidth}
    \includegraphics[width=\textwidth]{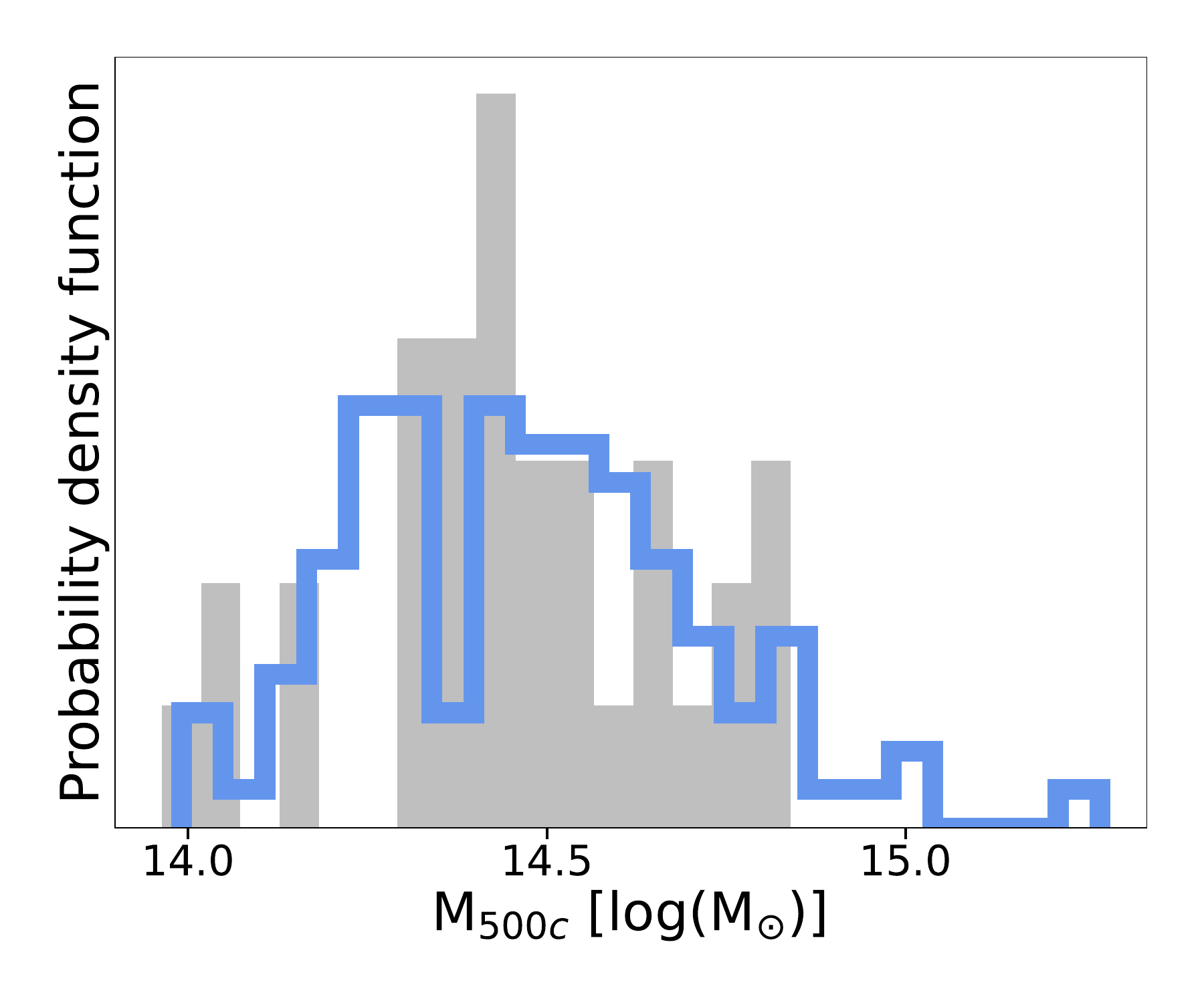}
  \end{subfigure}
  \caption{Sample selection of systems simulated within TNG-Cluster to match the observational Prunier+2025 sample introduced and used in this paper. \textit{Top:} Central cooling time plotted against the luminosity L$_\text{X500c}$. The 35 clusters in the volume-limited sample of observed systems are shown in grey, and the 105 matching  clusters from the TNG-Cluster simulation suite are shown in blue. The dashed lines delimit clusters classified as non-cool-cores, weak cool-cores, or strong cool-cores. In the bottom panels, from left to right, we compare the distributions of X-ray luminosity, central cooling time, and total halo mass for the two samples. The cooling times are taken within a small $<20$ kpc aperture for each cluster, whereas the X-ray luminosity is measured at the cluster scale, i.e. within a 3D $R_\text{500c}$ aperture.
  In this paper, we select and analyze TNG-Cluster systems whose central cooling time and X-ray luminosity are as close as possible to those of the observational Prunier+2025 sample; notably, also the M$_\text{500c}$ distributions are consistent between the two samples.}
  \label{fig:sample_selection_panel}
\end{figure*}

We have hence at our disposal 35 images of galaxy clusters observed with Chandra, depicted in Figure~\ref{fig:exp_count_mass} with triangular markers, in terms of number of X-ray photons within the innermost 50 kpc, exposure time, and total halo mass.
 The grey areas in the bottom panels of Figure~\ref{fig:sample_selection_panel} show the X-ray luminosities of the clusters, ranging from \( 10^{41} \) to \( 10^{44.8} \, \mathrm{erg \, s^{-1}} \), and their diverse central cooling states. These include strong cool-cores (SCCs, t$_\text{cool} < 1$ Gyr) clusters (22/35 clusters) to weak cool-cores (WCCs, t$_\text{cool}$ between 1 and 7.7 Gyr) clusters (10/35 clusters) and a few non-cool-cores (NCCs) (3/35 clusters). The cluster masses (M$_\text{500c}$ span \( 10^{14} \) to \( 10^{14.8} \, \mathrm{M_\odot} \). Some systems are well studied, such as Hydra A or the merging cluster Coma, for which high total exposure time observations are available. Other less-studied clusters in our sample have lower total exposure times, such as A1668 with 9.9 kiloseconds (ks).

After downloading the data from the Chandra Data Archive, we use the cleaning pipeline developed by \cite{rhea2020xtra}\footnote{\url{https://github.com/XtraAstronomy/AstronomyTools/tree/master/DataCleaning}} to process the level 1 event files and produce the final level 2 event files for each ObsID. The pipeline identifies periods with background flares exceeding the 3-$\sigma$ threshold, and removes these intervals from the data, it corrects any streaks, eliminates bad pixels, and processes the data using the \texttt{CIAO} tool \texttt{acis\_process\_events} with \texttt{VFAINT=True} since we are interested in the diffuse extended emission in the cluster. Background files are created using the \texttt{CIAO} tool \texttt{blanksky}. We then subtract the background from the observations and combine the ObsIDs corresponding to the same cluster. We identify point sources using the \texttt{CIAO} implementation of a wavelet source detection method \citep[\texttt{wavdetect,}][]{2002Freeman} and exclude them from the images; we hence use \texttt{dmfilth} to replace the source pixels with values interpolated from background region.

\subsection{TNG-Cluster, SMBH feedback model, and X-ray cavities}\label{subsec:meth_tng-cluster}

TNG-Cluster \citep{2024Nelson} is a set of 352 zoom-in simulations of massive galaxy clusters, expanding the TNG magnetohydrodynamic suite \citep[][and references therein]{2019NelsonPublicReleaseTNG}. With the same baryonic mass resolution as TNG300 ($m_\text{gas} \sim 10^{7}$\msun), it boosts the number of clusters with M$_\text{cluster,500c}> 10^{14}$\msun \hspace{0.1cm} from 30 to 352. Adaptive gas cells give an average hydrodynamic spatial resolution in the cluster core regions ($<0.5$R$_\text{500}$) from 1 to 8 kpc, but individual cells can be smaller than $< 1$ kpc, allowing to capture small-scale structures. The TNG-Cluster model, identical to TNG's \citep{2017Rainer,2018Pillepich}, includes gas cooling and heating, star formation, metal enrichment, stellar feedback, SMBH seeding and growth, and AGN feedback, as described below.

In TNG -- and consequently in TNG-Cluster -- SMBH physics is modeled using a subgrid approach, where SMBHs grow via Bondi accretion and mergers, releasing feedback energy either in thermal or kinetic forms, alongside continuous radiative-like feedback \citep{2013Vogelsberger}. At high accretion rates, SMBHs enter a thermal feedback mode, akin to a ‘quasar’ state, injecting energy isotropically as thermal energy. As the accretion rate decreases, SMBH feedback shifts to a kinetic mode, corresponding to the observed ‘radio’ state \citep{2006Croton}, releasing unidirectional bursts of kinetic energy that imparts momentum to the gas cells without immediately increasing thermal energy.

This kinetic feedback distributes energy isotropically through randomly directed injections of \(10^{41} - 10^{45} \, \text{erg s}^{-1}\) in clusters at low redshift \citep{Prunier2024} or even higher in other regimes \citep{2021PillepichErosita}. The kinetic feedback energy released is parametrized as \begin{math} \Delta \dot{E}_{\text{kin}} = \epsilon \dot{M}_{\text{SMBH}} c^2 \end{math}, where the maximum coupling efficiency $\epsilon$ is 0.2 \citep{2017Rainer,2018Pillepich}. This kinetic feedback does not reflect narrow, collimated AGN jets, nor incorporate cosmic ray physics \citep[although see][]{2024Ramesh}. Instead, it approximates the broad impact of AGN feedback within the constraints of cosmological simulation resolution. Physically, the TNG model is motivated by high-velocity accretion-disk winds or small-scale jets from low-luminosity SMBHs \citep{2014Yuan}. This feedback quenches star formation in massive galaxies by driving strong outflows \citep{2019Nelsonb}. Feedback imprints in the simulation are seen both on large scales \citep{TNG_stellar_mass,TNG_bimodality} and small scales, with emerging eRosita-like bubbles in Milky-Way like galaxies \citep{2021PillepichErosita} or X-ray cavities in BCGs \citep[e.g.,][]{2024Truong,Prunier2024}.

X-ray cavities arise naturally in the simulations due to SMBH feedback, demonstrating its capacity to reproduce feedback-induced features despite not explicitly modeling jet physics. In \cite{Prunier2024}, we demonstrate that X-ray cavities in the TNG-Cluster simulation exhibit a wide range of morphologies and evolutionary stages, including single, paired, and multiple X-ray cavities within the same cluster. These X-ray cavities are typically $\sim$ ten kpc in size, characterized by X-ray depleted, underdense regions filled with hot gas at \(\sim 10^{8}\) K, and are roughly in pressure equilibrium with the surrounding ICM. About $25$ per cent of them are surrounded by an X-ray bright and compressed rim often associated with a weak shock (Mach number $<2$). We refer the reader to \cite{Prunier2024} for additional notions on the origin and diversity of X-ray cavities in TNG-Cluster.

\subsection{Matching the clusters: the TNG-Cluster matched sample} \label{subsec:meth_selection_tng}

To minimize the possible effects of selection biases, we match the samples of observed and simulated clusters. For each of the 35 observed Prunier+2025 clusters, we identify three TNG-Cluster analogs based on two key properties: (i) the central cooling time, which indicates the rate at which gas in the cluster core cools, and (ii) the X-ray luminosity L$_\text{X,500c}$. In practice, we choose among the TNG-Cluster objects from the $z=0$ snapshot (number 99) those whose central cooling time and X-ray luminosity are within the error bars -- or as close as possible -- to those of their observational counterparts.

For the central cooling times of the observational sample, we use the quantities computed in \cite{2014Panagoulia_vol}, which were derived from spectral fitting in the innermost spectral bin (corresponding to an aperture $\sim$15 kpc). For the TNG-Cluster analogs, we extract the mass-weighted central cooling time from the simulation directly in a similar aperture\footnote{We verified that cooling times extracted from the simulation are consistent with those derived from spectral fitting on mock Chandra observations. To assess the impact on sample selection, we applied the selection pipeline again, excluding halos where spectrally fitted t$_\text{cool}$ values ($\pm$ errors) fall outside the observational range from \cite{2014Panagoulia_vol} and find the majority of the matched clusters remain, with the excluded subset showing no bias.}

For the X-ray luminosity, the observations come from the ROSAT or NORAS surveys and correspond to the $0.1-2.4$ keV band within R$_\text{500c}$. For each TNG-Cluster system, we estimate such a luminosity using projections of the soft band (0.5-2.0 keV) intrinsic X-ray luminosity of the cluster \citep[derived with an APEC model, in post-processing;][]{2024Nelson} within an aperture of R$_\text{500c}$. 

Our final sample of 105 simulated clusters from TNG-Cluster has similar characteristics to the observational Prunier+2025 sample, as shown in Figure~\ref{fig:sample_selection_panel}. In particular, there we highlight the demographics of the two samples, showing that the distributions and relationships of  X-ray luminosities, central cooling times, and total masses are very similar. For central cooling times, we have a slight under-representation of TNG-Cluster systems with t$_\text{cool}$ < 0.3 Gyr, but the fraction of cool-core clusters is the same as that in the observational Prunier+2025 sample ($\sim$ 90 per cent).

\begin{figure*}
    \centering
    \begin{subfigure}[b]{0.48\textwidth}
        \centering
        \includegraphics[width=\textwidth]{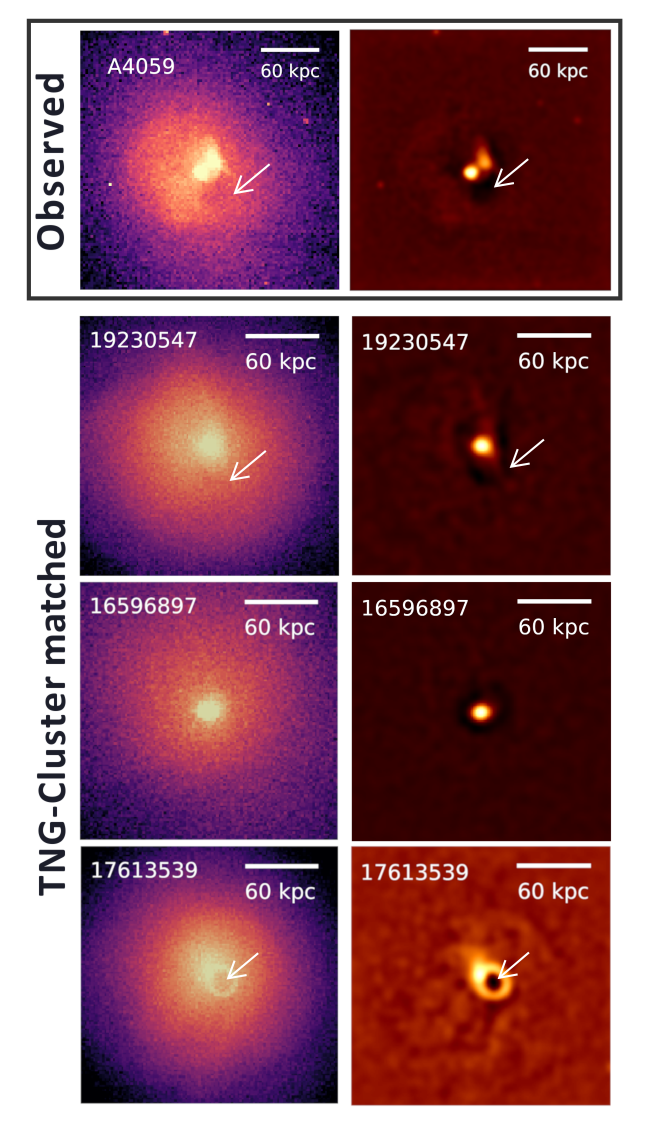} 
    \end{subfigure}
    \hspace{0.7cm}
    \vspace{1.4cm}
     \begin{subfigure}[b]{0.472\textwidth}
        \centering
        \includegraphics[width=\textwidth]{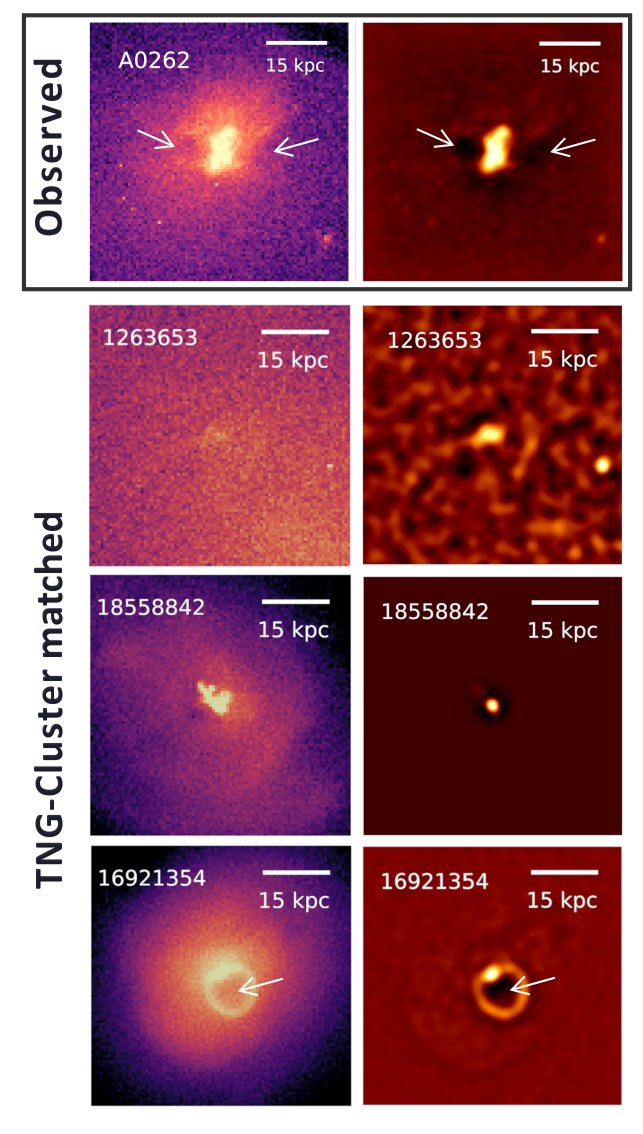}
    \end{subfigure}
    \vspace{-1.5cm}
    \caption{Examples of two clusters from the observational Prunier+2025 sample, each paired with three matched analogs from the TNG-Cluster simulation, selected based on similar L$_\text{X500c}$ and cooling times. \textit{Left:} Abell 0262 alongside its analogs: TNG-Cluster halos with IDs 1263653, 18558842, and 16921354. \textit{Right:} Abell 4059 with its analogs: TNG-Cluster halos with IDs 19230547, 16596897, and 17613539. The unsharp-masked images are convolved with (2, 10) pixels. Although analogs are selected based on their comparable thermodynamic properties, the presence and nature of cavities vary, highlighting diversity across systems even at fixed cluster cooling times and luminosity. Further discussion on TNG-Cluster cavities diversity can be found in \citealt{Prunier2024}.}
    \label{fig:chandra_ex}
\end{figure*}

\subsection{Mock Chandra X-ray images of TNG-Cluster} \label{subsec:meth_mocks}

Once selected, we produce mock Chandra X-ray observations of each TNG-cluster matched system. To do so, we add an observational layer to the simulated clusters taking into account the specific characteristics of the instrument used to image their observed counterpart. Since the observed clusters were imaged with Chandra's ACIS-S or ACIS-I instruments (pixel resolution 0.492 arcsec and, respectively, 8.4 and 16.9 arcmin field of view), and have various total exposure times, we create corresponding mock Chandra X-ray observations of the TNG-Cluster matched systems using pyXSIM \citep{pyXSIM_ZuHone} and SOXS \citep{SOXS_ZuHone}. The procedure is similar to that of \cite{Prunier2024} (see the method section therein for details). 

Each simulated cluster is placed at the same distance as its observed counterpart to ensure the field of view and pixel scale match. With SOXS we model instrumental effects, mainly the point spread function blurring and the effect of the instrumental response of the ACIS-I/-S detector. We choose the response files from Cycle 19 to account for the contamination buildup on Chandra's optical filters, which has degraded detection capabilities, particularly at low energies. Since our sample of Chandra observed clusters spans multiple observation cycles, Cycle 19 provides a consistent level of degradation that is representative of this range of cycles or more conservative. 

Each simulated cluster is integrated with a total exposure time that corresponds to that of their respective matching observed cluster, and restricted to an energy range of [0.5 - 7.0] keV also to match that of the observations -- the distributions of exposure times and photon counts of the systems from TNG-Cluster suite used in this work are reported in Figure~\ref{fig:exp_count_mass}, alongside the observed ones introduced above. 

We additionally create a blank sky background file with the SOXS routine \texttt{make\_background\_file} by writing the background/foreground events to a separate event file so that it can be subtracted in the same way in observations. Finally, we identify and remove point sources from the mock images, as in Section~\ref{subsec:meth_selection_obs}. 

\subsection{X-ray cavity identification} \label{subsec:_meth_det}

To identify X-ray cavities in both samples of clusters, we apply unsharp masking (UM) to the X-ray images. UM is a spatial frequency filtering technique that enhances contrast and small-scale features in images. This method, commonly used in previous X-ray cavity studies \citep[e.g.,][]{2014Panagoulia_vol,2015Hlv,2016Shin}, involves convolving images with Gaussian kernels of different sizes (e.g. 2, 4, 6, 10, 12 and 16 pixels) and subtracting images smoothed with larger kernels (10, 12 or 16 pixels) from those smoothed with smaller kernels (2, 4 or 6 pixels). Using filters of different sizes enables us to better visualize the large and small X-ray cavities that can coexist in a single cluster. 

We identify X-ray cavities by visually inspecting the original image and the UM images for each cluster, selecting only the ones that are clearly visible in both, or that are unambiguously present in the UM images.

\subsection{X-ray cavity properties} \label{subsec:met_power}

For each identified observed or simulated X-ray cavity, we measure a series of properties. 

\paragraph*{X-ray cavity sizes and areas} 
The physical extent of each X-ray cavity is manually estimated by superimposing an ellipse at the location of the X-ray cavity on both the original and UM images. The size of each X-ray cavity is taken as the average of the ellipse axes in kpc. Their projected area is computed as \begin{math}\pi r_a r_b \end{math}, where $r_a$ and $r_b$ are the axes along and perpendicular to the line that connects the SMBH to the center of the X-ray cavity. These are all projected quantities, i.e. projected in the plane of the sky, and not the actual 3D cavity sizes and volumes in the simulation (see implications in the discussion \ref{subsec:discu_limit})

While this method may lack precision, it aligns with the current practice among observers, who typically do not use automated methods for detecting and measuring X-ray cavity sizes \citep[though pioneering efforts in this direction have been made recently in][]{Plsek}. Due to these simplifying assumptions of ellipsoidal shape and symmetry, as well as projection effects, errors on radii are large and assumed to be of the order of $20$ per cent \citep[e.g.,][]{2015Hlv}.

\paragraph*{X-ray cavity ages} 
We estimate their ages assuming that X-ray cavities are not expanding much faster than the local sound speed in the ICM. This assumption is based on the fact that strong shocks around the X-ray cavities are rarely detected \citep[neither in observed X-ray cavities nor in those identified in the TNG-Cluster simulation,][]{Prunier2024}. Therefore, t$_\text{cav}$ is computed as the sound crossing time, which in turn is the time required for the X-ray cavity to rise the projected distance ($d$) from the SMBH to its present location at the speed of sound in the medium ($c_s$),
\begin{equation}
t_{\text{cav}} = \frac{d}{c_s} \quad \text{where} \quad c_s = \sqrt{\frac{\gamma k_B T}{\mu m_\mathrm{H}}},
\end{equation}
where $T$ is the temperature of the gas surrounding the X-ray cavity (see below),  $\gamma = 5/3$ is the adiabatic index for the ICM surrounding the X-ray cavity (assumed non-relativistic), $m_\mathrm{H}$ is the mass of the hydrogen atom, and $\mu$ is equal to 0.62.

\paragraph*{ICM properties at X-ray cavity height} 
In both the observational Prunier+2025 and the TNG-cluster matched samples, we determine the deprojected gas temperature and density profiles to compute the ICM sound speed, $c_s$, and thermal pressure (p$_\text{th}$) at the X-ray cavity height. As in most observational studies, we assume that X-ray cavities are in pressure equilibrium with their surroundings \citep[e.g.,][]{2006Rafferty}, and therefore the pressure of the X-ray cavities is measured as the azimuthally-averaged value at their location with 
\begin{math}\label{eq:pth}
p_\text{th} = (\mu_e/\mu) n_e k_B T
\end{math}, where we assume $\mu_e/\mu$ = 1.92 with $\mu_e$ the mean molecular weight per electron. 

In particular, we generate radial deprojected profiles of temperature and electron number density, from spectral fitting, for each cluster in a series of concentric annuli around the cluster center, ensuring that each annulus has a high signal-to-noise ratio for accurate measurements (in each annulus the number of counts is $>30^2$). To account for projection effects in the radial profiles we use the Deproject CIAO Sherpa package allowing for the deprojection of two-dimensional circular annular X-ray spectra to recover the three-dimensional source properties. The fits are performed over the energy range 0.5-7.0 keV in each annulus with XSPEC with a single-temperature plasma model (MEKAL) and foreground absorption (WABS) using the model \texttt{wabs*apec}. The free parameters in our spectral fits are temperature, metallicity, and normalization. The local absorption is fixed to the galactic neutral hydrogen column density (n$_\mathrm{H}$) using the NASA HEARSC tool that calculates n$_\mathrm{H}$ based on the sky position of the system \citep{Kalberla2005,HI4PI2016}, unless the best fit is found to be significantly different. The redshift is fixed to the values displayed in Table~\ref{tab:obs_sample} for each cluster.

Finally, we measure p$_\text{th}$ for each X-ray cavity taking the temperature and density values at the cavity height and using the formula above.

\paragraph*{X-ray cavity power} To determine the X-ray cavity power P$_\text{cav}$, as done in observational studies such as \cite{2014Panagoulia_vol}, we estimate the rate at which energy from an X-ray cavity would be dissipated in the ICM. The X-ray cavity energy is a combination of two key components: the work done to displace the intracluster gas, represented by the product of pressure and volume, and the thermal energy within the X-ray cavity. We assume that the X-ray cavities are filled with an ideal gas dominated by relativistic particles, with an adiabatic index $\gamma =4/3$ following standard observational practices \citep[e.g.,][]{2002Churazov_calorimeter}. While this does not reflect the nature of the gas cells within X-ray cavities in our simulation, which are non-relativistic, it ensures a consistent comparison with observational studies where $\gamma$ is a choice. See further discussion on the implications of this choice in Section \ref{subsec:discu_limit} and Appendix~B.\footnote{We note that this assumption would not be valid for estimating total AGN energy injection from the simulated X-ray cavities.} We calculate their power as:
\begin{equation}\label{eq:pcav}
P_{\mathrm{cav}} = \frac{E_{\mathrm{cav}}}{t_{\mathrm{cav}}} = \frac{\gamma}{\gamma-1} \frac{p_{\mathrm{th}} V}{t_{\mathrm{cav}}} = \frac{4 p_{\mathrm{th}} V}{t_{\mathrm{cav}}},
\end{equation}
where p$_{\mathrm{th}}$ is the thermal pressure of the gas surrounding the X-ray cavity, and t$_{\mathrm{cav}}$ is its age. The volume $V$ of the X-ray cavity assuming it is a prolate ellipsoid with \begin{math} V = 4/3 \pi r_a r_b^2 \end{math}. 

\begin{figure}
\centering
\includegraphics[trim=1.1cm 5cm 10cm 2cm, clip, width=0.48\textwidth]{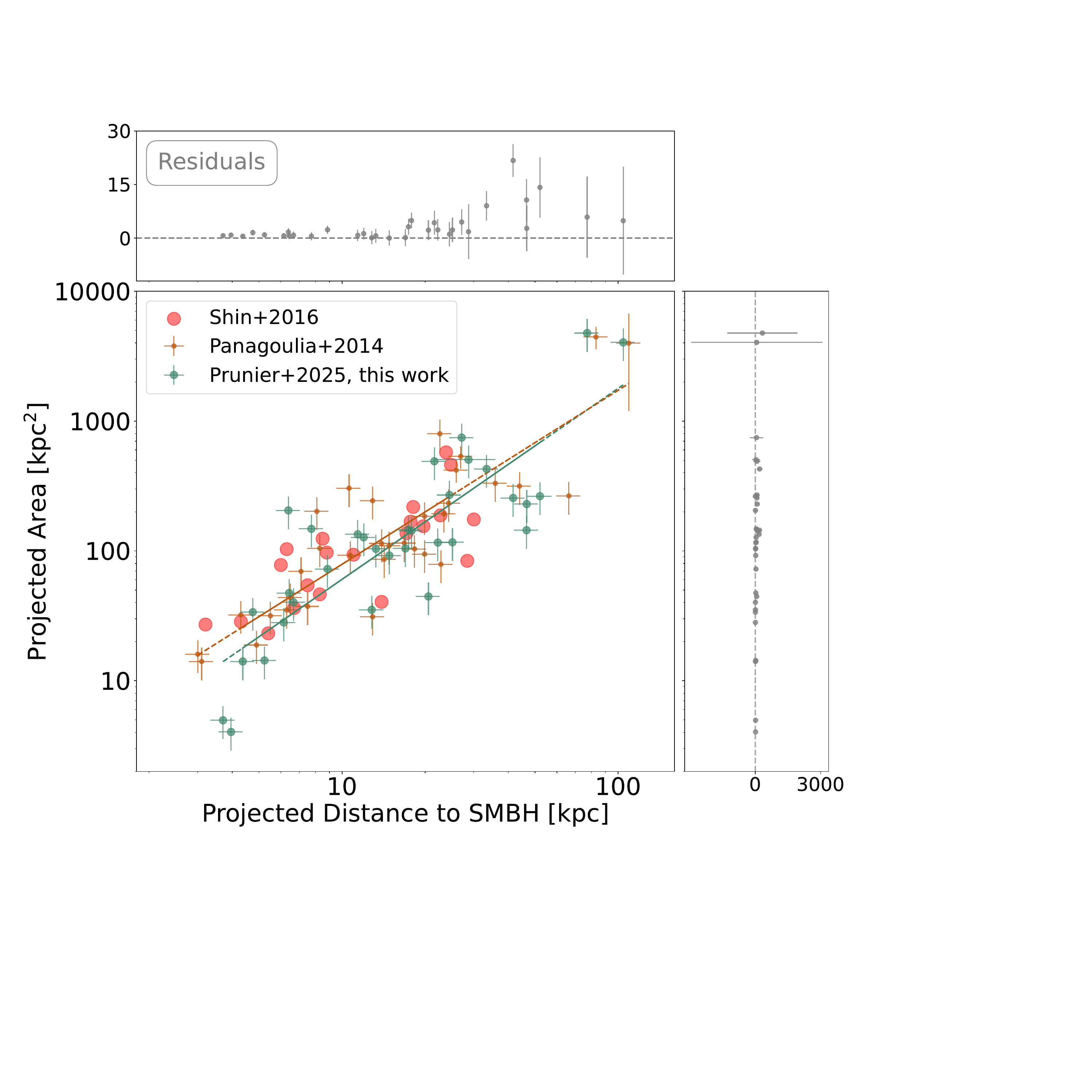}
\vspace{-1.8cm}
  \caption{Comparison of Prunier+2025 measurements with those reported in \citealt{2014Panagoulia_vol} for X-ray cavity area vs. distance from the central SMBH. Residuals are shown with dashed lines at zero in the top and right subpanels. In addition, we display in red the X-ray cavities identified in the clusters common to both our Prunier+2025 sample and the study by \citealt{2016Shin}. Our measurements are fully consistent with those of \citealt{2014Panagoulia_vol} and, for the common clusters, with \citealt{2016Shin}.}
  \label{fig:area_distance_pana_comp}
\end{figure}

\paragraph*{Cooling luminosity}\label{subsec:met_cool_lum}
In both samples, we also calculate the cooling luminosity L$_{\mathrm{cool}}$ in the 0.5 - 7.0 keV range, which represents the luminosity of the gas within the cooling radius r$_{\mathrm{cool}}$ -- defined as the radius within which the gas has a cooling time of 3 billion years. We first compute a radial profile for t$_{\mathrm{cool}}$,
\begin{equation}
t_{\text{cool}} = \frac{3 p_\text{th}}{n_e n_
\mathrm{H} \Lambda(Z, T)} = \frac{3 p_\text{th} V}{2 L_X},
\end{equation}
where $\Lambda(Z, T)$ is the cooling function for gas at a specific abundance $Z$ and temperature $T$ and where X-ray luminosity L$_X$ is fitted in each annuli region. We identify the radius where t$_{\mathrm{cool}}$ < 3Gyr and compute the final  L$_{\mathrm{cool}}$ within this aperture. In the observational sample, the central point source, when present, is excised during the spectral analysis. This approach is consistent with methodologies used in previous studies \citep{2009ApJS..182...12C,2014Panagoulia_vol,2017Hogan}.


\begin{figure*}
    \centering
  \begin{subfigure}[b]{0.60\textwidth}
    \includegraphics[width=\textwidth]{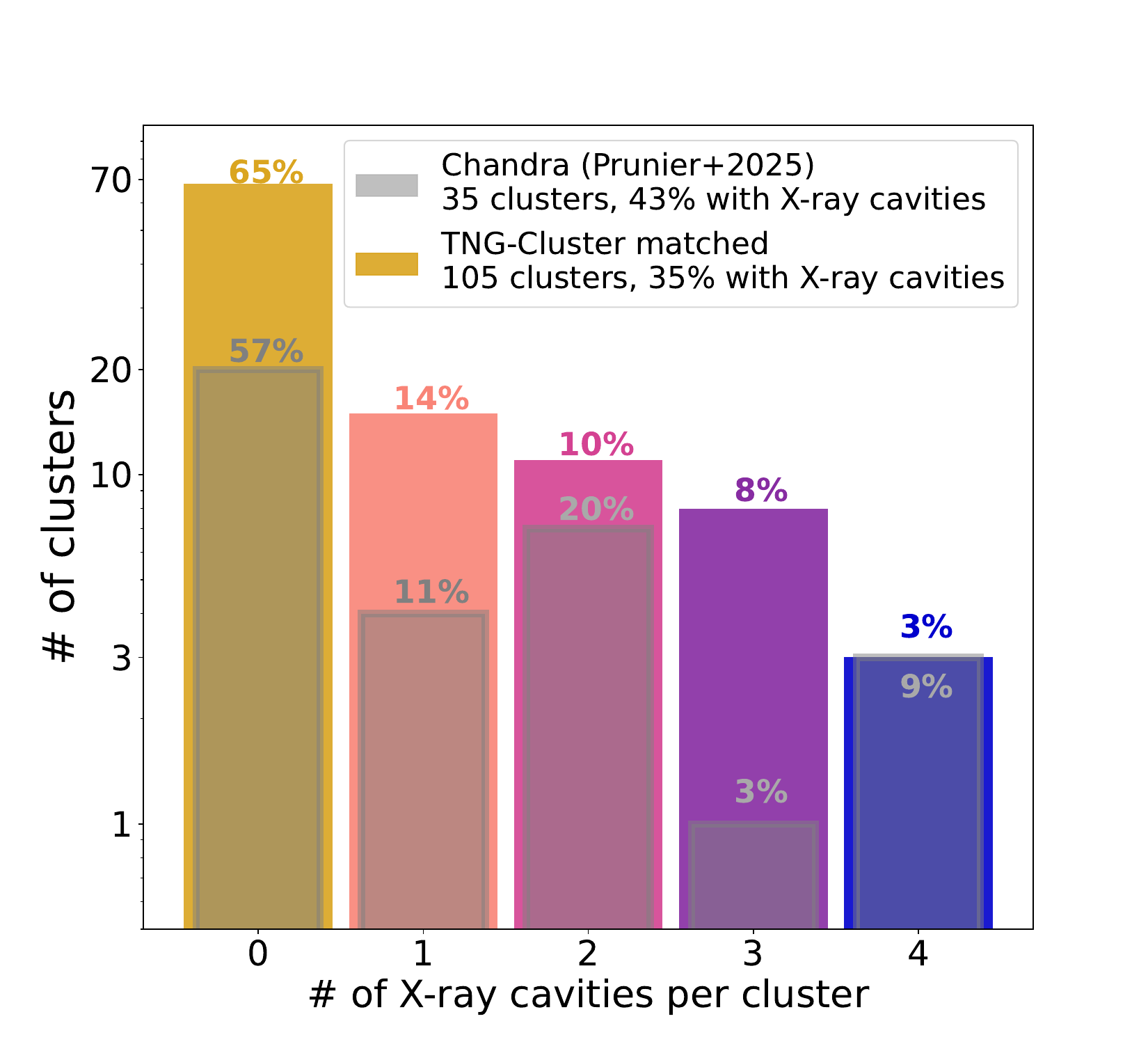}
  \end{subfigure}
    \begin{subfigure}[b]{0.38\textwidth}
    \includegraphics[width=\textwidth]{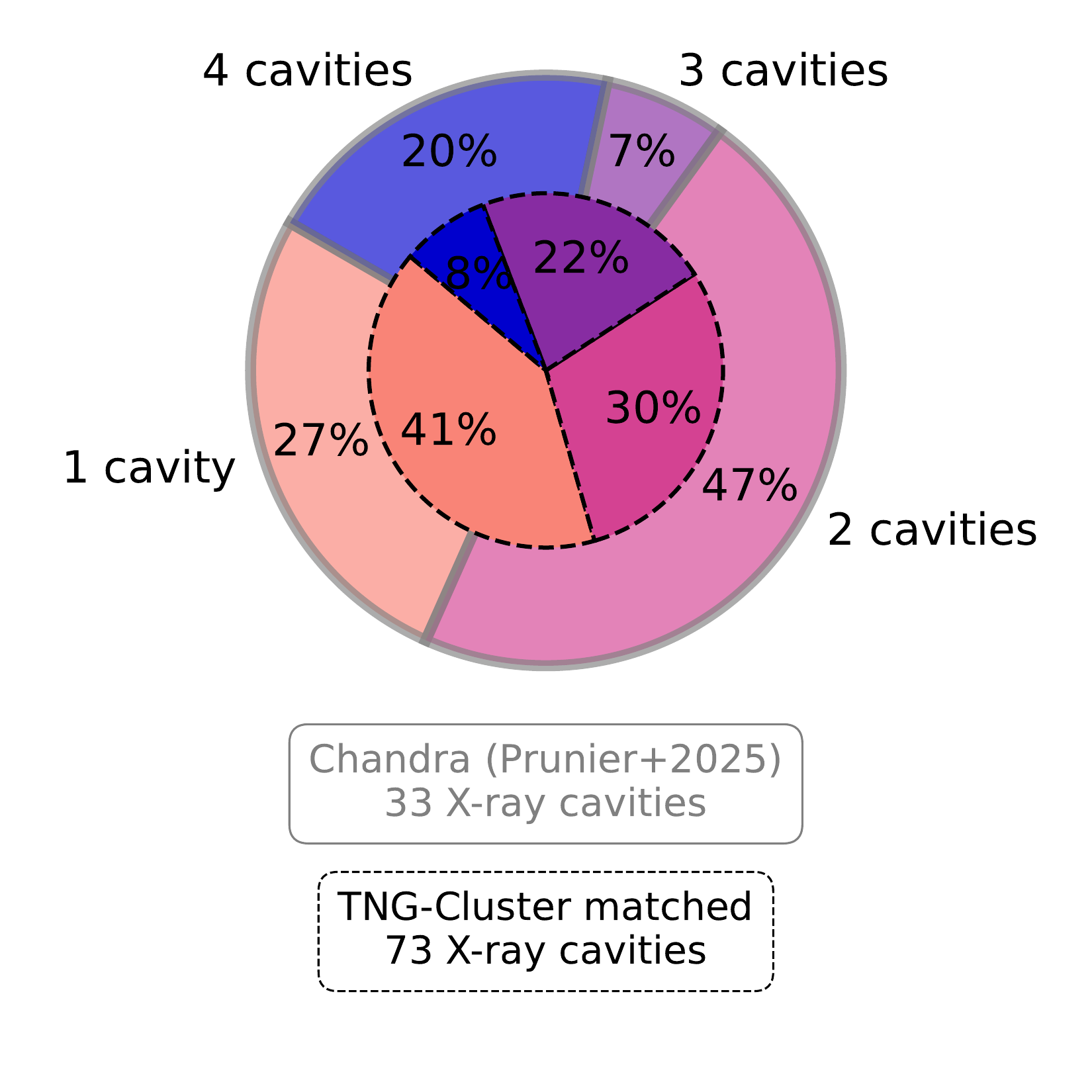}
    \vspace{0.5cm}
  \end{subfigure}
  \caption{Demographics of clusters with and without identified X-ray cavities in the observational and TNG-Cluster samples. \textit{Left:} per centage of clusters with identified X-ray cavities, with colored bars for TNG-Cluster and grey for the observational sample. \textit{Right:} Distribution of clusters with one, two, three, or more X-ray cavities, with black-edged (grey-edged) slices for TNG-Cluster (for observations). Both samples have a comparable detection rate. Clusters from the observational sample show a majority of pairs, while TNG-Cluster systems most often host a single X-ray cavity.}
  \label{fig:demographics}
\end{figure*}

\begin{figure*}
    \centering
    \includegraphics[width=\textwidth]{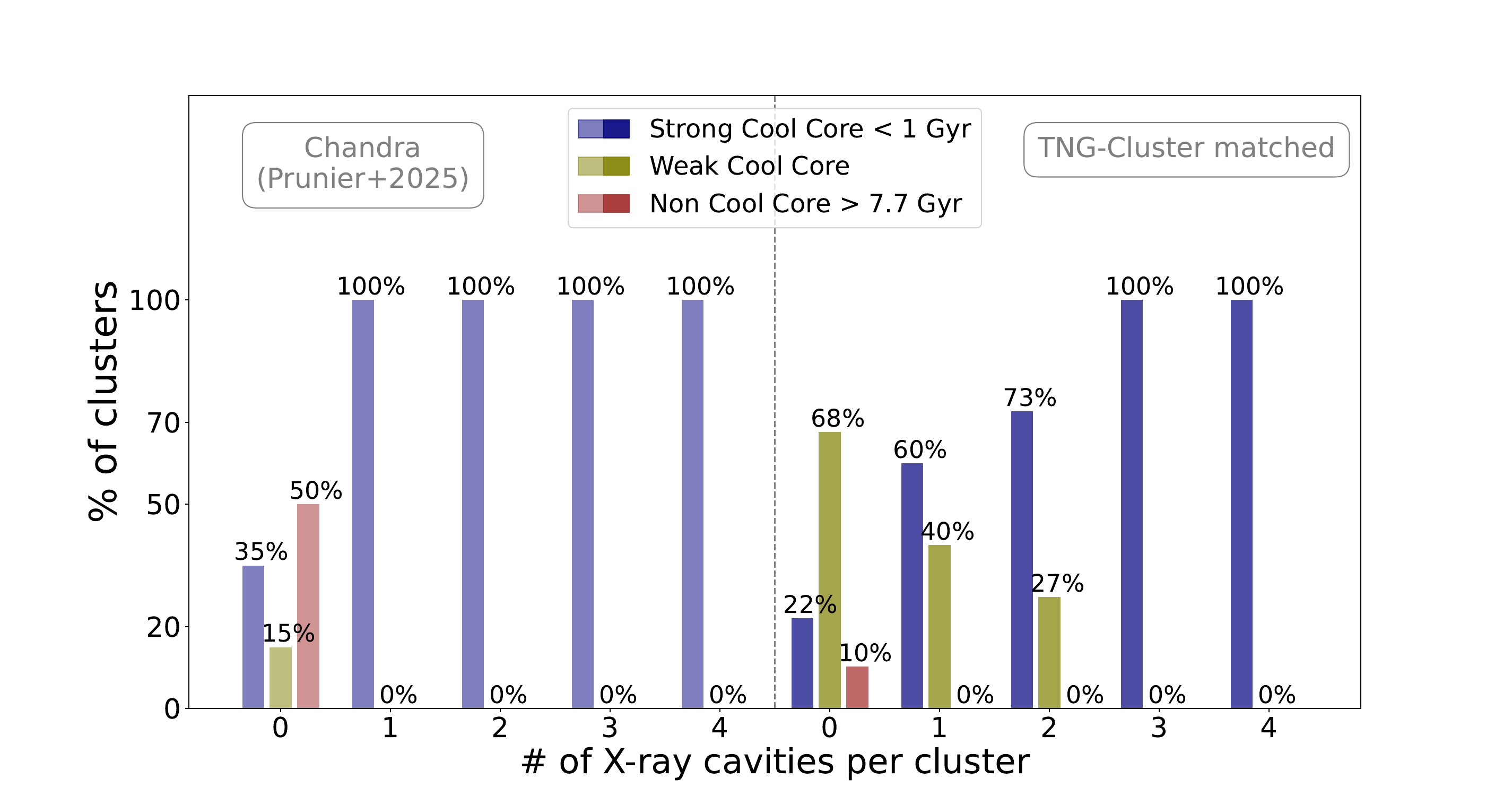}
  \caption{Fraction of strong cool core (SCC), weak cool core (WCC), and non-cool core (NCC) clusters (using a cooling time criteria), for clusters with or without X-ray cavities. The plot reads as follows: e.g. for clusters with two X-ray cavities (x-axis label ``2''), 100 per cent of the Prunier+2025 clusters are SCCs; for the TNG-Cluster matched systems, 73 per cent are SCCs, 27 per cent are WCCs and 0 are NCCs. In both samples, most X-ray cavities are found in strong and weak cool-core clusters.}
  \label{fig:demographics_CC_alt}
\end{figure*}

\section{Results}\label{sec:results}

\subsection{X-ray cavities in Chandra data and in TNG-Cluster}

We inspect for X-ray cavities in both the observational Prunier+2025 and TNG-Cluster matched samples by following the methodology described above. All clusters that have at least one X-ray cavity are listed in Tables~\ref{tab:results_pana} and \ref{tab:results_tng} for the two samples, respectively, including all their properties measured as described in Section~\ref{sec:meth}.

\begin{figure*}
    \begin{subfigure}[b]{0.48\textwidth}
    \hspace{-0.7cm}
    \includegraphics[width=\textwidth]
    {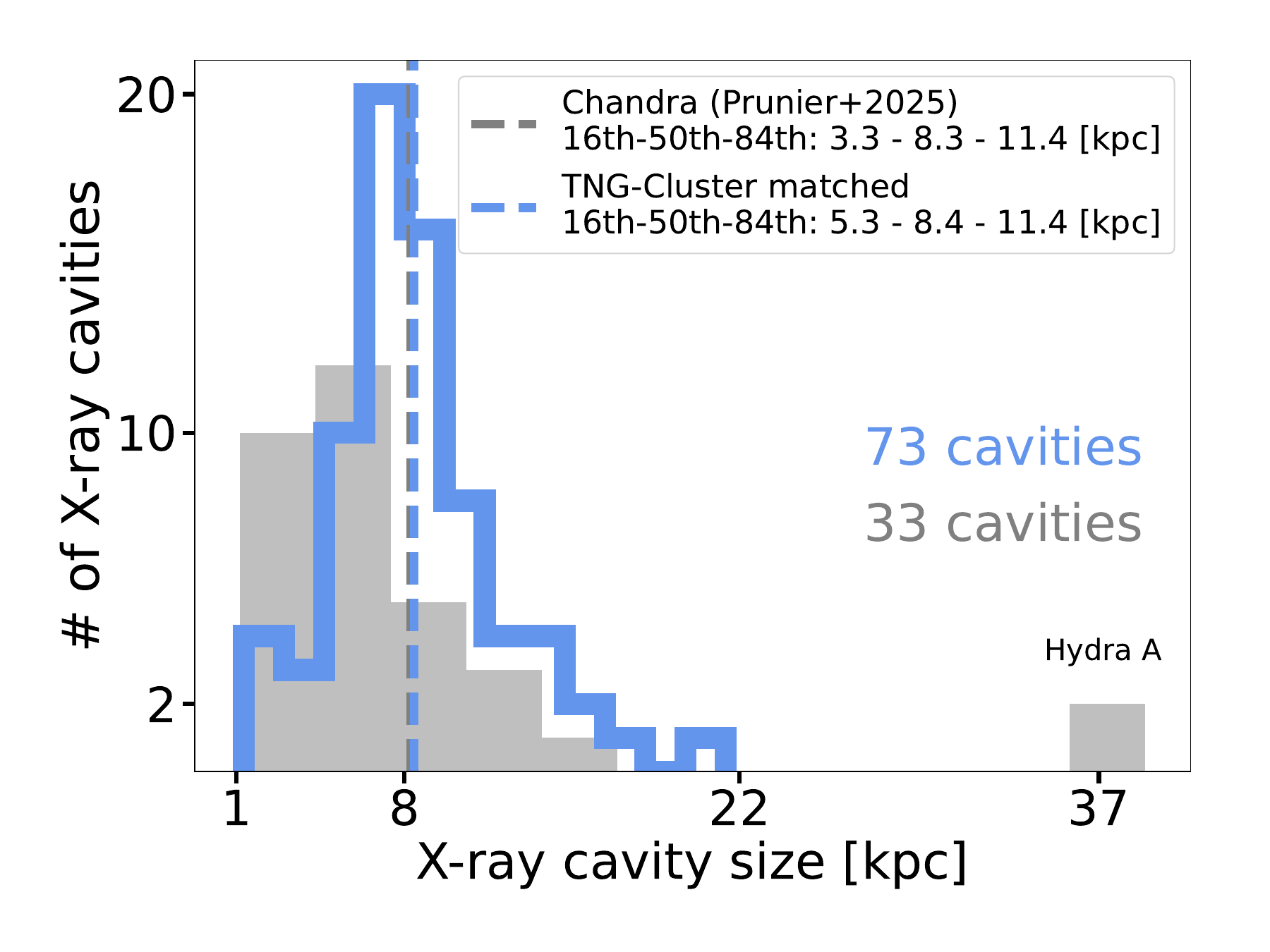}
    \vspace{0.8cm}
    \end{subfigure}
    \begin{subfigure}[b]{0.47\textwidth}
    \includegraphics[width=\textwidth]{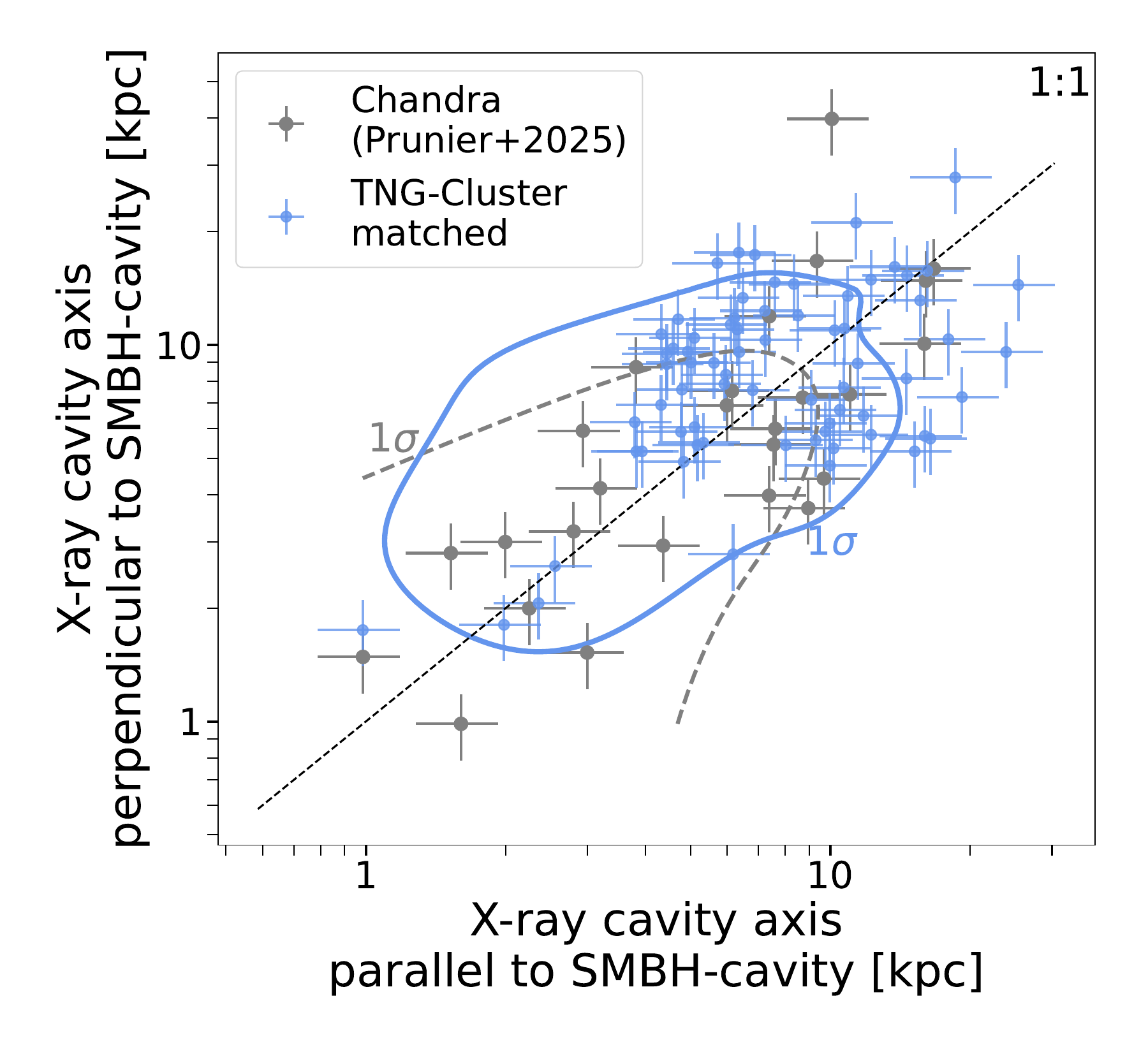}
    \end{subfigure}    
    \hfill
    \begin{subfigure}[b]{0.476\textwidth}
    \hspace{-0.68cm}
    \includegraphics[width=\textwidth]{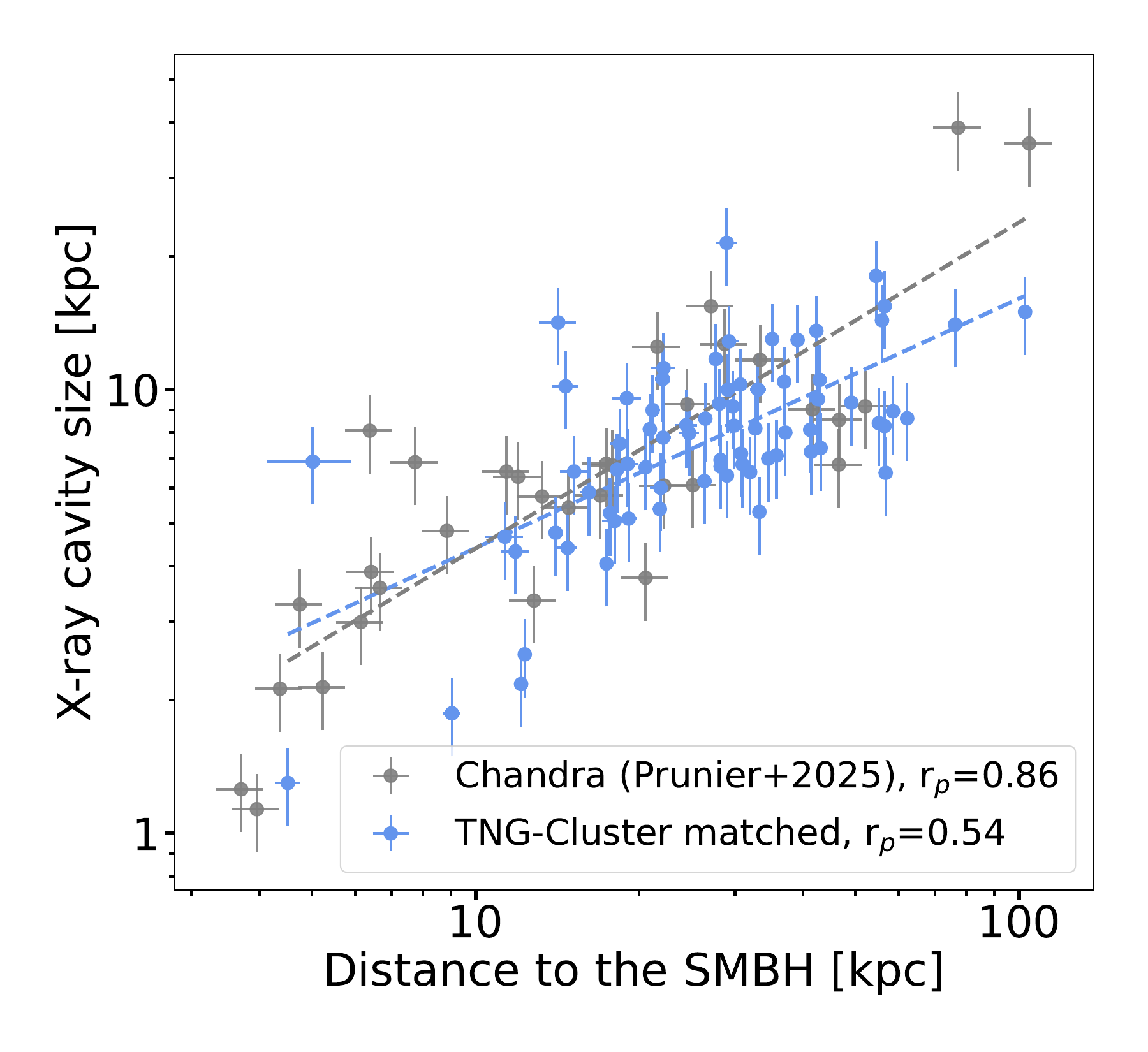}
    \end{subfigure}
    \begin{subfigure}[b]{0.48\textwidth}
   \includegraphics[width=0.98\textwidth]{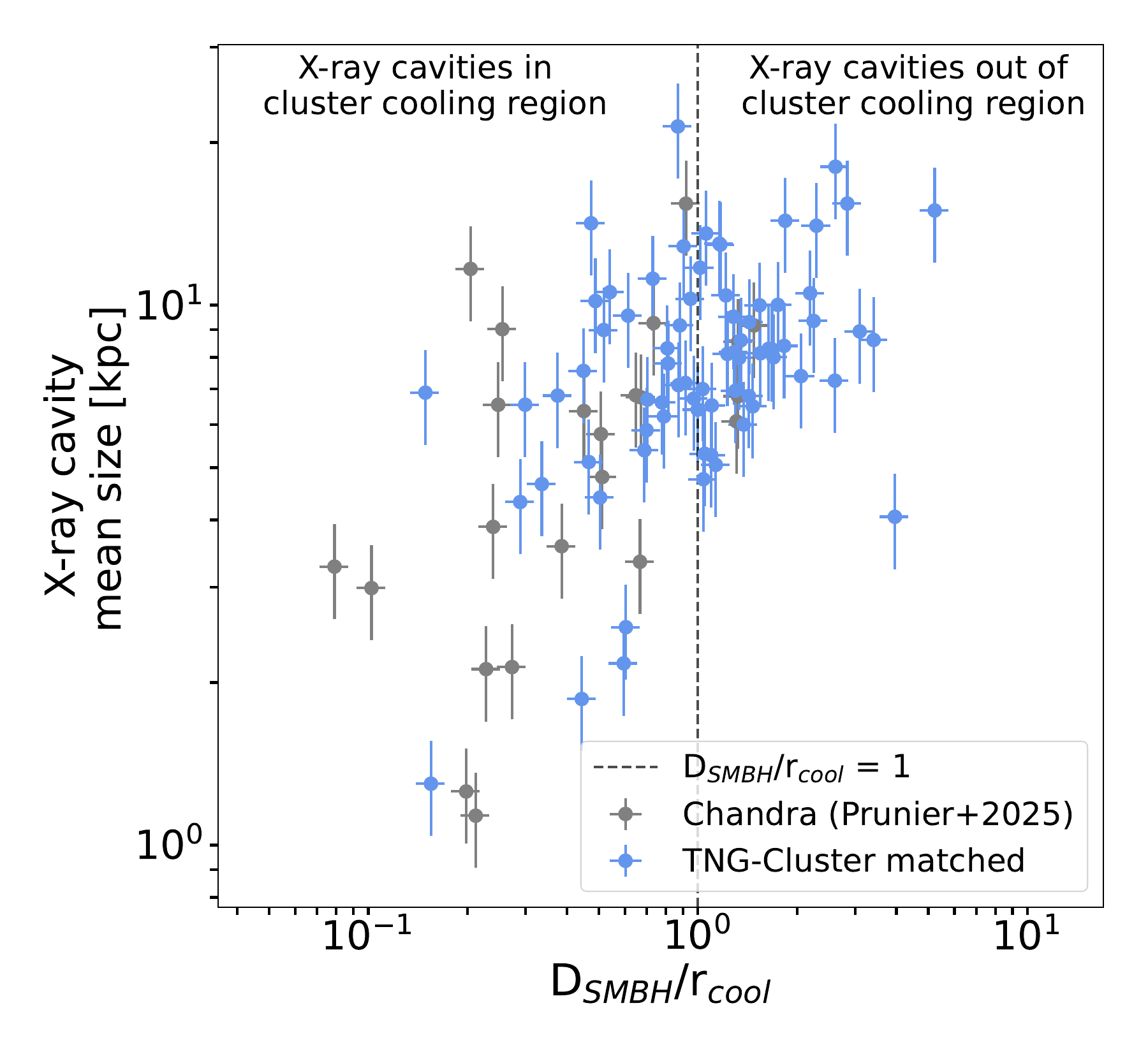}
   \end{subfigure}

  \caption{The size, location, and shape of X-ray cavities in the observed and simulated systems. In all panels, we compare the properties of clusters of the observational Prunier+2025 sample, in grey markers, with those from the TNG-Cluster matched sample, in blue. \textit{Top left:} Distribution of the mean X-ray cavity size (average of the ellipse axes in kpc) in both samples, with 50th and 16th-84th percentiles. \textit{Top right:} Morphologies, i.e. parallel axis vs. the perpendicular axis with respect to the SMBH-cavity center direction with 1-$\sigma$ contours. 1:1 correlation line in dotted black. \textit{Bottom left:} Mean X-ray cavity size as a function of distance from the central SMBH. \textit{Bottom right:} X-ray cavity size as a function of distance to the SMBH scaled by the cluster's cooling radius $r_{\mathrm{cool}}$, defined as the radius within which the gas has a cooling time of 3 Gyr. Overall, TNG-Cluster X-ray cavities exhibit similar sizes and size-distance trends to the observational sample and similar morphologies. X-ray cavities can be either inside or outside the cooling radius, but with a larger fraction of X-ray cavities outside of the cooling radius in TNG-Cluster than in the observed sample.}
  \label{fig:radius_dist_fraction}
\end{figure*}

\begin{figure*}
    \centering
    \includegraphics[width=\textwidth]{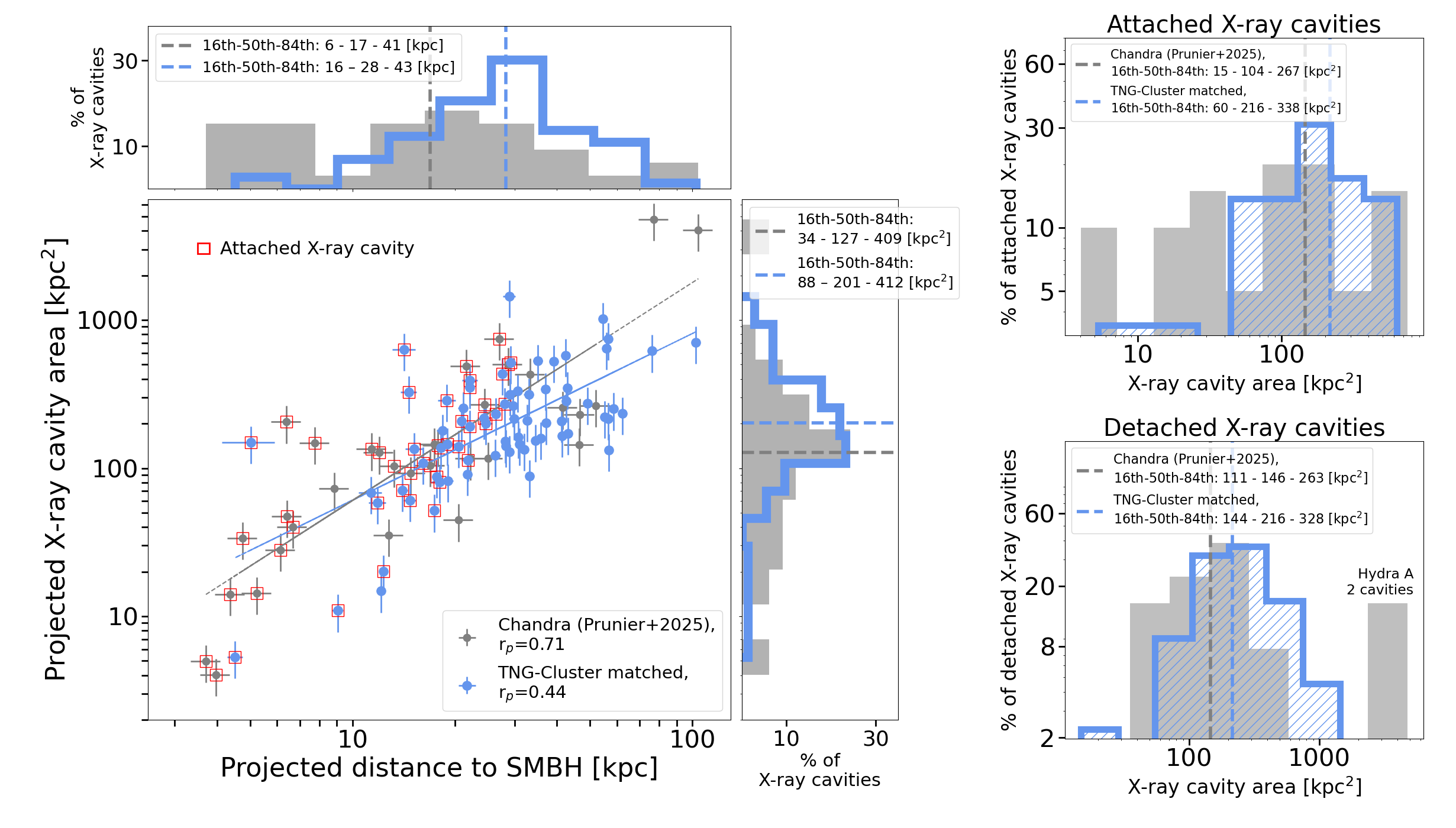}
  \caption{Comparison of X-ray cavity areas and distances from the central SMBH in the observed Prunier+2025 and TNG-Cluster matched samples. \textit{ Main left panel:} X-ray cavity areas vs. distance from the central SMBH (projected on the plane of the sky), with the Pearson coefficient quantifying the area-distance correlation for each sample. Blue and grey lines represent the best logarithmic fits. Red squared points denote the \textit{attached} X-ray cavities - i.e. for which size is equivalent to their distance from the central SMBH - in contrast to detached ones that are rising in the ICM. Histograms in the subpanels show the distribution of distances and areas, with 50th (dashed lines) and 16th-84th percentiles.
  \textit{Right panels}, top and bottom histograms: distribution of X-ray cavity areas for attached and detached cavities, with 50th (dashed lines) and 16th-84th percentiles quoted in the legend.
  Both the observed and TNG-Cluster X-ray cavities show similar slopes and scatter in the area-distance relation, where X-ray cavities expand as they rise buoyantly from the central SMBH. We find a lack of small ($\lesssim$ 100 kpc$^2$) {\it attached} X-ray cavities in the TNG-Cluster sample compared to the observed one.}
  \label{fig:area_distance}
\end{figure*}

\begin{figure*}
\hspace{2cm}
\includegraphics[width=0.88\textwidth]{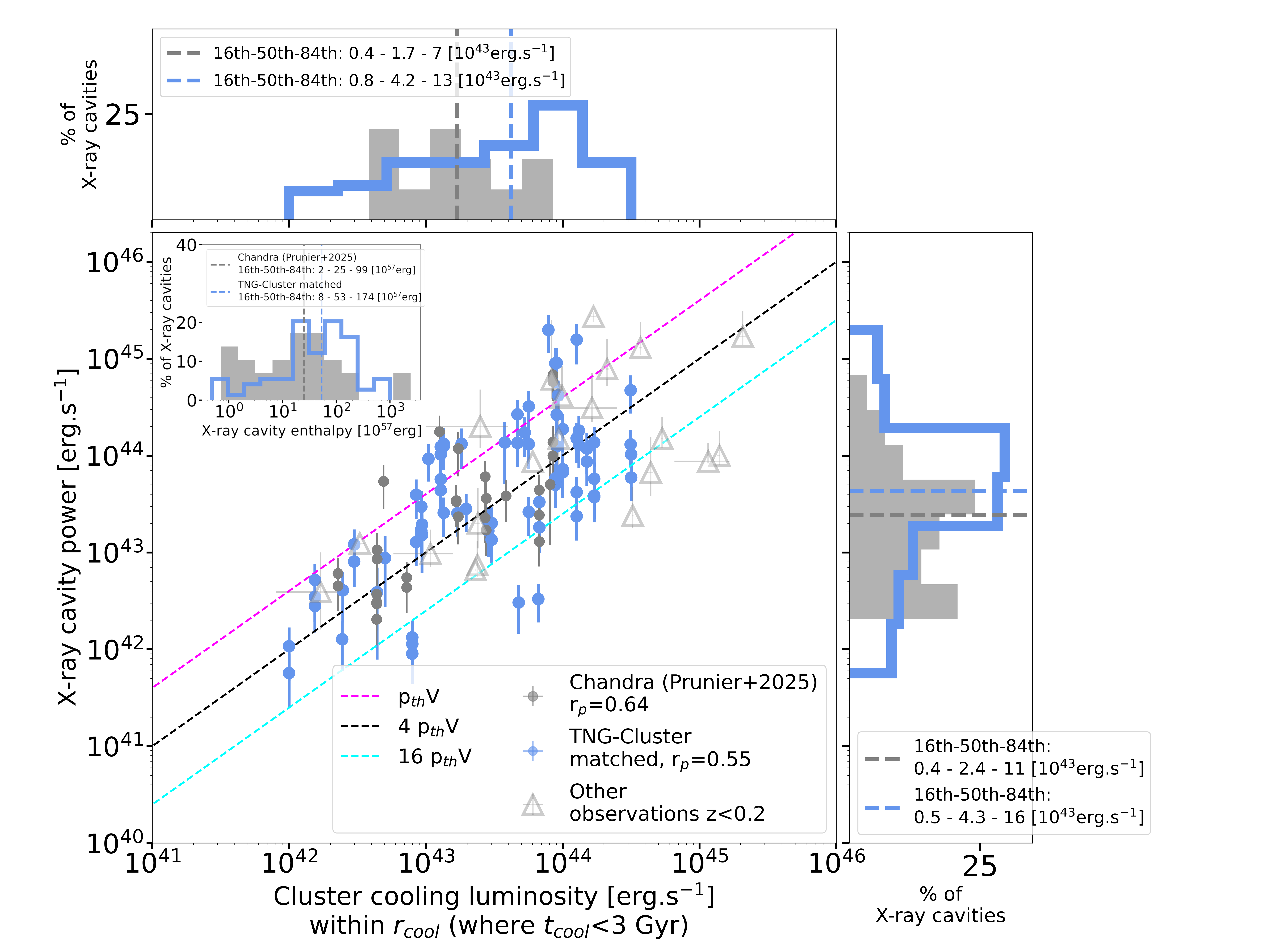}
    \caption{X-ray cavity power (P$_\text{cav}$) vs. cooling luminosity (L$_\text{cool}$) for clusters with identified X-ray cavities in the TNG-Cluster sample (blue) and our observational sample (grey). Each point represents the power inferred for a single X-ray cavity; pairs are not summed up. The pink, black, and blue dashed lines represent equality between cooling and heating, assuming energy inputs of p$_\text{th}$V, 4p$_\text{th}$V, and 16p$_\text{th}$V per X-ray cavity, respectively. 
    The side histograms display the distribution of P$_\text{cav}$ and L$_\text{cool}$, with 50th (dashed lines) and 16th-84th percentiles quoted in the legend. The light-grey triangles show the total X-ray cavity power in clusters of other observational studies with $z<0.2$ \citep{2006Rafferty,2008DursiPfrommer, 2012Hlavacek,cav_2013Pandge,cav_2015Sonkamble,cav_2017Vagshette,cav_2019Kadam,cav_2019Pandge,cav_2019Vagshette,cav_2020Liu,cav_2020Marylou,cav_2021Kyadampure}. The inset histogram shows the enthalpy distribution of the two X-ray cavity populations. In both samples, the power and enthalpy of X-ray cavities, and the cooling luminosity distributions are generally consistent (similar median and width). Importantly, TNG-Cluster and observed X-ray cavities exhibit well-consistent scaling relations, in terms of slope, normalization, and scatter.}
    \label{fig:Pcav_Lcool}
\end{figure*}

Firstly, as implied above, here we put forward an independent analysis of existing Chandra archival data, which can then be contrasted to existing observational results in the literature. By comparing our detection statistics to that of \cite{2014Panagoulia_vol}, we find a similar detection rate, with their study identifying 35 X-ray cavities and ours detecting 33; both yield the same numbers of X-ray cavities per cluster. Our results diverge only for MKW3s, where we did not identify the two X-ray cavities reported by \cite{2014Panagoulia_vol}. We also compare our measurements in terms of area and distance from the central SMBH with those by \cite{2014Panagoulia_vol} in Figure~\ref{fig:area_distance_pana_comp} and find that they align with a similar trend and no significant offset ($\leq$ 0.05 dex), and similar scatter. Although we could not retrieve the precise values for a quantitative comparison, our Prunier+2025 measurements of the X-ray cavity power and the ICM cooling luminosity also appear consistent with those reported by \citealt{2014Panagoulia_vol} (their Fig 4). We can hence confidently proceed by comparing our measurements of observed X-ray cavities with those from the TNG-Cluster simulation.

Before proceeding with the quantitative analysis, in Figure~\ref{fig:chandra_ex} we show two examples from the observational Prunier+2025 sample, Abell 4059 and Abell 0262, along with their analogs in the TNG-Cluster simulation: arrows point at the identified X-ray cavities. Whereas the analogs are identified based on thermodynamic summary statistics (see Section~\ref{subsec:meth_selection_tng}), their spatially-resolved features may be varied: most systems exhibit X-ray cavities, with a range of sizes and overall morphology or evolutionary stages -- we have further discussed the diversity of TNG-Cluster cavities in our previous work \citep[][Section 3.1]{Prunier2024}.

\subsection{Statistics and demographics}\label{subsec:res_demographics}
In the TNG-Cluster matched sample, we find at least one X-ray cavity in 37 out of 105 clusters (35 per cent), with a total of 73 X-ray cavities, whereas in the observational sample, X-ray cavities are identified in 15 out of 35 clusters (43 per cent), resulting in 33 X-ray cavities. The two panels of Figure~\ref{fig:demographics} summarize the detection statistics for both samples: overall, the fractional numbers of the two samples are in the right ballpark, but with one exception. The observational Prunier+2025 sample shows a higher fraction of clusters with pairs and quadruples: 47 per cent (7/33) of clusters with detections have a pair of X-ray cavities, and 20 per cent (3/33) contain two pairs of X-ray cavities (specifically, Hydra A, A3581, and A2052). The TNG-Cluster matched sample shows a lower proportion of X-ray cavity pairs (30 per cent) (see pie chart in Figure~\ref{fig:demographics}). Overall, the TNG-Cluster matched sample has fewer clusters with multiple simultaneous X-ray cavities (see Discussion \ref{disc:sect:agn}).

Figure~\ref{fig:demographics_CC_alt} shows the link between the cluster's cool-coreness and the number of identified X-ray cavities in both samples. Throughout, clusters with short central cooling times are classified as SCCs (t$_\text{cool} < 1$ Gyr), or WCCs (t$_\text{cool} $ < 1 Gyr and > 7.7 Gyr). Previous works suggest that cool-core clusters preferentially host X-ray cavities, while NCCs generally do not \citep[e.g.,][]{2005Dunn}. We find a clear trend between the presence of X-ray cavities and the cool-coreness of the host cluster in both samples: clusters with identified X-ray cavities are either SCCs or WCCs and the fraction of SCCs increase with the number of X-ray cavities hosted by the cluster. Specifically, in the TNG-Cluster sample, systems with identified X-ray cavities are predominantly cool-cores, with SCCs making up more than 88 per cent of clusters and hosting more than two. TNG-Cluster matched systems where no X-ray cavities are identified are mostly WCCs. This preference of X-ray cavities for SCC clusters also applies to the whole TNG-Cluster sample at $z=0$ \citep{Prunier2024}. Similarly, albeit not identically, all Prunier+2025 observed clusters hosting X-ray cavities are SCCs, while those without are a mix of WCCs, SCCs, and 50 per cent NCCs.

\subsection{Sizes, locations, and shapes}\label{subsec:res_sizes}
In the top left panel of Figure~\ref{fig:radius_dist_fraction}, we show the distribution of the mean size of X-ray cavities, calculated as the average of the fitted ellipse axes in kpc. X-ray cavity sizes in both samples range from a few kpc to a few tens of kpc, with a median of 8.4 kpc in TNG-Cluster matched and 8.3 kpc in the observational Prunier+2025. The width of the distributions, characterized by the 16th and 84th percentiles, is consistent between Prunier+2025 (3.3-11.4 kpc) and TNG-Cluster matched (5.3-11.4 kpc). However, the lower end of the TNG distribution is less skewed toward small X-ray cavity sizes compared to Prunier+2025. We discuss this further in Sec.~\ref{sec:discussion}.

\begin{table*}
    \caption{Results from the comparison of the properties of the X-ray cavities populations identified in this work across two samples of galaxy clusters: an observed one (which we have named Prunier+2025) and a simulated one (from TNG-Cluster), matched to the observational sample. Each row reports a different property or summary statistic. A checkmark in the right column indicates broad similarity and consistency, while a question mark indicates a possible disagreement.}
    \label{tab:disc_summary}
    \renewcommand{\arraystretch}{1.2} 
    \begin{tabular}{c c c c}
        \hline
        \textbf{X-ray cavities properties}  & \textbf{agreement} & \textbf{comments}  & \textbf{reference} \\
        \hline
        Fraction of clusters with X-ray cavities   & \checkmark & $\sim$ 35-43 per cent. & Fig.~\ref{fig:demographics}  \\
        Number of X-ray cavities per cluster       & ?          &  slight scarcity of pairs in TNG-Cluster.  & Fig.~\ref{fig:demographics} \\
        Prevalence of X-ray cavities in cool-core clusters  & \checkmark & clusters hosting X-ray cavities are CC or WCC. & Fig.~\ref{fig:demographics_CC_alt} \\
        Distributions of morphology i.e. ellipsoidal shape            & \checkmark  &  consistent, with varied elongation and average axis ratios near 1:1. & Fig.~\ref{fig:radius_dist_fraction} \\
        
        \multirow{2}{*}{Distributions of size/area: \it{attached} X-ray cavities} & \multirow{2}{*}{?} & Comparable median and width, but lack of a tail at the lower end of & \multirow{2}{*}{Figs~\ref{fig:radius_dist_fraction} and \ref{fig:area_distance}} \\  
        & & the area range; fewer X-ray cavities in TNG-Cluster with $\lesssim$\,100 kpc$^2$. & \\
                
        Distributions of size/area: {\it detached} X-ray cavities &  \checkmark & consistent, comparable median and width. & Fig.~\ref{fig:area_distance} \\

        Distributions of enthalpy & \checkmark & consistent, comparable median and width. & Inset Fig.~\ref{fig:Pcav_Lcool} \\
        Distributions of power & \checkmark & consistent, comparable median and width. & Fig.~\ref{fig:Pcav_Lcool} \\
        Scaling relation of area vs. distance to SMBH & \checkmark & consistent slope, normalization, and scatter. & Fig.~\ref{fig:area_distance} \\
        Scaling relation of power vs. cooling luminosity      & \checkmark & consistent slope, normalization, and scatter. & Fig.~\ref{fig:Pcav_Lcool} \\
        \hline
    \end{tabular}
\end{table*}

In the bottom left panel of the same figure, we show the mean size vs. distance from the central SMBH for the two samples. A positive correlation is present in both samples (with Pearson index > 0.5), and both follow a similar trend and scatter. A similar relationship is in place when the distance from the SMBH is scaled by the cluster cooling radius (bottom right panel of Figure~\ref{fig:radius_dist_fraction}): observed and simulated systems follow similar trends. However, in both scaling relations subtle differences are noticeable, as the two X-ray cavity populations appear to occupy somewhat different regions of the parameter space.

Even though the morphologies of X-ray cavities are approximated as ellipsoids, their actual shapes are likely more complex \citep[see][]{2006Rafferty,2012Hlavacek}. We quantify the ellipse axes of the X-ray cavities (along and perpendicular to the line connecting the SMBH to the center of each X-ray cavity), in the top right panel of Figure~\ref{fig:radius_dist_fraction}. We find that both populations of X-ray cavities display similar shapes, showing varying degrees of elongation without a discernible preference, and with the average distribution of the axes centered around 1:1 ratio (i.e. spherical shape).

\subsection{Areas and distances to the SMBH}\label{subsec:res_areadist}
Since the X-ray cavity shape is generally elliptical, the X-ray cavity area, rather than the mean size, may better represent its extent. Figure~\ref{fig:area_distance} shows the projected X-ray cavity areas vs. the distance (left main panel), a relationship supported by observations \citep[e.g.][]{2016Shin}, which indicate that X-ray cavities expand in size as they rise away from the cluster, i.e., the BCG center. We find that the resulting trend and scatter of the observed and simulated samples are similar. However, as the histograms at the top and side of the main panel show, the marginalized distributions for both samples indicate that TNG-Cluster X-ray cavities are on average slightly larger in area and are farther from their central SMBH compared to the observational Prunier+2025 sample. Within TNG-Cluster, there are proportionally fewer X-ray cavities located at a distance $\lesssim$ 10 kpc from the central SMBH, i.e. fewer $\lesssim$ 100 kpc$^2$ cavities than in the observed sample. Specifically, in terms of area distributions, TNG-matched cavities have a larger median (201 kpc$^2$) compared to Prunier+2025 (127 kpc$^2$) and also exhibit larger 16th percentiles (88 kpc$^2$ versus 34 kpc$^2$).

Our preliminary comparison of X-ray cavities between observations and TNG-Cluster, supplemented by considerations on the other available simulation analysis based on Hyenas \citep[see Figure 3 in][]{Prunier2024}, suggests that the qualitative agreement reported above is far from trivial. Notably, the TNG feedback model was neither designed nor calibrated to produce X-ray cavities, making this agreement an emergent prediction of the model rather than an expected outcome. We discuss the significance and implications of this in Section~\ref{sec:discussion}, while here we expand on different types of X-ray cavities. 

Namely, in the two histograms on the right-hand side of Figure~\ref{fig:area_distance}, we provide the distributions of X-ray cavity areas by distinguishing between attached (still connected to the central SMBH, top) and detached ones (rising in the intracluster gas, bottom). The {\it attached} population of X-ray cavities in TNG is notably less skewed toward small, $\lesssim$ 100 kpc$^2$ X-ray cavity areas, compared to Prunier+2025. Meanwhile, the {\it detached} population appears broadly consistent between the observed and simulated samples, except for the two large outer X-ray cavities in Hydra, which have no counterparts in the TNG-Cluster matched systems.

\subsection{X-ray cavity power vs. cooling luminosity}\label{subsec:res_Pcav_Lcool}
Finally, we quantitatively compare X-ray cavities from TNG-Cluster with observed ones in terms of energetics. Observational studies suggest that the energy released by X-ray cavities can effectively counteract the cooling of the hot ICM driven by bremsstrahlung emission. This balance is reflected in the commonly observed near 1:1 correlation between the estimated X-ray cavity power and the cooling luminosity in the cluster cores. We estimate the power of all the X-ray cavities in our two samples, P$_\text{cav}$, and the cooling luminosity, L$_\text{cool}$, of the corresponding clusters by using the methodology detailed in Section~\ref{subsec:met_power}. These results, along with uncertainties, are provided too in the already-mentioned Tables~\ref{tab:results_pana} and ~\ref{tab:results_tng}. 

Figure~\ref{fig:Pcav_Lcool} shows our correlation between X-ray cavity power and cooling luminosity for each cluster. The diagonal lines indicate the 1:1 correlation between cooling and heating, for energy inputs of p$_\text{th}V$, 4p$_\text{th}V$, and 16p$_\text{th}V$ per X-ray cavity, which correspond to different assumptions about the equation of state of the gas inside the X-ray cavity -- 4p$_\text{th}V$ representing a relativistic ideal gas. Additionally, we include power estimates from other observational studies of X-ray cavities in clusters at low redshifts ($z\leq0.2$), all of which align well with TNG-Cluster's. In fact, the X-ray cavities of the observed Prunier+2025 sample and of the TNG-Cluster matched systems appear to be in remarkable agreement: they both exhibit a similar trend (in both slope and normalization), where the power of X-ray cavities is similar to the cooling luminosity, and they also display a similar level of scatter. The top and bottom histograms show that P$_\text{cav}$ and L$_\text{cool}$ values follow similar distributions, exhibiting similar medians $\sim 2-4 \times 10^{43}$ erg s$^{-1}$, and 16th-84th percentiles (width) for both quantities.

As the power measurement can be influenced by the method used to compute the X-ray cavity age -- the denominator in the P$_\text{cav}$ formula eq.~\ref{eq:pcav} -- we also include in the inset of Figure~\ref{fig:Pcav_Lcool} the distribution of cavity enthalpy (E$_\text{cav}$ = 4p$_\text{th}$V). The observed and TNG-Cluster X-ray cavity populations exhibit a similar energy distribution with consistent median and width.

\section{Discussion and implications}\label{sec:discussion}

Our analyses and comparisons presented above indicate strong similarities between X-ray cavities in the TNG-Cluster simulation and those observed with Chandra in the Prunier+2025 sample constructed and analyzed in this work. We explicitly take stock of our results by summarizing them in Table~\ref{tab:disc_summary}.

According to our analysis, the TNG-Cluster simulation not only reproduces the observed ranges and scaling relations of X-ray cavity size, distance, and power but also naturally recovers the locus and scatter of these relations. This suggests that the simulation model is capable of capturing fundamental aspects of the interplay between AGN feedback and the ICM, even though none of the model designs were calibrated to produce them.
Interestingly, amid the agreement, we do find two differences between the simulated and observed populations of X-ray cavities: in the TNG-Cluster matched sample we identify (i) fewer pairs, and (ii) a scarcity of small, \textit{attached} X-ray cavities, i.e. those still connected to the central SMBH.

Given that our methods for identifying and characterizing X-ray cavities are consistent across both datasets, including as-similar-as-possible selection criteria and instrumental biases, these differences likely stem from the simulation itself. What drives these (dis)agreements, and what do they imply for the realism of the simulation and the fidelity of TNG's galaxy-formation model? Namely, can we use the revealed (dis)agreements to (in)validate the characteristics of the simulated AGN feedback, or of the ICM? 
And more specifically, do the characteristics of X-ray cavities in TNG-Cluster depend more on \textit{nature} -- the AGN feedback driving their inflation -- or on \textit{nurture}, such as the (thermo)dynamics and gas properties of the surrounding ICM? Both could play a role, as we discuss below, though the non-linear interplay among processes presents a significant challenge for disentangling their respective contributions.
 
Finally, the possibly lingering underlying selection biases behind our Chandra observational sample -- despite our best efforts to select a representative sample of clusters in Prunier+2025 and match it in the TNG-Cluster sample -- may still limit our ability to draw definitive conclusions about the TNG model. In the following, we discuss arguments and considerations on all these relevant directions.

\subsection{On the overlooked impact of ICM on X-ray cavities} \label{disc:sect:icm}

Traditionally, scaling relations such as X-ray cavity power vs. cooling luminosity have been used to infer characteristics of AGN feedback. However, as we have thoroughly discussed in \cite{Prunier2024}, the dynamical properties of the ICM also influence the appearance and manifestation of X-ray cavities, as well as their overall morphologies, radial ascent, and lifetime. For a complete discussion, we refer the reader to Section 5.2 of \cite{Prunier2024}, and here highlight some salient aspects and implications.

In TNG-Cluster, the dynamic state of the hot intracluster gas can disrupt X-ray cavities shortly after inflation via turbulence, sound waves, weak shocks from galaxy motions, and AGN activity, as well as cold fronts and merger-induced sloshing. Several numerical studies with idealized hydrodynamical simulations also suggest that bulk and chaotic motions of the ICM could disturb the jet-inflated cavities \citep[e.g.,][]{2012Mendygral,2019Yang,2020WittorGaspari}. 

\begin{figure}
 \hspace{-0.6cm}
  \begin{subfigure}[b]{0.26\textwidth}
    \includegraphics[width=\textwidth]{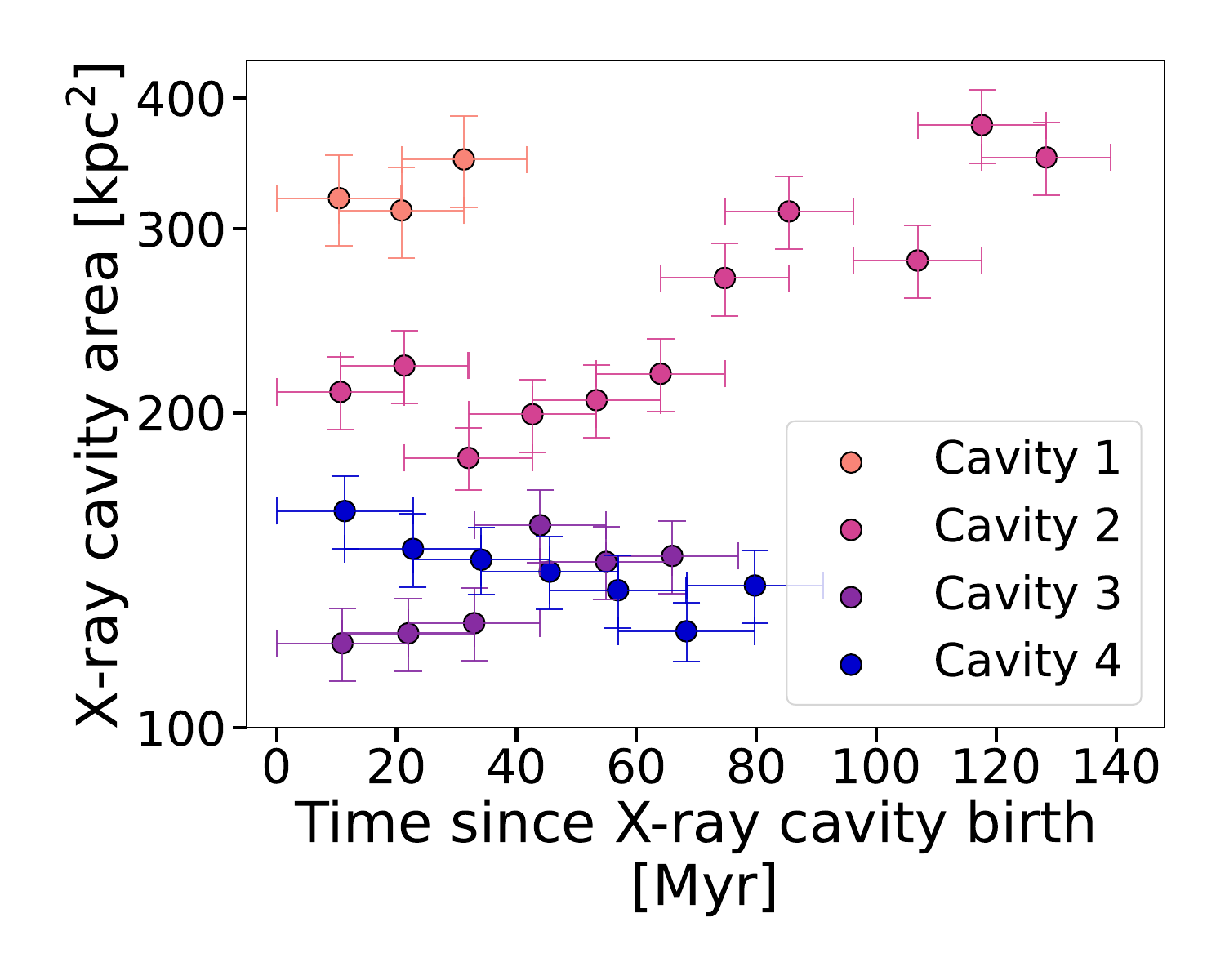}
  \end{subfigure}
  \hspace{-0.1cm}
    \begin{subfigure}[b]{0.26\textwidth}
    \includegraphics[width=\textwidth]{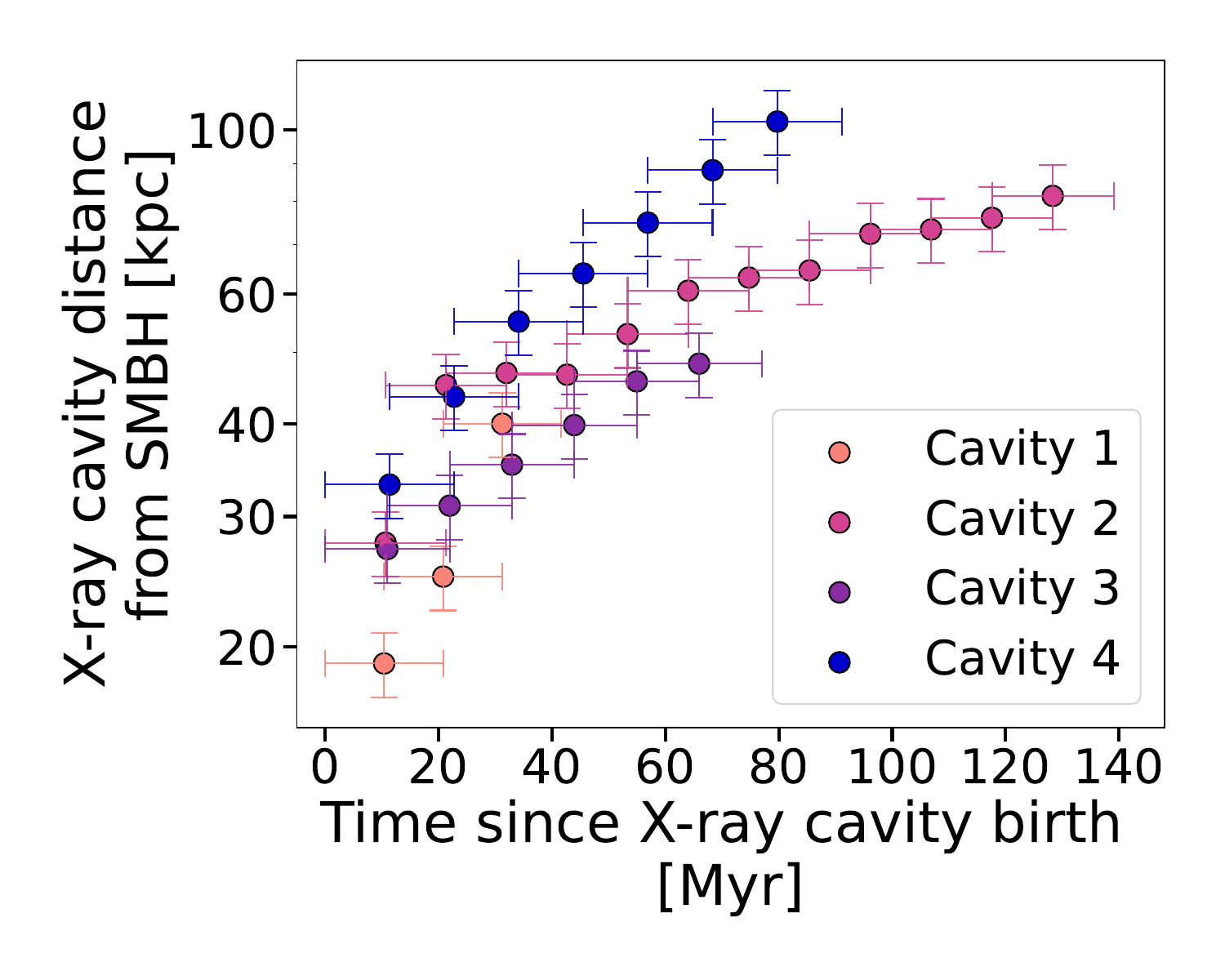}
  \end{subfigure}
\vspace{-0.7cm}
  \caption{Effects of the ICM, in simulations. We show the time evolution of four X-ray cavities that form and evolve in TNG300's most massive halo. Using a $\sim$10 Myr snapshot cadence, we track their evolution from birth to fading (no longer visible on X-ray maps). The x-axis shows time since their first identification. \textit{Left:} Projected distance from the SMBH for each X-ray cavity. \textit{Right:} Changes in projected area over time. The X-ray cavities show similar evolutionary patterns, but differing slopes and lifetimes reflect their dynamic interaction with the ICM.}
  \label{fig:study_case_subbox}
\end{figure}

This disruption is evident when tracking the spatial evolution of the X-ray cavities over time, as illustrated in Figure~\ref{fig:study_case_subbox}. Using the TNG300 subbox, with the same TNG model but with higher temporal resolution (snapshots every <10 Myr), we capture the birth and rise of four X-ray cavities over 140 Myr harbored by the same cluster. Their trajectories depend on the ICM's state -- calm, turbulent, or sloshing, from visual inspection --  which affects their ascent rates and path (left panel Fig.~\ref{fig:study_case_subbox}). Steeper slopes indicate rapid, radial ascent through undisturbed ICM, while gentler slopes reflect interactions with the surrounding gas motion. X-ray cavities generally expand as they rise, though their areas may visually decrease as they become fainter and blend with the ICM X-ray emission, or via projection effect (e.g., X-ray cavity 4 in the right panel of  Fig.~\ref{fig:study_case_subbox}). Lifetimes are also affected by the ICM dynamics, with cavity 1 lasting only 30 Myr before being erased by a sound wave or weak shock front.

\begin{figure}
\includegraphics[width=0.48\textwidth]{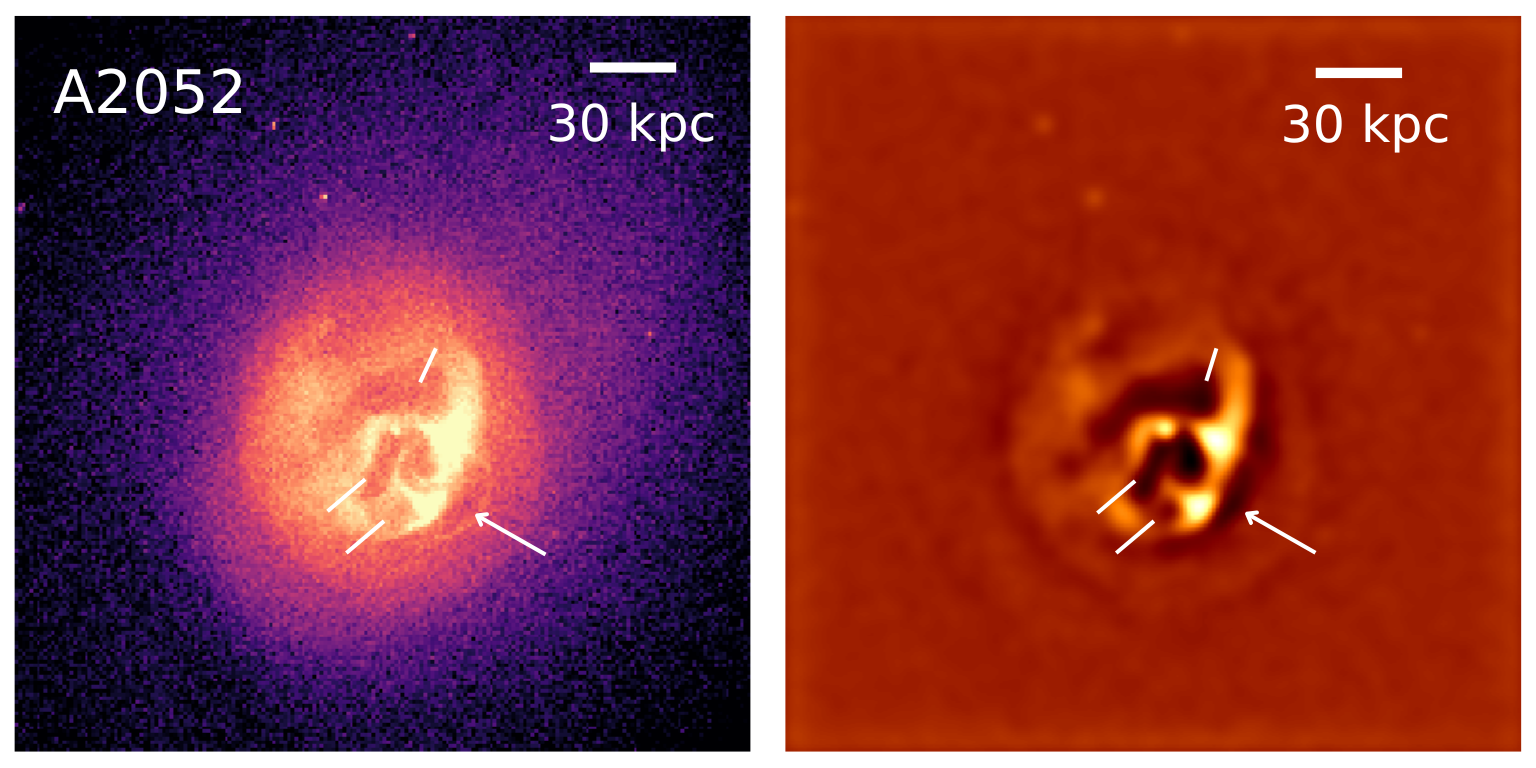}
    \caption{Effects of the ICM, in observations. \textit{Left:} Chandra X-ray image of A2052. White lines indicate the three X-ray cavities, with the arrow pointing to the, fourth, outermost cavity. \textit{Right:} (2,8) unsharp-masked X-ray image. The outer X-ray cavity appears elongated and is likely being disrupted by sloshing and/or by the expansion of the two southern cavities.}
    \label{fig:A2052}
\end{figure}

These findings suggest that X-ray cavities are influenced and shaped by the ICM kinematics in the cluster core. In the observational Prunier+2025 sample of clusters we also find a few suggestive examples: the visibly disturbed core of A2052 hints that the cluster has recently undergone a merger event with a subgroup, initialing gas sloshing that disrupts the outer cavity (Figure \ref{fig:A2052}). A detailed study of this system in \citealt{2010Randall} shows that sloshing patterns can have a significant effect on the rise trajectories of the buoyant X-ray cavity \citep[see Figure 19][]{2010Randall}. Similar patterns are also clearly observed in other systems such as M87 \citep{2007Forman} and Perseus \citep{2011Fabian_perseus}. 

While the importance and impact of the ICM dynamics are often overlooked in the interpretation of observational studies of X-ray cavities \citep[though see][]{2022Fabian,2023Olivares}, our results indicate that the dynamics of the ICM may in fact have a significant influence on the evolution and visibility of these structures.

\subsection*{Nature vs. nurture: is the ICM of TNG-Cluster realistic?} 

Building on the considerations above, we set out the task of discussing the realism of the ICM simulated within TNG-Cluster. As a starting point, we note that the comparable detection rates of clusters with X-ray cavities and the similar number of cavities per cluster (excluding the fraction of pairs) suggest that the dynamic state of the ICM may affect in similar manners the observed and simulated samples. This qualitative consistency may imply that the scarcity of small, attached, X-ray cavities in TNG-Cluster is not due to overly turbulent or unrealistic ICM dynamics in the simulation. To investigate further the realism of the cavity-ICM interactions in TNG-Cluster, we quantitatively examine the thermodynamic properties of the simulated clusters, both globally and in the vicinity of the X-ray cavities.

\begin{figure*}
    \centering
    \hspace{-0.3cm}
    \includegraphics[width=0.6\textwidth]{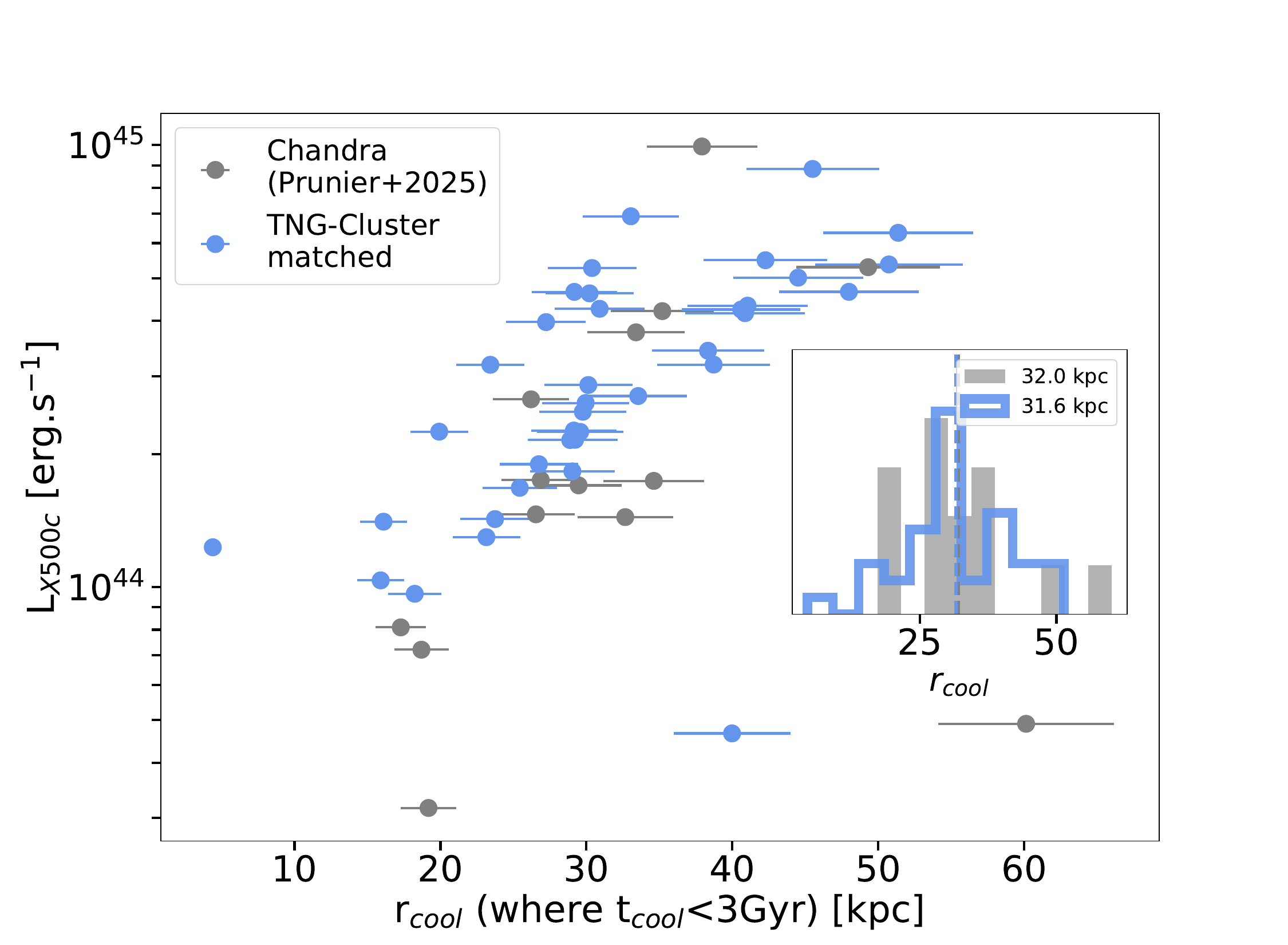}
    \hfill
    \includegraphics[width=0.8\textwidth]{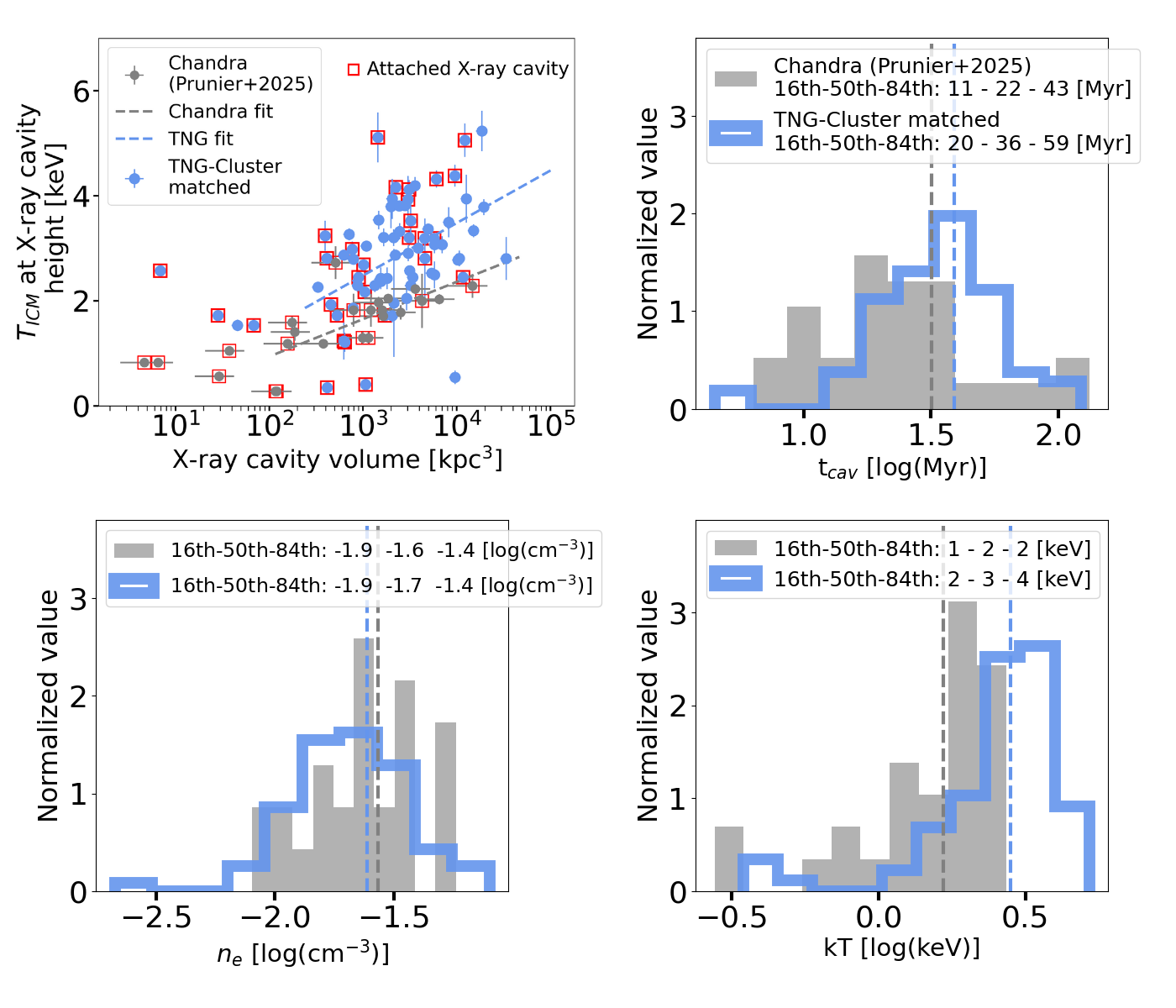}
    \vspace{-0.4cm}
    \caption{Is the ICM of TNG-Cluster realistic?
    \textit{Top:} The relationship between X-ray luminosity, measured within R$_{\text{500c}}$ in the soft X-ray band, and the cooling radius, defined as r$_\text{cool}$ where t$_\text{cool}$ < 3 Gyr. The inset displays the normalized distribution of cooling radii, with mean values indicated by dashed lines, for both the TNG-Cluster matched sample and observed clusters. Both samples follow a similar trend, with more luminous clusters exhibiting larger cooling regions. This highlights that the thermodynamic properties of the TNG-Cluster simulated clusters are consistent with those of observed clusters in their cores, at least from a zeroth-order qualitative comparison. 
    \textit{Bottom four panels:} Distribution of thermodynamic properties of the ICM at the locations of X-ray cavities, measured from spectral fitting on the deprojected profiles. \textit{Top left:} ICM temperature at the X-ray cavity azimuthal height compared to the X-ray cavity volume. The positive correlation (down to 1.5 keV) suggests that hotter gas environments are associated with larger X-ray cavities. \textit{Top right:} X-ray cavity age (sound crossing time) in Myr. \textit{Bottom left:} Electron density at the X-ray cavity location. \textit{Bottom right:} Temperature measured at the X-ray cavity location. The temperature gradient at TNG-Cluster cavity heights is slightly higher (+$\sim$1 keV), which may influence their expansion dynamics.}
    \label{fig:combined_thermo}
\end{figure*}

\paragraph*{Fidelity of the global ICM properties in TNG-Cluster}
Many global physical properties of halos in the TNG-Cluster simulation have been shown to align reasonably well with observations \citep{2024Nelson,2024Ayromlou,2024Lee,2024Truong,2024Lehle,2024Rohr}. The thermodynamical profiles of the 352 clusters -- density, temperature, entropy, and pressure -- are in the ballpark of observational inferences, along with the fraction of cool-core clusters at $z=0$ \citep{2024Lehle}. Their multiphase ICM kinematics broadly matches observations across various scales, from the cluster cores to their outskirts \citep{2024Ayromlou}. In the central regions of Perseus-like clusters, subsonic turbulence in the gas measured from XRSIM mocks in \cite{2024Truong} are consistent with the Hitomi result. 

We expand upon previous studies and present our comparison in Figure~\ref{fig:combined_thermo} (top). For the observed Prunier+2025 clusters hosting X-ray cavities and their TNG-Cluster counterparts, we show the X-ray luminosity extracted from the simulation and the literature for the simulated and observed clusters, respectively (see Method~\ref{sec:meth}), along with the size of the cooling region, derived from spectral fitting on both mock and real Chandra observations. We show only the matched sample of clusters considered in this study, sharing as closely as possible the same selection bias. Keeping this in mind, we see that both datasets have a similar trend and scatter and occupy similar loci; the cooling radii distributions are also consistent, as shown in the inset. X-ray luminous clusters tend to have larger cooling regions, suggesting comparable thermodynamic properties in the cores of the simulated and observed cluster population.

These considerations suggest that the global and core thermodynamic properties of the ICM in TNG-Cluster are quantitatively consistent with observations. However, local differences in the ICM properties at X-ray cavity height may influence cavity expansion and growth.

\paragraph*{ICM thermodynamics at the X-ray cavity height}

If the thermodynamics of the ICM influence the characteristics of X-ray cavities, then it may be important to verify more specifically the ICM properties at their azimuthal location i.e. height. We examine the temperature ($k_BT$) and the density ($n_e$) of the ICM gradient at the location of X-ray cavities by fitting the deprojected spectra in an annulus centered on the SMBH with radii equal to their height. We also compare the age distributions of X-ray cavities, relying on ICM sound speed estimates, as this quantity effectively probes the ICM thermodynamics (see Section~\ref{sec:meth} for details). We show these distributions in the bottom four panels of Figure~\ref{fig:combined_thermo} for both the simulated and observed clusters.

Again, the properties of the TNG-Cluster ICM at the radii probed by the X-ray cavities fall very well within the ballpark of the observational data. In particular, the ages and $n_e$ distributions (middle right and bottom left panels) are consistent between the simulated and observed clusters, and both samples exhibit a similar relationship between ICM temperatures at the X-ray cavity sites and their volume (middle left panel). All this represents a non-trivial outcome of the simulation and further validation of the TNG model at the cluster scales.

However, one difference also emerges. In TNG-Cluster the spectral-fitting background ICM temperatures at these sites are slightly higher (mean $k_BT$ \( \approx +1 \text{ keV}\)) compared to observations. This suggests a higher thermal pressure, p$_\text{th} \propto n_e (k_BT)$, at the X-ray cavity height (bottom right panel of Figure~\ref{fig:combined_thermo}). The higher pressure gradient of the ICM should slow down the area growth of X-ray cavities, as TNG cavities are in approximate pressure equilibrium with the surrounding ICM \citep[as shown in][]{Prunier2024}. 
This is reflected in the dependence of the X-ray cavity volumes on the surrounding gas temperature. We find a positive correlation between the volume of X-ray cavities and the ICM temperature at their height for both samples, suggesting that larger cavities are associated with a hotter ICM (top left panel of Figure~\ref{fig:combined_thermo}). 
This is consistent with \cite{2014Panagoulia_vol}, who noted that X-ray cavity size loosely depends on the ambient ICM temperature down to 1.5 keV, but differs from \cite{2016Shin}, who found little evidence for temperature dependence. The correlation is more pronounced for TNG-X-ray cavities than for observed ones and a shift towards higher temperature is manifest for the TNG-Cluster matched sample compared to the observational Prunier+2025 sample. 

However, the top left panel of Figure~\ref{fig:combined_thermo} illustrates that this trend is constant across the entire sample, not limited to attached X-ray cavities, and therefore might not explain their scarcity in TNG-Cluster.

\subsection{Nature vs. nurture: is the SMBH feedback of TNG-Cluster realistic?} 

In \cite{Prunier2024}, we demonstrated that X-ray cavities in the TNG simulations are caused by AGN feedback. In particular, these features naturally emerge from episodic and purely kinetic (no cosmic rays) feedback outbursts, injected unidirectionally by the SMBH. How could the TNG SMBH feedback prescription influence X-ray cavity appearance and expansion? Conversely, how can the results of this paper (Table\ref{tab:disc_summary}) be used to (in)validate or refine the TNG SMBH feedback model?

First of all, the very nature of the SMBH kinetic feedback of TNG is primarily responsible for the lower fraction of symmetric pairs in TNG-Cluster than those that are typically observed in real clusters. By design, the energy injections in TNG-Cluster are unidirectional and are not meant to replicate bipolar, ultra-relativistic, and collimated jets. As a consequence, the TNG kinetic feedback in TNG-Cluster predominantly produces either isolated X-ray cavities or asymmetric pairs formed by two closely timed, consecutive energy injections. However, this is less direct or obvious than it might seem. Even consecutive energy injections in random directions can emerge as symmetric or bipolar manifestations on larger scales, due to the influence of non-spherically-symmetric gas distributions in the vicinity of the SMBHs. These distributions create a path-of-least-resistance effect, as found in the star-forming disk of Milky Way-like galaxies in TNG \citep{2021PillepichErosita} and presumably in a fraction of TNG-Cluster BCGs that remain star-forming \citep{Rohr2024_cool,2024Nelson}.

In future generations of TNG-like galaxy formation models, the inclusion of bipolar collimated jets is certainly desirable, because these are manifestly observed in the Universe: however, the statistics on X-ray cavities quantified in this paper and in \cite{Prunier2024} would not justify per se the substitution, or a change of the functioning, of the current SMBH kinetic feedback model, but rather an addition of yet another mechanical radio-like mode.

In the following, we discuss more subtle aspects of the implementation of SMBH feedback in TNG and of the implications thereof from the results uncovered in this paper. In fact, in TNG, feedback episodes are self-regulated, i.e. both the frequency of energy injections and the amount of energy released over a given timescale depend on the SMBH accretion rate and thus on the properties of the surrounding gas \citep{2017Rainer}. However, the parameterization of this model involves several choices that may have an influence, directly or indirectly, on X-ray cavity characteristics, whether they are in agreement or disagreement with observations.

\begin{figure*}
    \centering
    \hspace{-0.4cm}
    \includegraphics[width=0.95\textwidth]{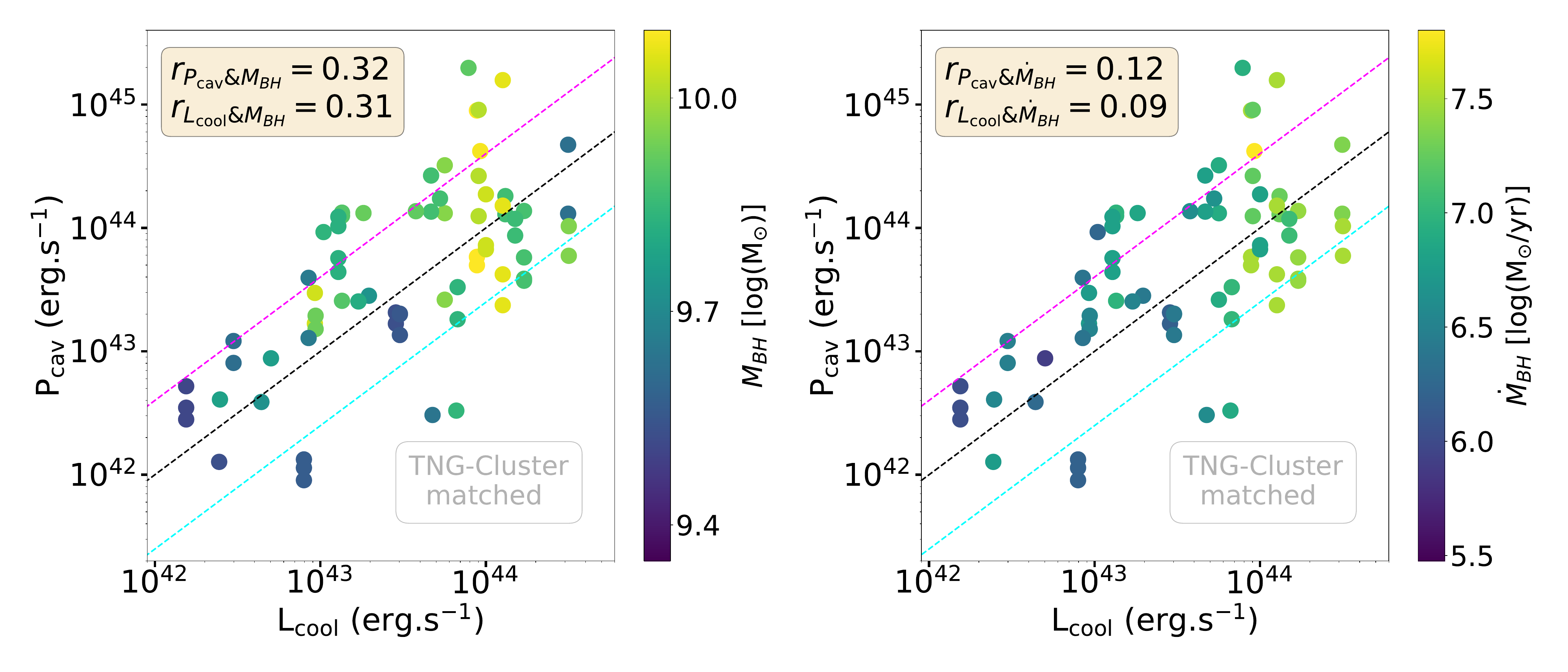}
    \caption{Connection between X-ray cavity power, cooling luminosity, and SMBH properties in TNG-Cluster X-ray cavities. The panels show X-ray cavity power versus cluster cooling luminosity within the cooling radius, color-coded by SMBH mass (right) and instantaneous accretion rate (left). Pearson correlation coefficients, indicated in the legends, suggest weak positive correlations, implying that more powerful X-ray cavities are powered by massive SMBHs accreting at higher rates and associated with stronger cooling clusters.}
    \label{fig:pcav_lcool_correlation}
\end{figure*}

\paragraph*{X-ray cavities in connection to SMBH properties}

Despite the fact that no specific model parameters are explicitly designed to create X-ray cavities in TNG-Cluster, notable and observed scaling relations naturally emerge, as shown in the previous Sections.

In the TNG-Cluster matched sample, we measure a best-fit scaling between $\log$  P$_\text{cav} \propto$ 0.67 $\log$ L$_\text{cool}$, fitted in log-log space using linear regression, and a Pearson correlation index of 0.64, meaning that the power of X-ray cavities is nearly proportional and correlated to the cooling luminosity. In fact, only a few TNG X-ray cavities have weak powers, below the 16p$_\text{th}$V line, pointing towards a significant contribution of X-ray cavities to balancing cooling within clusters.

TNG X-ray cavities resulting from different energy injections within the same cluster can exhibit widely varying power levels, spanning orders of magnitude. This variability is a direct consequence of the dynamic nature of feedback events, where the energy released can fluctuate significantly from one event to another, even within the same cluster. As a result, we expect that this variability contributes to the scatter observed in the P$_{\text{cav}}$-L$_{\text{cool}}$ relation in Figure~\ref{fig:Pcav_Lcool} and hints at the direct connection between the gas accretion, the AGN, and the self-regulation of kinetic feedback events in the TNG model. 

Can we really see a direct relationship between X-ray cavity power and e.g. SMBH mass and accretion rate? Figure~\ref{fig:pcav_lcool_correlation} shows that higher SMBH masses are associated with more powerful X-ray cavities and higher L$_\text{cool}$ values, indicating a positive correlation. Similarly, there is a weak positive correlation between the SMBH’s instantaneous accretion rate at $z=0$ and cavity power. These results suggest that clusters hosting more massive and rapidly accreting SMBHs tend to produce more energetic feedback, leading to higher P$_\text{cav}$. Notably, in \cite{Prunier2024}, we found no correlation between cavity size or distance and SMBH properties, at least at the time of inspection i.e. $z=0$, and across all available 352 clusters. In contrast, Figure~\ref{fig:pcav_lcool_correlation} suggests that P$_\text{cav}$ may be more directly influenced by the SMBH properties, integrated or instantaneous, as typically expected: as cooling ICM gas infalls toward the SMBH, it drives accretion, leading to more energetic feedback events that carve powerful X-ray cavities.

However, whether the diversity of SMBHs and their activity is the culprit for the scatter in the P$_{\text{cav}}$-L$_{\text{cool}}$ relation is not immediately clear from the mostly vertical trends of the color coding of Figure~\ref{fig:pcav_lcool_correlation}: more massive clusters are expected to host more massive SMBHs and to have higher cooling luminosities. A more direct proof of the SMBH-related origin of such a scatter would be higher cavity powers at fixed cooling luminosities for more massive or more active SMBHs, i.e. horizontal color trends in Figure~\ref{fig:pcav_lcool_correlation}: for TNG-Cluster, this is the case only towards the mid-lower end of the L$_\text{cool}$ distribution.

\paragraph*{Is the TNG-SMBH feedback too powerful?}
How could the TNG feedback prescription explain a scarcity of small attached X-ray cavities? One possibility is that if SMBH outbursts in TNG-Cluster are consistently or occasionally too powerful, this could manifest in a scarcity of small {\it attached} X-ray cavities.

In \cite{Prunier2024}, using the TNG300 subbox, we evaluated that X-ray cavity formation is associated with SMBH kinetic energy releases that range between \(10^{42-45}\) erg s$^{-1}$. These intrinsic estimates are of the same order as the inferred power values for the X-ray cavities in TNG-Cluster (Fig.~\ref{fig:Pcav_Lcool}, Table~\ref{tab:results_tng}). In turn, the latter for TNG-Cluster matched sample are in striking agreement with the observational sample and other studies. Therefore, assuming P$_\text{cav}$ as an upper limit for the outburst energy used to create the X-ray cavity, the SMBH feedback modeled in TNG and studied in TNG-Cluster does not appear to be excessively powerful compared to observational inference. 

In this context, although not shown here, we have examined how the inferred powers of attached and detached cavities compare, both within and between the simulated and observed samples. We find that within each sample, attached cavities are typically somewhat more powerful than detached ones. This difference in the power of attached X-ray cavities appears slightly larger in the simulated samples, however, given the width of the distributions, these differences remain marginal.
We stress that these arguments rely on power estimates using observational methods, which assume sound speed propagation and pressure equilibrium of X-ray cavities within the gaseous medium. While suitable for detached cavities that have risen away from the SMBH, these assumptions are less applicable to attached cavities, whose dynamics are influenced by ongoing AGN activity. Therefore, we cannot conclude that power arguments may be the culprit for the lack of small attached X-ray cavities in TNG-Cluster, but additional focus studies may be granted.

\paragraph*{AGN feedback timing in TNG-Cluster and observations} \label{disc:sect:agn}

\begin{figure}
    \includegraphics[width=0.45\textwidth]{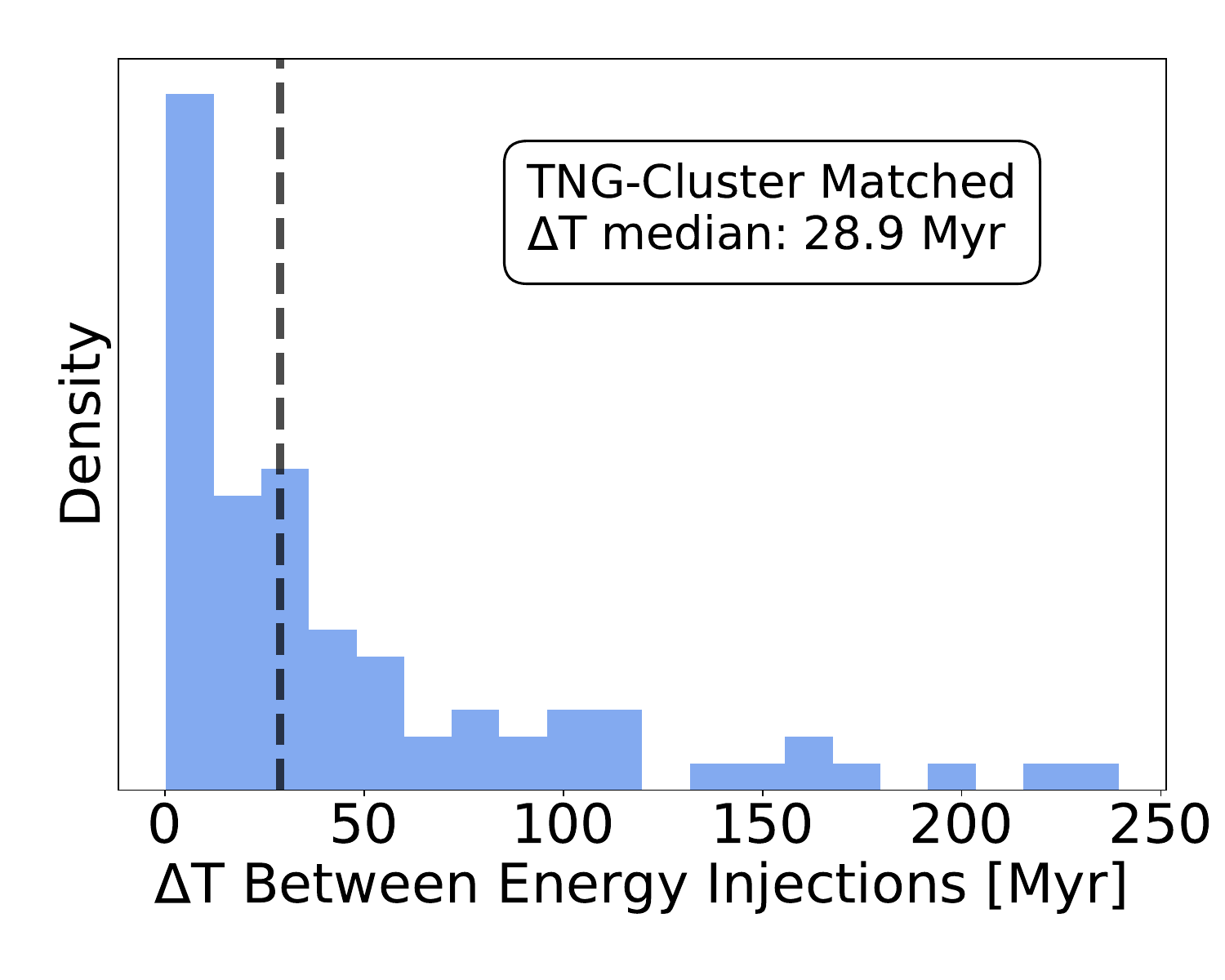}
    \caption{Timing between subsequent SMBH feedback events in the TNG-Cluster sample (for the last three energy injections of each central SMBH). On average the timing distribution between SMBH energy injections is a few tens of Myr: these are qualitatively similar to the ones derived in observational studies.}
    \label{fig:timing_inj}
\end{figure}

In \cite{Prunier2024}, we showed that feedback events in TNG are typically spaced by a few tens of million years (Myr). In Figure~\ref{fig:timing_inj}, we further illustrate the distribution of feedback event timings in the TNG-Cluster matched sample of clusters. By examining the time intervals between subsequent energy injections from central SMBHs, we find them to range from a few Myr to several hundred Myr, with a median of  $\sim$ 30 Myr. These rates are in the ballpark of estimates from observational studies, which suggests that central SMBHs typically trigger new cycles of activity every ten to hundreds of Myr, as indicated by the buoyancy timescale between different generations of X-ray cavity pairs in the same cluster (first in \cite{2013Babul}, see also \cite{2014Vantyghem,2016Sanders_centaurus,2021Biava,2022aOlivares}). Additionally, the number of X-ray cavities per cluster, typically one to four, is in agreement with both the TNG-Cluster sample and observational data. These findings suggest that the frequency of kinetic energy injection in TNG-Cluster falls within the range inferred in real clusters. We warn the reader that this consistency, although interesting and favorable for the TNG model, needs to be taken with a grain of salt. In fact, in the TNG feedback model, each outburst delivers energy in a single, short time-scale, ``instantaneous'' event\footnote{More precisely, SMBH energy injection occurs over a fixed time interval corresponding to the local timestep of the gas cells.}, rather than lasting for tens of millions of years, as inferred in observations of clusters like Perseus or M87, or as implemented in many idealized jet simulations \citep{2018Cielo}. Indeed, observed attached X-ray cavities are often linked to radio emission, indicating active or recent SMBH influence \citep{2007Forman}. Similarly, in the Prunier+2025 sample, all clusters hosting X-ray cavities are associated with radio emission \citep{2014Panagoulia_vol}. TNG uses ‘single-shot’ feedback events in contrast to a longer-lasting and/or precessing jet event, which could impact the characteristic expansion timescale of X-ray cavities.

\paragraph*{X-ray cavity formation: a rapid inflation phase in TNG}
Finally, the scarcity of small, attached X-ray cavities in TNG-Cluster, could be related to their inflationary phase occurring faster in the simulation than in observations. This would reduce the likelihood of capturing inflating cavities at a fixed snapshot\footnote{The region where the SMBH kinetic energy is initially injected is defined by a prescribed number of neighboring gas cells in an SPH-like kernel-weighted fashion \citep[see][]{2017Rainer}.}. Indeed, only 40 per cent of TNG-Cluster cavities are classified as attached, compared to 60 per cent in the observational Prunier+2025 sample. Furthermore, inspecting the high time cadence ($<$ 10 Myr) subbox we find that the attached phase of X-ray cavities is rapid, typically under 10 Myr, before they detach and rise through the ICM, where they remain visible up to ten times longer, increasing their probability to be identified in mock Chandra X-ray observations. 

Therefore, even if the SMBH kinetic energy injection were perfectly realistic, we might not be able to catch X-ray cavities at a given snapshot due to the short timescale over which they inflate (in comparison to the longer time interval in between consecutive TNG-Cluster snapshots: about 150 Myr). However, whereas the TNG model assumes an ``instantaneous'' energy injection, followed by rapid inflation of cavities, observational studies suggest a more continuous inflation process associated with longer-lasting energy injections, as discussed above. These differences might help explain the lower prevalence of attached X-ray cavities in the simulation and could be the direction for future investigations.

\subsection{Nature vs. nurture: final thoughts}
To conclude, based on our analyses of both simulated and observed clusters, we argue that the size/area of X-ray cavities likely depends on both the characteristics of the injected SMBH feedback energy as well as the thermodynamic state of the gas in the core region. Given the complex interplay between these two factors, where each influences the other, it remains challenging to pinpoint the exact driver behind the scarcity of small attached X-ray cavities in TNG-Cluster.

However, we argue that detached X-ray cavities may be more influenced by the ICM in which they expand, retaining less memory of the feedback injection event that created them compared to attached cavities. This is because detached cavities represent a later stage in their evolution and generally cover larger areas, distances, and ages. Despite the good consistency between the X-ray cavity populations of TNG-Cluster and observations, the attached cavities present a point of tension: future research, both from simulations and observations, should focus more on attached cavities to better constrain AGN feedback models.

\subsection{Considerations on the present analysis}  \label{subsec:discu_limit}

Our analysis shows that X-ray cavities in TNG-Cluster largely align with those observed with Chandra in Prunier+2025, with only a few discrepancies (see Table~\ref{tab:disc_summary} and Sections above). It is worth considering how different choices or assumptions in the analysis process might have influenced the outcomes of the present work. Below, we examine key methodological factors that could influence either the comparative nature of our study or, more generally, the broader investigation of X-ray cavities in clusters.

\begin{itemize}

    \item \textbf{Sample selection.} The selection process for the Prunier+2025 sample adopted throughout this study (Section~\ref{subsec:meth_selection_obs}) begins with a parent volume-limited sample presented in \cite{2014Panagoulia1}, comprising 289 sources located within 300 Mpc, that was reduced to 65 sources based on the availability of observations from Chandra or XMM-Newton and the ability to derive radial deprojected profiles of the ICM. We further refined the sample to 35 objects by selecting only galaxy clusters observed with Chandra. While the parent sample itself is unbiased, the selection process involving cross-matching with Chandra archival data and ensuring sufficient data quality for deprojection introduces a bias toward brighter, well-studied systems. 
    Although this bias should not critically impact the comparative aspect of this work, it is important to consider for two reasons. First, the underlying selection bias in Prunier+2025 limits the representativeness of our study for the entire X-ray cavity population in clusters. Second, as we replicate the sample selection by choosing analogs in TNG-Cluster, it also affects the assessment of the realism, in a more generalizable sense, of the TNG model outcomes. It would be informative to investigate whether any of the X-ray cavity-related analyses presented here are sensitive to the specific matched sample or whether they remain robust across variations in sample selection criteria.

    \item \textbf{X-ray cavity detectability: impact of photon counts.} The detectability of X-ray cavities in observed clusters depends on the quality of the available data (see discussion and Figure~\ref{fig:exposure_times} in Annex \ref{annex:exp_time}). Clusters in the Prunier+2025 sample with fewer than $10^{4}$ photon counts in the core typically lack identified cavities \citep[see also][]{2014Panagoulia_vol}. However, in the TNG-Cluster matched sample, the dependence between photon counts and cavity identification, while present, has a less sharp cut-off: this requires further investigation. 

    \item \textbf{Uncertainties on the X-ray cavity volume and projected quantities.} The X-ray cavity volume was estimated using the ellipse formula (Section~\ref{subsec:met_power}), approximating cavities as prolate or oblate spheroids with symmetry assumed along the axis nearest the galactocentric direction \citep[][]{2014Panagoulia_vol,2006Allen,Plsek}. However, X-ray cavities can exhibit various shapes, introducing uncertainties that likely contribute to the scatter in the P$_\text{cav}$-L$_{\text{cool}}$ relation across the cluster samples. In addition, we do not evaluate the impact of projecting quantities such as distances and size on X-ray cavity energetics. As we apply the same methodology in both the Prunier+2025 and TNG-Cluster matched sample, these effects should not impact the conclusions from the comparative aspect of our work.
    
    \item \textbf{Ideal equation of state.} In the TNG model, gas throughout the simulated domain follows the ideal monatomic gas equation of state with \(\gamma= 5/3\); namely, X-ray cavities in the simulation are not filled with relativistic gas with a different $\gamma$. We recompute the TNG X-ray cavity power using the formula for a non-relativistic filling gas P$_\text{cav}$ = 2.5p$_\text{th}$V (as shown in annex Figure~\ref{fig:2.5pV_gamma_nn_relat}). This adjustment only leads to a systematic $\sim$ 0.2 dex downward shift of P$_\text{cav}$: however, the ensuing scaling relation from TNG-Cluster is still well consistent with the observational data.

\end{itemize}

\subsection{Considerations for future work}  \label{subsec:discu_futur}

Looking ahead, the realism of TNG X-ray cavities offers a valuable pathway for further explorations of AGN feedback in clusters within a realistic cosmological context. Several areas of further study may be promising.

\begin{itemize}
    
    \item \textbf{Primary heating channel.} The connection between TNG-X-ray cavities and a self-regulated feedback loop that offsets gas cooling and mitigates star formation has not yet been explored in this work. Additional mechanisms potentially related to AGN, such as sound waves \citep[e.g. in Perseus,][]{2006Fabian_cond}, weak shocks \citep[e.g. RBS 797,][]{2023Ubertosi_rbs}, and turbulence \citep[e.g.][]{2014Zhuravleva}, may also contribute to heating and might be necessary to fully offset cooling in some clusters. Similar questions have arisen in studies of Abell 2052 \citep{2009Blanton}, and Centaurus \citep{2016Sanders_centaurus}, where pressure waves may transport energy from the nucleus, providing the distributed heating needed to counter high cooling rates. TNG-Cluster can be used to probe the specific contribution of these processes to balance radiative losses from the ICM.

    \item \textbf{Hidden cooling flows.} Recent studies have inferred potentially ``hidden'' cooling flows in the innermost regions of ellipticals, groups, and clusters cores \citep[e.g.,][]{Fabian_hiddend1,Fabian_hiddend2,Fabian_hiddend3,Fabian_hiddend4}, where gas may in fact cool below 1 keV but its soft X-ray emission may be absorbed by dusty gas, re-emitting energy in the far-infrared. While AGN feedback suppresses cooling in outer regions, soft X-ray cooling persists in the cores of systems. This effect has been suggested to be more important in lower-mass elliptical galaxies and groups \citep[][]{Fabian_hiddend3}. TNG-Cluster, occasionally returns cool, i.e. <10$^4$ K disky structures in BCG cores that are actively star-forming: a dedicated study of the cold gas content in TNG-Cluster systems can be found in \cite{Rohr2024_cool}. To fully understand these structures, their implications, and if they are eventually linked to hidden cooling flows, a dedicated study is required.
\end{itemize}

\section{Summary and conclusions}\label{sec:conclusion}

In this paper, we have performed a direct comparative study between X-ray cavities that we identify in galaxy clusters observed with Chandra versus those in the TNG-Cluster suite of cosmological magnetohydrodynamical simulations. In a previous paper \citep{Prunier2024}, we demonstrated that TNG-Cluster returns a diverse population of X-ray cavities in about 40 per cent of its clusters at $z=0$, and demonstrated that these are due to episodic, high velocity, kinetic energy injections from the central SMBH. Here we set out to undertake a quantitative like-for-like comparison between observed and simulated X-ray cavities to evaluate the realism of the former.

We first selected a volume-limited sample of 35 clusters from the Chandra Archive, which we refered to as ``Prunier+2025'' and which includes both cool-core and non-cool-core clusters. We then identified 105 matching clusters in TNG-Cluster (``TNG-Cluster matched'') and generated equivalent, mock Chandra X-ray observations for each (Figs.~\ref{fig:sample_selection_panel} and \ref{fig:chandra_ex}). In both samples, we applied an identical methodology for the detection and analysis of X-ray cavities, by following techniques that are typically adopted with observational X-ray data (Section~\ref{subsec:met_power}) and enabling an apples-to-apples comparison. As a preliminary methodology check, we confirm consistency between our analysis of our observational sample and that from previous observational works (Fig.~\ref{fig:area_distance_pana_comp}).

We find striking similarities between the populations of simulated and observed X-ray cavities, as their frequencies, demographics, morphologies, area-distance trends, and energetics all show good agreement. More specifically, our key findings are:

\begin{itemize}

    \item[$\blacksquare$] The total fraction of galaxy clusters harboring X-ray cavities is similar across the two samples, being 15/35, i.e. 43 per cent for the observational Prunier+2025 set and 37/105, i.e. 35 per cent for TNG-Cluster matched systems (Fig.~\ref{fig:demographics}). Clusters hosting cavities are either weak or strong cool-cores in both samples (Fig.~\ref{fig:demographics_CC_alt}).\\

    \item[$\blacksquare$] Both samples of clusters exhibit X-ray cavities at various evolutionary stages. These can be attached to the central SMBH, or detached and rising in the hot cluster atmosphere. Observed Prunier+2025 clusters show a higher abundance of X-ray cavity pairs and quadruples (67 per cent) compared to TNG-Cluster matched X-ray cavities (38 per cent), which are predominantly single (Fig.~\ref{fig:demographics}). This difference demonstrates that details in the AGN feedback model clearly manifest in the spatial distribution of X-ray cavities.\\

    \item[$\blacksquare$] Simulated and observed X-ray cavities have similar morphologies, sizes, and also consistent area vs. distance-to-SMBH trends and scatter (Figs.~\ref{fig:radius_dist_fraction} and \ref{fig:area_distance}). However, we find a scarcity of small-sized attached X-ray cavities in the TNG-Cluster matched sample (right panel of Fig.~\ref{fig:area_distance}) in comparison to the observed ones. This difference is challenging to explain due to the complex interactions between the environment (i.e. the intracluster gas properties) and the engine (i.e. SMBH energy release) and we find no smoking gun as culprit.\\
        
    \item[$\blacksquare$] The power of TNG-Cluster X-ray cavities closely balances the cooling luminosity of their host cluster. They exhibit a clear scaling relation between X-ray cavity power and cluster cooling luminosity, as observed. This suggests that, as obtained and interpreted in observational studies, most simulated X-ray cavities provide sufficient energy to counterbalance cooling, consistent with our observational sample and other studies (Fig.~\ref{fig:Pcav_Lcool}).\\

    \item[$\blacksquare$] In TNG-Cluster, X-ray cavities within the same cluster, generated by distinct energy injections, display inferred powers that vary by orders of magnitude. This variability contributes to scatter in the P$_\text{cav}$-L$_{\text{cool}}$ relation (Fig.~\ref{fig:Pcav_Lcool}) and may illustrates the link between SMBH accretion and feedback self-regulation feedback in the TNG model, where the energy of each feedback episode is modulated by the physical conditions of the gas accreted by the SMBH (see also correlation plots Fig.~\ref{fig:pcav_lcool_correlation}).\\

    \item[$\blacksquare$] In both samples, the dynamics of the intracluster gas can significantly influence the longevity and behavior of X-ray cavities (Figs.~\ref{fig:study_case_subbox} and \ref{fig:A2052}). We hence stress the importance of accounting for the surrounding gas dynamical state when interpreting diagnostics based on X-ray cavities and their implications on SMBH feedback.
    
\end{itemize}

In practice, in this work, we have demonstrated that AGN feedback in the IllustrisTNG model not only generates a diverse population of X-ray cavities at the current cosmic epoch, but also that these cavities closely align with observations (Table~\ref{tab:disc_summary} for a summary). Crucially, TNG-Cluster broadly reproduces the shape and the scatter of observed relations. This realism of TNG X-ray cavities constitutes a non-trivial, and somewhat unexpected, validation of the TNG galaxy formation model and provides support for novel avenues to studying AGN feedback in clusters within a highly realistic cosmological context. In fact, the similarities between TNG-Cluster and observed X-ray cavities are even more striking given that, in the simulations, these arise from a relatively simple AGN feedback model of episodic kinetic energy injections that differs fundamentally from the collimated jets observed in many clusters.
In fact, beyond TNG, various AGN feedback implementations have been shown to produce X-ray cavities, including the SIMBA model \citep{2024Jennings}, the model of Romulus-C \citep{2019Tremmel}, and higher-resolution jet models of idealized cluster simulations \citep[e.g.][]{2022Huvsko}. Therefore, the mere presence of X-ray cavities is not necessarily a strong constraint on AGN feedback models. A direct and quantitative comparison between observed and simulated cavity populations remains crucial. In particular, the \citet{2024Jennings} study of Hyenas galaxy groups shows that a different feedback model can similarly reproduce observed scaling relations of cavity area–distance to SMBH and P$_\text{cav}$-L$_{\text{cool}}$. However, it is informative to understand whether the positions of simulated systems within the scaling relations align with observed cavities in real halos in the same mass regime, rather than just with the overall trends. Our finding that X-ray cavities in the \textit{matched} TNG-Cluster systems occupy a similar locus in the cavity area–distance to SMBH and P$_\text{cav}$-L$_{\text{cool}}$ relations as observed clusters is a significant result.

Whether the two lingering points of tension between TNG-Cluster and Chandra X-ray cavities are significant enough to require changes to the underlying SMBH feedback model in TNG is not obvious. On the one hand, the relative scarcity of bipolar X-ray cavities in TNG-Cluster, alongside the observation of bipolar jets in the Universe (regardless of their direct connection to X-ray cavities), supports the need to include another mechanical radio-like mode, rather than modifying the current low-accretion kinetic mode. On the other hand, the scarcity of small attached X-ray cavities offers much less clear implications for the SMBH feedback implementation of TNG-Cluster, with prolonged rather than ``instantaneous'' energy releases being a possible exploration avenue.

It remains to be seen how the inclusion of cosmic ray physics may affect the quantitative and population-wide X-ray manifestations of cavities in clusters. Nonetheless, given that attached X-ray cavities have arguably better chances to retain ``memory'' of the SMBH feedback event that has generated them, whereas detached cavities may be more modulated by the surrounding ICM and its dynamics, we suggest that future studies aiming to use X-ray cavities to constrain AGN feedback should focus on attached X-ray cavities. Careful and accurate comparisons between {\it large} populations of simulated and observed systems are however imperative, and their correct execution and interpretation require, in turn, a deep understanding of observational biases and selection functions.

\section*{Acknowledgements}
MP thanks Carter Rhea for insightful discussions, as well as Nicolas Esser for comments on the manuscript. MP acknowledges funding from the Physics Department of the University of Montreal (UdeM) and the Centre for Research in Astrophysics of Quebec (CRAQ). MP and AP acknowledge funding from the European Union (ERC, COSMIC-KEY, 101087822, PI: Pillepich). JHL acknowledges funding from the Canada Research Chairs and from the Discovery grant program from the Natural Sciences and Engineering Research Council of Canada (NSERC). DN acknowledges funding from the Deutsche Forschungsgemeinschaft (DFG) through an Emmy Noether Research Group (grant number NE 2441/1-1). 

The TNG-Cluster simulation has been executed on several machines: with compute time awarded under the TNG-Cluster project on the HoreKa supercomputer, funded by the Ministry of Science, Research and the Arts Baden-Württemberg and by the Federal Ministry of Education and Research; the bwForCluster Helix supercomputer, supported by the state of Baden-Württemberg through bwHPC and the German Research Foundation (DFG) through grant INST 35/1597-1 FUGG; the Vera cluster of the Max Planck Institute for Astronomy (MPIA), as well as the Cobra and Raven clusters, all three operated by the Max Planck Computational Data Facility (MPCDF); and the BinAC cluster, supported by the High Performance and Cloud Computing Group at the Zentrum für Datenverarbeitung of the University of Tübingen, the state of Baden-Württemberg through bwHPC and the German Research Foundation (DFG) through grant no INST 37/935-1 FUGG. 

All the analysis and computations associated to this paper have been realized on the Vera cluster of the MPCDF.

\section*{Data Availability}
The TNG simulations, including TNG-Cluster, are publicly available and accessible at \url{www.tng-project.org/data}, as described in \cite{2019NelsonPublicReleaseTNG}. The mock Chandra images generated for this paper are accessible there. Other data directly related to this publication are available on request from the corresponding author.

\bibliographystyle{mnras}
\bibliography{TNG-Cav-Xray-Comp}

\appendix

\section{On the impact of photon counts for X-ray cavity detectability}\label{annex:exp_time} 

\begin{figure*}
    \centering
    \includegraphics[width=0.87\textwidth]{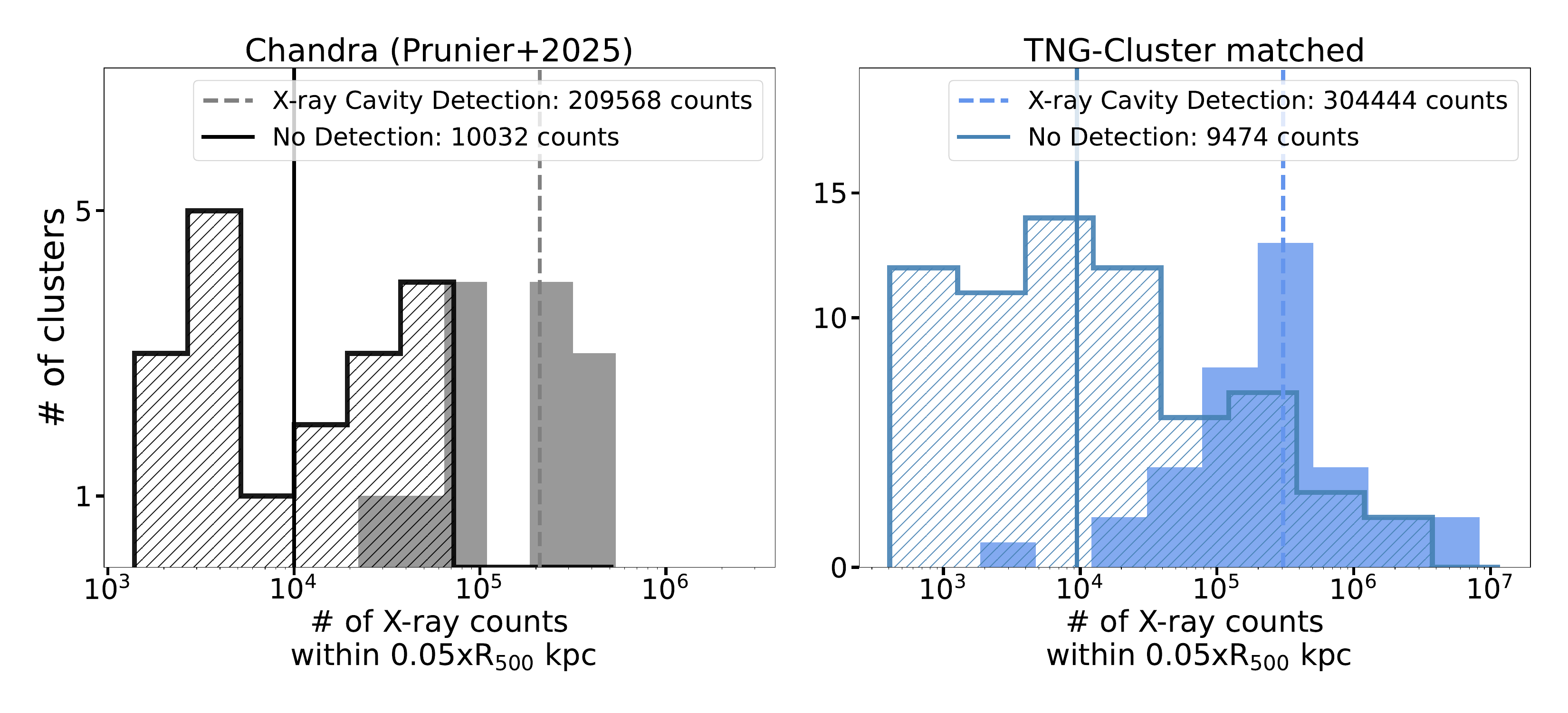}
    \caption{Influence of the photon count on the detection of X-ray cavities. \textit{Left:} Histogram of the total number of counts within the central 50 kpc of each source. The hatched black histogram represents clusters without identified X-ray cavities, while the filled light grey histogram represents clusters with at least one identified X-ray cavity. The legend specifies the median of each distribution. \textit{Right:} Same histogram for the TNG-Cluster matched clusters. In both samples, clusters with lower photon counts have fewer X-ray cavity detections.}
    \label{fig:exposure_times}
\end{figure*}
We investigate how the detectability of X-ray cavities in one cluster depends on the quality of the data available for that source. In the Prunier+2025 sample, we find a clear link between the number of X-ray photon counts within a 50 kpc aperture (with background-subtracted images and point sources have been removed) and the detection of X-ray cavities: see left panel of Figure~\ref{fig:exposure_times}. Clusters without identified X-ray cavities typically have <$10^{5}$ photon counts within their central regions and, in fact, below <$10^{4}$ photon counts, no Prunier+2025 observed cluster exhibits X-ray cavities. A notable exception is the well-studied galaxy cluster merging Coma, which has high exposure times but no identified X-ray cavities.

In the TNG-Cluster matched sample, similar, albeit not identical, trends are in place. The median photon counts for clusters without X-ray cavities is consistent with the observational sample (vertical solid lines). However, there is no absolutely sharp cut-off below the $10^{4}$ value of the observational data (right panel); one cluster has lower photon counts but still shows detected X-ray cavities. Overall, the photon count distributions across which X-ray cavities are detected or not are broader in TNG-Cluster than in the Prunier+2025. Whether these differences are significant and whether they can tell us something about the simulation needs to be investigated further.

To explore the impact of photon counts on X-ray cavity detection we make use of the systematic search for X-ray cavities across all 352 TNG-Cluster systems made in \cite{Prunier2024}, where all mock Chandra observations were standardized to a high (200 ks) exposure time, and a fixed distance of 200 Mpc, reaching a photon count >$10^{5}$ in each cluster. Restricting to the TNG-Cluster matched sample of this study, we obtain a 40 per cent detection statistic (42 out of 105 clusters), which is 5 per cent higher than in our current analysis at lower exposure times. Therefore, the increased number of counts in the central regions of the systems in \cite{Prunier2024} likely improves the visibility of X-ray cavities, thereby enhancing the detection rate. 

\cite{2014Panagoulia_vol}, achieving nearly identical detections to ours (except for Mk3W), also emphasized that the detection of X-ray cavities depends heavily on the number of counts in the source's core, reflecting the data quality. This finding aligns with \cite{2010Dong}, who highlighted that sufficient X-ray photon counts are essential for detecting X-ray cavities.

\vspace{-0.4cm}
\section{On different assumptions to compute X-ray cavity power in TNG-Cluster} \label{annex:relat}

In the TNG model, X-ray cavities are filled with non-relativistic gas cells following the ideal equation of state. We recompute the TNG X-ray cavity power using the non-relativistic gas formula P$_\text{cav}$ = 2.5p$_\text{th}$V and keep the relativistic assumption for the observations. The scaling relation remains consistent with observational data despite a systematic shift of $\sim 0.2$ dex downward in P$_\text{cav}$. With this more realistic assumption for the gas inside TNG X-ray cavities, 39 X-ray cavities fall below the 1:1 correlation, instead of 30 in Figure~\ref{fig:Pcav_Lcool}.

\begin{figure}
    \hspace{-0.4cm}
    \includegraphics[width=0.5\textwidth]{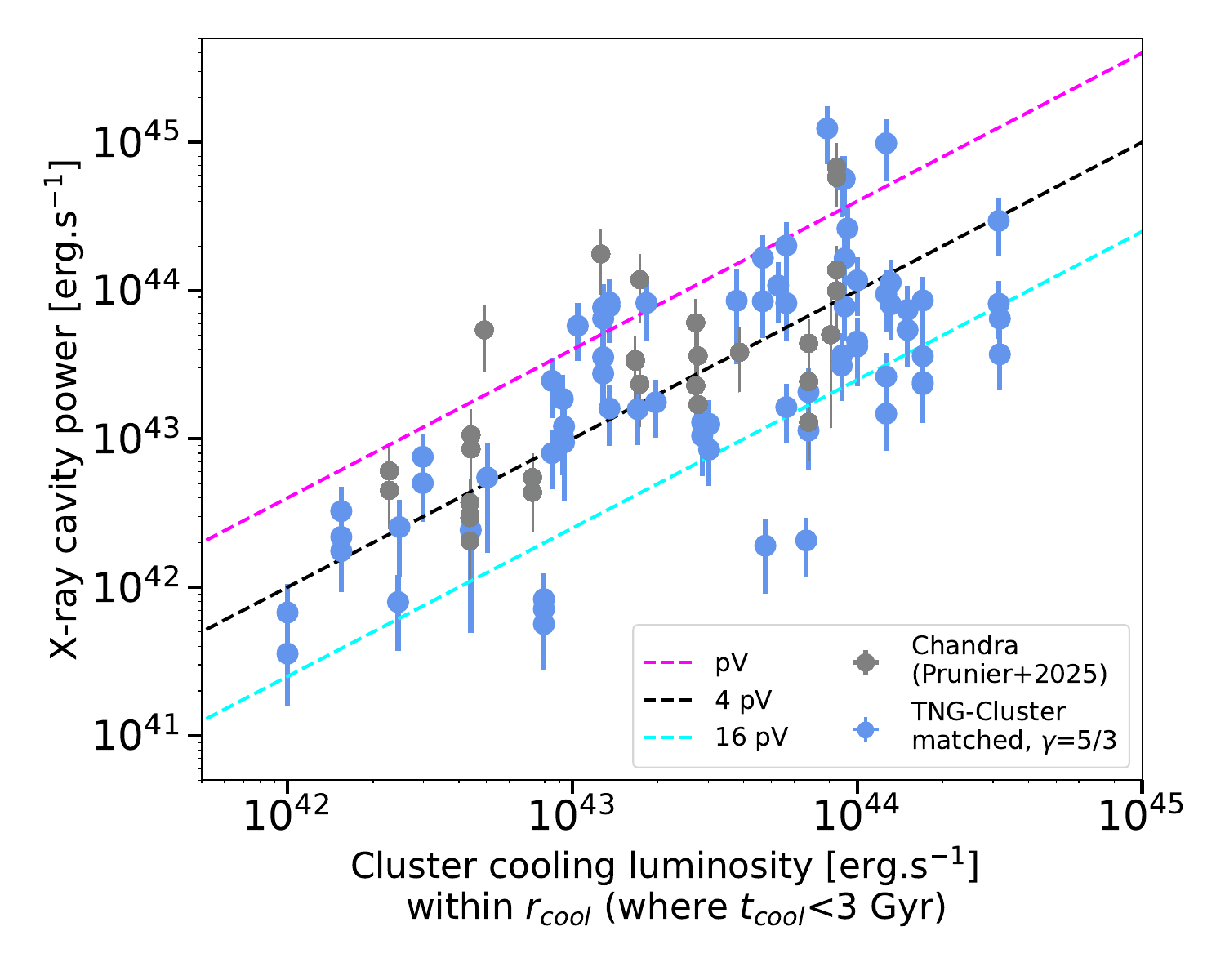}
    \caption{X-ray cavity power vs. cooling luminosity for the TNG-Cluster systems, assuming a non-relativistic equation of state with $\gamma$=5/3 for the gas filling the TNG-X-ray cavities and keeping $\gamma$=4/3 for the Prunier+2025 X-ray cavities. Detailed assumptions on the nature of the cavity-filling gas do not qualitatively change the conclusions from the comparison between samples.}
    \label{fig:2.5pV_gamma_nn_relat}
\end{figure}

\newpage

\section{Collection of measurements} \label{annex:coll_data}

\begin{table*}
    \centering
    \caption{List of the 35 sources in our observational volume-limited sample and their properties. (1) Source name as in the REFLEX or NORAS catalog, (2) alternative source name (sometimes referring to the central dominant galaxy), (3) and (4) source right ascension and declination in epoch 2000 coordinates, (5) source redshift, (6) source X-ray luminosity in the soft band, (7) central cooling time in Gyr, (8) total exposure time, (9) Chandra ObsIDs used in the analysis}
    \begin{tabular}{lllllllll}
        \hline
        \hline
        \makecell{Source name \\ (1)} & \makecell{Alt. source \\ name \\ (2)} & \makecell{RA \\ (2000) \\ (3)} & \makecell{DEC \\ (2000) \\ (4) } & \makecell{Redshift \\ (5)} & \makecell{L\(_X\) \\ (×10\(^{44}\)\\  erg s\(^{-1}\)) \\ (6)} & \makecell{t\(_{\mathrm{cool}}\) \\ (Gyr) \\ (7)} & \makecell{Exposure \\ Time (ks) \\ (8)} & \makecell{ObsIDs \\ (9)} \\
        \hline
        \hline
        \\
        RXCJ1248.7-4118 & NGC 4696 & 192.2000 & -41.3078 & 0.0114 & 0.721 & $0.09^{+0.004}_{-0.003}$ & 198.0 & \makecell{4954, 4955, \\  504, 5310.} \\
        RXCJ0152.7+3609 & A 0262 & 28.1948 & 36.1513 & 0.0163 & 0.81 & $0.13 \pm 0.02$ & 138.6 & \makecell{2215, 7921.} \\
        RXCJ1407.4-2700 & A 3581 & 211.8667 & -27.0153 & 0.0230 & 0.316 & $0.29^{+0.02}_{-0.04}$ & 84.5 & \makecell{12884.} \\
        RXCJ1715.3+5724 & NGC 6338 & 258.8414 & 57.4074 & 0.0276 & 0.49 & $0.22^{+0.05}_{-0.02}$ & 45.9 & \makecell{4194.} \\
        RXCJ1628.6+3932 & A 2199 & 247.1582 & 39.5487 & 0.0299 & 3.77 & $0.48^{+0.06}_{-0.09}$ & 119.7 & \makecell{10748, 10803, \\ 10804, 10805.} \\
        RXCJ0433.6-1315 & A 0496 & 68.4083 & -13.2592 & 0.0326 & 1.746 & $0.33^{+0.04}_{-0.09}$ & 53.2 & \makecell{4976.} \\
        RXCJ0338.6+0958 & 2A0335+096 & 54.6699 & 9.9745 & 0.0347 & 4.21 & $0.16 \pm 0.02$ & 102.0 & \makecell{919, 7939, 9792.} \\
        \\
        RXCJ1516.7+0701 & A 2052 & 229.1834 & 7.0185 & 0.0353 & 2.58 & $0.21^{+0.06}_{-0.11}$ & 599.7 & \makecell{5807, 10477, \\ 10478, 10479,\\ 10480, 10879, \\ 10914, 10916, \\ 10917.} \\
        \\
        RXCJ1521.8+0742 & MKW3s & 230.4583 & 7.7088 & 0.0442 & 2.70 & $0.09 \pm 0.03$ & 57.3 & \makecell{900.} \\
        RXCJ0041.8-0918 & A 85 & 10.4583 & -9.2019 & 0.0555 & 5.293 & $0.40^{+0.11}_{-0.03}$ & 38.4 & \makecell{904.} \\
        RXCJ2313.9-4244 & AS 1101 & 348.4958 & -42.7339 & 0.0564 & 1.738 & $0.34^{+0.08}_{-0.05}$ & 96.5 & \makecell{11758.} \\
        RXCJ0102.7-2152 & A 0133 & 15.6750 & -21.8736 & 0.0569 & 1.439 & $0.18 \pm 0.03$ & 102.9 & \makecell{2203, 9897.} \\
        RXCJ1454.5+1838 & A 1991 & 223.6309 & 18.6420 & 0.0586 & 1.46 & $0.23 \pm 0.01$ & 37.3 & \makecell{3193.} \\
        RXCJ1348.8+2635 & A 1795 & 207.2207 & 26.5956 & 0.0622 & 9.93 & $0.27^{+0.12}_{-0.06}$ & 37.1 & \makecell{493, 494.} \\
        RXCJ2357.0-3445 & A 4059 & 359.2583 & -34.7606 & 0.0475 & 1.698 & $0.65^{+0.05}_{-0.06}$ & 107.1 & \makecell{897, 5785.} \\
        RXCJ0918.1-1205 & Hydra A & 139.5292 & -12.0933 & 0.0539 & 2.659 & $0.53^{+0.06}_{-0.10}$ & 165.9 & \makecell{4969, 4970.} \\
        RXCJ0425.8-0833 & \makecell{RBS 0540} & 66.4625 & -8.5592 & 0.0397 & 1.008 & $0.33^{+0.14}_{-0.05}$ & 10.0 & \makecell{4183.} \\
        RXCJ1252.5-3116 & - & 193.1417 & -31.2678 & 0.0535 & 0.861 & $0.32^{+0.12}_{-0.04}$ & 9.9 & \makecell{12275.} \\
        
        RXCJ2205.6-0535 & A 2415 & 331.4417 & -5.5933 & 0.0582 & 1.135 & $0.48^{+0.13}_{-0.04}$ & 9.9 & \makecell{12272.} \\
        RXCJ1604.9+2355 & AWM4 & 241.2377 & 23.9206 & 0.0326 & 0.55 & $0.70^{+0.16}_{-0.06}$ & 74.3 & \makecell{9423.} \\
        RXCJ2336.5+2108 & A 2626 & 354.1262 & 21.1424 & 0.0565 & 1.55 & $0.60^{+0.4}_{-0.3}$ & 24.4 & \makecell{3192.} \\
        RXCJ1303.7+1916 & A 1668 & 195.9398 & 19.2715 & 0.0643 & 1.79 & $0.80 \pm 0.1$ & 9.9 & \makecell{12877.} \\
        RXCJ2347.7-2808 & A 4038 & 356.9292 & -28.1414 & 0.0300 & 1.014 & $1.07^{+0.4}_{-0.5}$ & 39.1 & \makecell{4992, 4188.} \\
        RXCJ1347.4-3250 & A 3571 & 206.8667 & -32.8497 & 0.0391 & 3.996 & $1.07^{+3.0}_{-0.4}$ & 33.9 & \makecell{4203.} \\
        RXCJ1523.0+0836 & A 2063 & 230.7724 & 8.6025 & 0.0355 & 1.94 & $1.88^{+2.1}_{-0.8}$ & 49.6 & \makecell{5795,4187, \\6262, 6263.} \\
        RXCJ0721.3+5547 & A 0576 & 110.3426 & 55.7864 & 0.0381 & 1.41 & $1.51^{+3.4}_{-0.6}$ & 29.1 & \makecell{3289.} \\
        RXCJ1254.6-2913 & A 3528S & 193.6708 & -29.2233 & 0.0544 & 1.064 & $2.68^{+0.8}_{-0.5}$ & 10.0 & \makecell{10746.} \\
        RXCJ2350.8+0609 & A2665 & 357.7109 & 6.1611 & 0.0562 & 1.81 & $4 \pm 1.0$ & 9.9 & \makecell{12280.} \\
        RXCJ0011.3-2851 & A2734 & 00 11 20.7 & -28 51 18 & 0.0620 & 1.089 & $5 \pm 1.2$ & 19.9 & \makecell{5797.} \\
        RXCJ0342.8-5338 & A3158 & 03 42 53.9 & -53 38 07 & 0.0590 & 2.951 & $8^{+3.8}_{-8.1}$ & 55.7 & \makecell{3201, 3712.} \\ 
        RXCJ2012.5-5649 & A3667 & 20 12 30.5 & -56 49 55 & 0.0550 & 5.081 & $6^{+2.9}_{-5.0}$ & 433.0 & \makecell{5751, 5752,  \\ 5753, 6292, \\ 6295, 6296.} \\ 
        RXCJ0246.0+3653 & A0376 & 41.5108 & 36.8879 & 0.0488 & 1.36 & $8 \pm 2.0$ & 10.4 & \makecell{12277.} \\ 
        RXCJ0909.1-0939 & A0754 & 09 09 08.4 & -09 39 58 & 0.0542 & 3.879 & $7^{+4.0}_{-5.1}$ & 43.6 & \makecell{577.} \\ 
        \\
        RXCJ1259.7+2756 & Coma & 94.9294 & 27.9386 & 0.0231 & 7.01 & $6^{+3.2}_{-5.9}$ & 535.4 & \makecell{13993, 13994,  \\ 13995, 13996,  \\ 14406, 14410, \\ 14411, 14415, \\ 9714.} \\
        \\
        RXCJ1440.6+0328 & MKW 8 & 220.1592 & 3.4765 & 0.0263 & 0.35 & $8 \pm 1.2$ & 23.1 & \makecell{4942.} \\ 
        \\
    \end{tabular}
    \label{tab:obs_sample}
\end{table*}

\begin{table*}
\caption{List of the properties of the X-ray cavities of the clusters in the observational sample. (1) source name, (2) X-ray cavity location, (3) and (4) X-ray cavity size along and perpendicular to the AGN, (5) distance of X-ray cavity from source center, (6) sound crossing time, (7) cooling luminosity within the cooling radius, (8) X-ray cavity power}
\centering
\begin{tabular}{llllllll}
\makecell{Source \\ Name \\ (1)} & \makecell{X-ray cavity \\ Location \\ (2)} & \makecell{X-ray cavity $r_a$ \\ (kpc) \\ (3)} & \makecell{X-ray cavity $r_b$ \\ (kpc) \\ (4)} & \makecell{Distance from \\ Core (kpc) \\ (5)} & \makecell{t$_{\text{rise}}$ \\ (Myr) \\ (6)} & \makecell{L$_{\text{cool}} \times 10^{42}$ \\ (erg s$^{-1}$) \\ (7)} & \makecell{P$_{\text{cav}}$ $\times 10^{42}$ \\ (erg s$^{-1}$) \\ (8)} \\ \hline
 \\
\multirow{3}{*}{2A0335+096} & NW & 5.5 & 8.3 & 46.5 & 51.3 & \multirow{3}{*}{$67.5 \pm 0.7$} & $44.1 \pm 19.7$ \\
 & NW & 6.2 & 11.8 & 46.6 & 56.8 &  & $24.4 \pm 12.0$ \\
 & N & 15.9 & 5.3 & 52.1 & 89.5 &  & $12.9 \pm 5.8$ \\ 
  \\
\multirow{2}{*}{NGC4696} & E & 1.6 & 1.0 & 3.7 & 6.4 & \multirow{2}{*}{$7.3 \pm 0.1$} & $5.5 \pm 2.5$ \\
 & W & 1.5 & 0.9 & 4.0 & 6.6 &  & $4.4 \pm 2.0$ \\ 
 \\
\multirow{2}{*}{A0262} & W & 4.4 & 2.9 & 6.7 & 13.3 & \multirow{2}{*}{$22.8 \pm 0.2$} & $4.5 \pm 2.0$ \\
 & E & 5.9 & 3.9 & 8.9 & 11.5 &  & $6.1 \pm 2.7$ \\ 
 \\
\multirow{4}{*}{A3581} & E & 2.3 & 2.0 & 4.4 & 9.2 & \multirow{4}{*}{$43.7 \pm 0.4$} & $3.1 \pm 1.4$ \\
 & W & 3.0 & 1.5 & 5.2 & 11.0 &  & $2.1 \pm 0.9$ \\
 & E & 2.8 & 4.0 & 12.8 & 21.0 &  & $2.9 \pm 1.4$ \\
 & W & 3.5 & 10.6 & 25.1 & 33.6 &  & $3.7 \pm 1.7$ \\
 \\
\multirow{2}{*}{NGC6338} & SW & 2.8 & 3.2 & 6.2 & 20.2 & \multirow{2}{*}{$44.1 \pm 0.4$} & $8.6 \pm 4.2$ \\
 & NE & 4.2 & 2.6 & 4.8 & 23.1 &  & $10.6 \pm 5.2$ \\ 
 \\
\multirow{2}{*}{A2199} & E & 3.8 & 8.7 & 17.0 & 24.0 & \multirow{2}{*}{$276.3 \pm 2.8$} & $17.1 \pm 7.9$ \\
 & W & 7.2 & 11.8 & 24.5 & 31.9 &  & $36.2 \pm 17.7$ \\
 \\
\multirow{2}{*}{A0496} & N & 6.1 & 7.6 & 17.4 & 19.5 & \multirow{2}{*}{$271.1 \pm 2.7$} & $22.8 \pm 10.3$ \\
 & S & 2.8 & 5.4 & 6.4 & 10.9 &  & $60.7 \pm 27.5$ \\ 
 \\
\multirow{4}{*}{A2052} & N & 8.7 & 5.4 & 7.7 & - & \multirow{4}{*}{-} & - \\
 & N & 5.7 & 2.5 & 20.6 &- &  & - \\
 & S & 16.0 & 4.1 & 6.4 & -&  & - \\
 & S & 7.4 & 5.0 & 22.2 & - &  & - \\ 
 \\
\multirow{1}{*}{A85} & S & 9.0 & 3.7 & 13.3 & 16.2 & $80.7 \pm 1.6$ & $50.3 \pm 38.5$ \\ 
 \\
\multirow{1}{*}{AS1101} & SE & 7.4 & 4.0 & 14.8 & 3.4 & $38.6 \pm 0.4$ & $38.3 \pm 17.7$ \\ 
 \\
\multirow{2}{*}{A0133} & NW & 11.0 & 7.4 & 41.7 & 28.8 & \multirow{2}{*}{$16.6 \pm 0.2$} & $34.3 \pm 15.5$ \\
 & SW & 11.9 & 11.4 & 33.4 & 32.9 &  & $33.4 \pm 15.0$ \\ 
 \\
\multirow{2}{*}{A1991} & N & 7.6 & 6.0 & 17.8 & 21.9 & \multirow{2}{*}{$17.3 \pm 0.2$} & $118.2 \pm 57.2$ \\
 & S & 6.9 & 5.9 & 12.0 & 18.1 &  & $23.4 \pm 11.3$ \\ 
 \\
\multirow{1}{*}{A1795} & N & 9.7 & 4.4 & 11.4 & 15.2 & $4.9 \pm 0.1$ & $54.3 \pm 25.9$ \\
 \\
\multirow{1}{*}{A4059} & N & 16.1 & 14.8 & 27.1 & 28.8 & $12.6 \pm 0.1$ & $176.5 \pm 82.4$ \\ 
 \\
\multirow{4}{*}{Hydra A} & NE & 9.4 & 16.7 & 21.6 & 31.5 & \multirow{4}{*}{$84.7 \pm 0.8$} & $99.7 \pm 45.4$ \\
 & NE & 38.1 & 39.8 & 77.2 & 98.6 &  & $679.1 \pm 310.1$ \\
 & SW & 11.5 & 16.3 & 28.7 & 32.8 &  & $137.8 \pm 62.8$ \\
 & SW & 34.1 & 37.3 & 104.3 & 132.1 &  & $579.7 \pm 50.1$ \\ 
 \hline
\end{tabular}
\label{tab:results_pana}
\end{table*}

\begin{table*}
\centering
\caption{List of the properties of the X-ray cavities of the clusters in the TNG-Cluster sample. (1) TNG-Cluster system ID, (2) and (3) X-ray cavity size along and perpendicular to the AGN, (4) distance of X-ray cavity from source center, (5) sound crossing time, (6) cooling luminosity within the cooling radius, (7) X-ray cavity power}
\begin{tabular}{llllllll}
\makecell{Source \\ Name \\ (1)}   & \makecell{X-ray cavity $r_a$ \\ (kpc) \\ (2)} & \makecell{X-ray cavity $r_b$ \\ (kpc) \\ (3)} & \makecell{Distance from \\ Core (kpc) \\ (4)} & \makecell{t$_{\text{rise}}$ \\ (Myr) \\ (5)} & \makecell{L$_{\text{cool}} \times 10^{42}$ \\ (erg s$^{-1}$) \\ (6)} & \makecell{P$_{\text{cav}}$ $\times 10^{42}$ \\ (erg s$^{-1}$) \\ (7)} \\ 
\hline
\\
\multirow{3}{*}{10545631} & 7.0 & 11.0 & 25.8 & 25.1 & \multirow{3}{*}{$56.4 \pm 0.6$} & $131.6 \pm 59.2$ \\
& 9.0 & 8.8 & 58.2 & 61.6 & & $26.2 \pm 11.3$ \\
& 18.9 & 9.9 & 37.1 & 36.0 & & $321.8 \pm 141.9$ \\
\\
\multirow{3}{*}{11659115} & 18.6 & 27.8 & 30.7 & 35.4 & \multirow{3}{*}{$88.2 \pm 0.9$} & $894.5 \pm 392.6$ \\
& 13.8 & 16.1 & 80.9 & 93.1 & & $58.3 \pm 24.4$ \\
& 8.5 & 12.0 & 45.2 & 51.2 & & $49.8 \pm 21.1$ \\
\\
\multirow{3}{*}{11767506} & 12.2 & 5.8 & 26.0 & 24.7 & \multirow{3}{*}{$18.2 \pm 0.2$} & $132.0 \pm 58.6$ \\
& 25.4 & 14.4 & 57.9 & 48.8 & & $419.3 \pm 174.8$ \\
\\
\multirow{3}{*}{12534413} & 19.2 & 7.3 & 23.4 & 21.6 & \multirow{3}{*}{$90.4 \pm 0.9$} & $905.6 \pm 394.8$ \\
& 8.3 & 14.5 & 39.0 & 36.7 & & $124.6 \pm 52.2$ \\
& 23.9 & 9.6 & 59.0 & 58.5 & & $263.9 \pm 112.8$ \\
\\
\multirow{2}{*}{12582631} & 6.9 & 17.3 & 22.8 & 24.6 & \multirow{2}{*}{$131.1 \pm 1.3$} & $181.2 \pm 75.7$ \\
& 6.5 & 13.3 & 21.9 & 23.6 & & $129.1 \pm 54.5$ \\
\\
\multirow{4}{*}{13634459} & 3.8 & 5.2 & 12.2 & 13.1 & \multirow{4}{*}{$169.9 \pm 1.7$} & $57.7 \pm 27.2$ \\
& 4.8 & 5.9 & 19.8 & 22.5 & & $37.3 \pm 16.8$ \\
& 10.2 & 5.3 & 37.0 & 40.9 & & $38.7 \pm 16.4$ \\
& 6.4 & 9.6 & 19.0 & 21.7 & & $137.2 \pm 60.2$ \\
\\
\multirow{2}{*}{13686751} & 6.1 & 11.3 & 38.5 & 52.0 & \multirow{2}{*}{$9.3 \pm 0.1$} & $16.8 \pm 7.6$ \\
& 5.7 & 16.5 & 51.0 & 62.2 & & $29.7 \pm 13.6$ \\
\\
\multirow{3}{*}{14038844} & 12.2 & 14.9 & 40.8 & 42.1 & \multirow{3}{*}{$100.0 \pm 1.0$} & $187.3 \pm 79.7$ \\
& 4.4 & 9.5 & 27.5 & 28.2 & & $72.5 \pm 33.1$ \\
& 5.1 & 10.4 & 36.0 & 35.0 & & $67.1 \pm 30.7$ \\
\\
\multirow{2}{*}{15304907} & 5.1 & 6.0 & 22.5 & 24.0 & \multirow{2}{*}{$67.4 \pm 0.7$} & $33.1 \pm 14.8$ \\
& 4.3 & 6.9 & 34.3 & 39.6 & & $18.2 \pm 8.3$ \\
\\
\multirow{1}{*}{15465516} & 4.6 & 9.8 & 58.6 & 82.8 & $66.2 \pm 0.7$ & $3.3 \pm 1.4$ \\
\\
\multirow{2}{*}{16731208} & 10.9 & 13.5 & 28.6 & 26.5 & \multirow{2}{*}{$46.6 \pm 0.5$} & $265.9 \pm 111.7$ \\
& 9.1 & 7.1 & 22.9 & 21.6 & & $135.5 \pm 58.9$ \\
\\
\multirow{1}{*}{16921354} & 11.6 & 10.3 & 10.0 & 12.8 & $78.5 \pm 0.8$ & $1980.7 \pm 836.1$ \\
\\
\multirow{2}{*}{16951023} & 18.0 & 10.3 & 31.0 & 38.2 & \multirow{2}{*}{$313.8 \pm 3.1$} & $472.3 \pm 200.5$ \\
& 14.6 & 8.1 & 32.6 & 40.1 & & $130.4 \pm 54.5$ \\
\\
\multirow{1}{*}{17222002} & 4.3 & 10.7 & 18.7 & 20.9 & $19.6 \pm 0.2$ & $28.2 \pm 12.0$ \\
\\
\multirow{2}{*}{17315353} & 6.2 & 2.8 & 17.8 & 58.3 & \multirow{2}{*}{$1.0 \pm 0.2$} & $1.1 \pm 0.6$ \\
&  4.2 & 5.0 & 16.1 & 59.8 & & $0.6 \pm 0.1$
\\
\\
\multirow{2}{*}{17382530} & 5.6 & 9.0 & 19.9 & 26.0 & \multirow{2}{*}{$316.6 \pm 3.2$} & $59.5 \pm 25.5$ \\
& 5.9 & 7.9 & 15.8 & 18.6 & & $103.4 \pm 44.7$ \\
\\
\multirow{3}{*}{17534833} & 7.6 & 14.7 & 30.7 & 33.8 & \multirow{3}{*}{$13.5 \pm 0.1$} & $125.5 \pm 54.7$ \\
& 6.4 & 17.6 & 34.9 & 38.4 & & $132.9 \pm 58.7$ \\
& 7.2 & 12.3 & 61.8 & 68.8 & & $25.7 \pm 11.2$ \\
\\
\multirow{1}{*}{17560743} & 16.2 & 15.7 & 58.5 & 61.8 & $10.4 \pm 0.1$ & $92.6 \pm 38.6$ \\
\\
\multirow{1}{*}{17613539} & 10.2 & 10.9 & 10.2 & 17.5 & $37.7 \pm 0.4$ & $136.6 \pm 85.4$ \\
\\
\end{tabular}
\label{tab:results_tng}
\end{table*}

\begin{table*}
\centering
\begin{tabular}{llllllll}
\makecell{Source \\ Name \\ (1)}   & \makecell{X-ray cavity $r_a$ \\ (kpc) \\ (2)} & \makecell{X-ray cavity $r_b$ \\ (kpc) \\ (3)} & \makecell{Distance from \\ Core (kpc) \\ (4)} & \makecell{t$_{\text{rise}}$ \\ (Myr) \\ (5)} & \makecell{L$_{\text{cool}} \times 10^{42}$ \\ (erg s$^{-1}$) \\ (6)} & \makecell{P$_{\text{cav}}$ $\times 10^{42}$ \\ (erg s$^{-1}$) \\ (7)} \\ 
\\
\hline
\\
\multirow{1}{*}{17672515} & 5.3 & 5.5 & 18.1 & 31.5 & $2.5 \pm 0.0$ & $4.1 \pm 2.2$ \\
\\
\multirow{2}{*}{17704708} & 9.3 & 5.6 & 29.3 & 36.3 & \multirow{2}{*}{$9.3 \pm 0.1$} & $19.4 \pm 11.6$ \\
& 6.0 & 8.3 & 32.1 & 39.8 & & $15.2 \pm 9.0$ \\
\\
\multirow{3}{*}{17812828} & 2.6 & 2.6 & 12.5 & 19.5 & \multirow{3}{*}{$7.9 \pm 0.1$} & $1.3 \pm 0.7$ \\
& 2.0 & 1.8 & 9.2 & 13.5 & & $1.1 \pm 0.6$ \\
& 2.4 & 2.1 & 12.3 & 19.2 & & $0.9 \pm 0.5$ \\
\\
\multirow{1}{*}{17838922} & 11.5 & 8.9 & 20.1 & 20.6 & $52.8 \pm 0.5$ & $173.0 \pm 75.5$ \\
\\
\multirow{2}{*}{18083780} & 10.5 & 6.7 & 42.6 & 42.1 & \multirow{2}{*}{$150.2 \pm 1.5$} & $86.8 \pm 37.9$ \\
& 4.7 & 11.7 & 31.8 & 31.4 & & $118.4 \pm 53.4$ \\
\\
\multirow{1}{*}{18143062} & 10.0 & 6.2 & 45.8 & 63.2 & $5.0 \pm 0.1$ & $8.8 \pm 6.0$ \\
\\
\multirow{2}{*}{18512862} & 4.8 & 7.6 & 16.6 & 50.4 & \multirow{2}{*}{$4.4 \pm 0.0$} & $3.9 \pm 3.1$ \\
& 4.9 & 9.6 & 21.1 & 26.1 & & $20.6 \pm 8.8$ \\
\\
\multirow{2}{*}{18624090}  & 4.9 & 9.6 & 21.1 & 26.1 & \multirow{2}{*}{$28.6 \pm 0.3$} & $20.6 \pm 8.8$ \\
& 3.9 & 5.2 & 15.2 & 17.5 &  & $16.7 \pm 7.7$ \\
\\
\multirow{3}{*}{18809506} & 9.8 & 5.9 & 43.3 & 63.8 & \multirow{3}{*}{$1.5 \pm 0.0$} & $5.2 \pm 2.4$ \\
& 5.2 & 5.4 & 18.9 & 27.7 & & $3.5 \pm 1.6$ \\
& 4.4 & 8.9 & 22.9 & 40.1 & & $2.8 \pm 1.3$ \\
\\
\multirow{1}{*}{18948062} & 1.0 & 1.7 & 4.6 & 5.5 & $47.6 \pm 0.5$ & $3.0 \pm 1.6$ \\
\\
\multirow{4}{*}{19119113} & 15.2 & 5.2 & 27.4 & 29.6 & \multirow{4}{*}{$12.8 \pm 0.1$} & $103.3 \pm 46.1$ \\
& 16.4 & 5.6 & 29.0 & 31.4 & & $123.0 \pm 54.5$ \\
& 10.7 & 11.0 & 44.4 & 46.7 & & $56.9 \pm 24.1$ \\
& 11.4 & 21.1 & 106.0 & 122.7 & & $44.0 \pm 18.1$ \\
\\
\multirow{1}{*}{19230547} & 6.2 & 11.8 & 14.2 & 36.8 & $17.0 \pm 0.2$ & $25.4 \pm 10.8$ \\
\\
\multirow{2}{*}{19313671} & 6.3 & 11.0 & 21.4 & 31.4 & \multirow{2}{*}{$3.0 \pm 0.0$} & $12.1 \pm 5.2$ \\
& 4.8 & 4.9 & 14.3 & 19.9 & & $8.1 \pm 3.6$ \\
\\
\multirow{3}{*}{19357750} & 5.0 & 9.0 & 32.9 & 41.2 & \multirow{3}{*}{$30.2 \pm 0.3$} & $20.0 \pm 9.0$ \\
& 8.0 & 5.4 & 29.8 & 38.0 & & $20.2 \pm 9.1$ \\
& 10.0 & 4.8 & 29.0 & 37.0 & & $13.6 \pm 5.8$ \\
\\
\multirow{1}{*}{19372389} & 15.6 & 13.1 & 44.6 & 116.9 & $2.4 \pm 0.0$ & $1.3 \pm 0.7$ \\
\\
\multirow{2}{*}{5348819} & 11.8 & 6.5 & 31.4 & 40.1 & \multirow{2}{*}{$8.5 \pm 0.1$} & $39.4 \pm 17.2$ \\
& 10.7 & 7.7 & 65.5 & 78.8 & & $12.8 \pm 5.5$ \\
\\
\multirow{4}{*}{8629805} & 6.8 & 7.6 & 5.2 & 4.5 & \multirow{4}{*}{$126.4 \pm 1.3$} & $1576.9 \pm 702.7$ \\
& 16.0 & 5.7 & 31.0 & 37.6 & & $151.5 \pm 67.3$ \\
& 7.2 & 10.3 & 58.9 & 58.4 & & $23.6 \pm 10.4$ \\
& 3.8 & 6.2 & 11.8 & 15.1 & & $42.1 \pm 18.5$ \\
\\
\hline
\end{tabular}
\end{table*}

\bsp	
\label{lastpage}
\end{document}